\documentclass[twocolumn,showpacs,aps,tightenlines]{revtex4}
\usepackage{graphicx}
\usepackage{color}
\usepackage{psfrag}
\usepackage{dcolumn}
\usepackage{bm}
\usepackage{xspace}

\usepackage{amssymb}
\usepackage{amsmath}
\usepackage{amsfonts}
\usepackage{longtable}

\newcommand{\beq}{\begin{equation}}
\newcommand{\eeq}{\end{equation}}
\newcommand{\KMS}{\rm km\,s^{-1}}

\newcommand{\be}{\begin{equation}}
\newcommand{\ee}{\end{equation}}
\newcommand{\bea}{\begin{eqnarray}}
\newcommand{\eea}{\end{eqnarray}}
\newcommand{\bes}{\begin{subequations}}
\newcommand{\ees}{\end{subequations}}

\newcommand{\hn}{\hat{{n}}}
\newcommand{\hL}{\hat{{L}}}

\newcommand{\hangup}{{\it hangup }\xspace}
\newcommand{\ang}[2]{\{#1, #2\}}

\newcommand{\hl}{\hat{{\lambda}}}
\newcommand{\J}[1]{{\bf J}_{#1}}
\usepackage{bm}

\begin{document}

\title{Black hole binary remnant mass and spin: A new phenomenological
formula}
\author{Carlos O. Lousto}
\author{Yosef Zlochower}
\affiliation{Center for Computational Relativity and Gravitation,
School of Mathematical Sciences,
Rochester Institute of Technology, 85 Lomb Memorial Drive, Rochester,
 New York 14623}

\date{\today}

\begin{abstract} 
We perform a set of 38 fully-nonlinear numerical
simulations of equal-mass black-hole binaries in a configuration where
the  two black-hole spins in the binary are equal in both magnitude
and direction, to study precession effects.  We vary the
initial direction of the total spin $\vec{S}$ with respect to the
orbital angular momentum $\vec{L}$, covering the 2 dimensional space
of orientation angles with 38 configurations consisting of 36
configurations distributed in the azimuthal angle $\phi$ and
polar angle $\mu=\cos\theta$, and two configurations on the poles.  In
all cases, we set the initial dimensionless black-hole spins to 0.8.
We observe that during the late-inspiral stage, the total angular
momentum of the system $\vec{J}$ remains within $5^\circ$ of its
original direction, with the largest changes in direction occurring
when the spins are nearly (but not exactly) counter-aligned with the orbital angular
momentum.  We also observe that the angle between $\vec{S}$ and
$\vec{L}$ is nearly conserved during the inspiral phase.  These two
dynamical properties allow us to propose a new phenomenological
formula for the final mass and spin of merged black holes in terms of
the individual masses and spins of the progenitor binary at far
separations.  We determine coefficients of this formula (in the
equal-mass limit) using a least-squared fit to the results of this
new set of 38 runs, an additional set of five new configurations with
spins aligned/counteraligned with the orbital angular momentum, and
over a hundred recent simulations. We find that our formulas reproduce
the remnant mass and spin of these simulations to within a relative
error of $2.5\%$.  We discuss the region of validity of this dynamical
picture for precessing unequal-mass binaries.  Finally, we perform a
statistical study to see the consequence of this new formula for
distributions of spin-magnitudes and remnant masses with applications
to black-hole-spin distributions and gravitational radiation in
cosmological scenarios involving several mergers.
\end{abstract}

 \maketitle

\section{Introduction and motivations}\label{sec:intro}

The 2005 breakthroughs in Numerical Relativity~\cite{Pretorius:2005gq,
Campanelli:2005dd, Baker:2005vv} allowed for accurate investigations
of the orbital dynamics of merging black-hole binaries (BHBs) in the
highly-nonlinear regime between the slow inspiral (which can be
modeled by post-Newtonian dynamics~\cite{Blanchet02}) and the post-merger phase
(which can be modeled by black-hole perturbation theory~\cite{Baker:2001sf}).
The current state-of-the-art simulations cover a broad range of astrophysical
parameters, including mass ratios as small as $q=1/100$
~\cite{Lousto:2010ut,
Sperhake:2011ik}, highly spinning black holes with near maximal
dimensionless spins
$\alpha\leq0.97$~\cite{Lovelace:2011nu}, 
and initial separations as large as $D=100M$~\cite{Lousto:2013oza}.

Some of the most remarkable findings were related to effects
of black-hole spin in the last stages of the merger.
 These include the
\hangup effect~\cite{Campanelli:2006uy}, which delays or expedites
the merger depending on whether the spins are aligned or
counter-aligned with the orbital angular momentum, and 
 recoil velocities
of several thousand km/s~\cite{Campanelli:2007cga, Gonzalez:2007hi, Healy:2008js}  
when the spins lie in opposite directions in the orbital plane.
A configuration \cite{Lousto:2011kp}
combining both effects leads to even larger recoils, up to 5000 km/s.

Those effects were discovered by simulating highly symmetric
configurations of equal-mass, nonprecessing BHBs. While it was
possible to evolve generic configurations soon after the breakthrough
\cite{Campanelli:2007ew, Campanelli:2008nk}, it has not been until
recently that systematic studies of precessing binaries have been
performed \cite{Hannam:2013pra, Lousto:2012gt, Hinder:2013oqa,
Mroue:2013xna}.  In a recent paper, Ref.~\cite{Lousto:2013vpa}, we
began the analysis and modeling of a set of runs of BHBs with
equal-mass and equal-spin BHs (i.e., in magnitude and direction).
These configurations, while symmetric, tend to maximize precession
effects.  Here, we provide extensive analysis of these and newer runs
to model the evolution of the orbital angular momentum $\vec{L}$, the
total spin of the system $\vec{S}$, and the radiation of the energy
and angular momentum of the binary system. 

The following are several highlights from our studies presented here.  
First, we find (see also
Ref.~\cite{Lousto:2013vpa}) that the angle between $\vec{L}$ and
$\vec{S}$ varies by $\lesssim 2^\circ$ during the inspiral.  This is a
remarkable feature in a fully dynamical precessing binary that has
immediate application to the modeling of the remnant since it connects
parameters of the binary at large separation with those at merger,
where most of the radiation takes place. While such a conservation of
direction is indeed predicted \cite{Apostolatos94,Racine:2008qv} by a
lower-order post-Newtonian analysis, the degree to which this
direction is conserved in the highly-dynamical (and highly-nonlinear)
merger regime is remarkable.  A second important observation for the
modeling of our set of runs  is the relatively small change in the
direction of the total angular momentum $\hat{J}$, despite the
significant continuous decrease in the magnitude of $\vec J$ as the
binary evolves.  Remarkably, in our runs we do not observe variations
larger than $5^\circ$ in the direction of $\hat J$. This observation
is consistent with the conclusions of~\cite{Barausse:2009uz}, which
where based on comparing initial and final directions of the angular
momentum for a diverse set of fully nonlinear BHB simulations.

Our ultimate goal is to develop and test an empirical formula for the remnant
mass and spin based on a {\it Taylor} expansion in terms of mass ratio
and spins, as was done in~\cite{Boyle:2007sz, Boyle:2007ru}, but with
different effective variables. 
Modeling the final spin and mass of the merger remnant is important
for the creation of gravitational wave templates
\cite{Hannam:2013oca} to assist
gravitational wave detectors, for studies of astrophysical scenarios
involving merger of galactic black holes, and for modeling cosmological
scenarios involving light versus heavy black holes seeds forming the
structure of the universe~\cite{Volonteri:2012yn, AmaroSeoane:2012zu}.
Importantly, our new formula reproduces the measured remnant mass and spin to
within $0.1\%$.

Here we will use the PN-inspired
spin variables  $\vec S = \vec S_1 + \vec S_2$ and $\vec \Delta = (M_1
+ M_2)(\vec S_2/M_2 - \vec S_1/M_1)$, where $\vec S_i$ and $M_i$ are
the spin and mass of BH $i$ (note that~\cite{Boyle:2007sz,
Boyle:2007ru} base their expansion on $\vec S_i$ directly).
 We found when  modeling
the recoil velocity in \cite{Lousto:2012gt} that $\vec S$ and $\vec
\Delta$ 
are better suited as expansion variables than $\vec S_1$ and $\vec
S_2$ because fewer terms are needed to get accurate fits.
Our new formula represents a more 
comprehensive framework for modeling remnant quantities than
our previous model~\cite{Lousto:2009mf}, which
 made use of analytic expressions for
the ISCO combined with numerical results available at the time
(See also \cite{Buonanno:2007sv}).
Other approaches to model remnants \cite{Tichy:2008du} also used
 lower-order spin expansions, and
successive reasonable assumptions about the dynamics of spinning binaries
\cite{Barausse:2009uz,Barausse:2012qz}.

In this paper, we use the convention
that $M_i$ is the BH mass as measured from a full numerical simulation
and $m_i$ is the BH mass according to post-Newtonian (PN) theory.
However, because it does not lead to confusion, and because we
use the symbol $M$ to denote the unit of mass, we use
$m$ to denote both the PN total mass $m = m_1+m_2$ and the full
numerical total BH mass $ m = M_1+M_2$.
Note that in terms of dimensionless quantities
\begin{eqnarray}
  \frac{\vec S}{m^2} &=& \frac{\vec \alpha_2 + q^2\vec \alpha_1}{(1+q)^2},\nonumber\\
  \frac{\vec \Delta}{m^2} &=& \frac{\vec \alpha_2 - q \vec
\alpha_1}{1+q},
\end{eqnarray}
where $q=M_1/M_2$ and $\vec \alpha_i = \vec S_i/M_i^2$.
In addition, we use $M_{\rm rem}$ and $\alpha_{\rm rem}$  to denote
the final remnant BH mass and dimensionless spin and $$
\delta {\cal M} = \frac{M_1 + M_2 - M_{\rm rem}}{M_1 + M_2}
$$ to denote the fraction of the initial mass radiated by the binary.
Finally, we denote unit vector in the direction of an arbitrary
vector $\hat v$ with a {\it hat} (i.e., $\hat v = \vec
v/ v$).

This paper is arranged as follows. In Sec.~\ref{sec:numsim}, 
we give an overview
of the numerical techniques used to generate initial data,
 evolve the BHBs, and measure the individual and remnant masses
and spins. We also describe the initial configurations of the new
BHB simulations presented here.
In Sec.~\ref{sec:results}, we present the results from our
BHB simulations described in Sec.~\ref{sec:numsim}.

In Sec.~\ref{sec:modeling}, we develop the framework for modeling
the final remnant mass and spin using PN-inspired variables
and symmetry considerations.
In Sec.~\ref{sec:fitting}, we use a least-squares fit of
the remnant mass and spin
of the new 38 precessing runs, 5 new
non precessing runs, and those of
Refs.~\cite{Lousto:2012su} and \cite{Lousto:2012gt} in the expansion
formulas derived in Sec.~\ref{sec:modeling}.
Our formulas are valid for the case of equal-mass binaries, and we
provide an ansatz for extending them to the unequal-mass case.

In Sec.~\ref{sec:discussion}, to assess some astrophysical 
consequences of these new formulas,
we find the distribution of spin-magnitudes after successive mergers
(see also Ref.~\cite{Lousto:2009ka}) of similar-mass BHBs
 assuming randomly-oriented
or partially-aligned spins, which are meant to model the effects of
dry mergers, as well as mergers where accretion partially aligns
the spins (note that we do not model the effects of accretion on the
spin magnitudes)
\cite{Bogdanovic:2007hp, Lousto:2012su, Dotti:2012qw}.

\section{Numerical Simulations}\label{sec:numsim}

\begin{figure}
  \includegraphics[width=0.9\columnwidth]{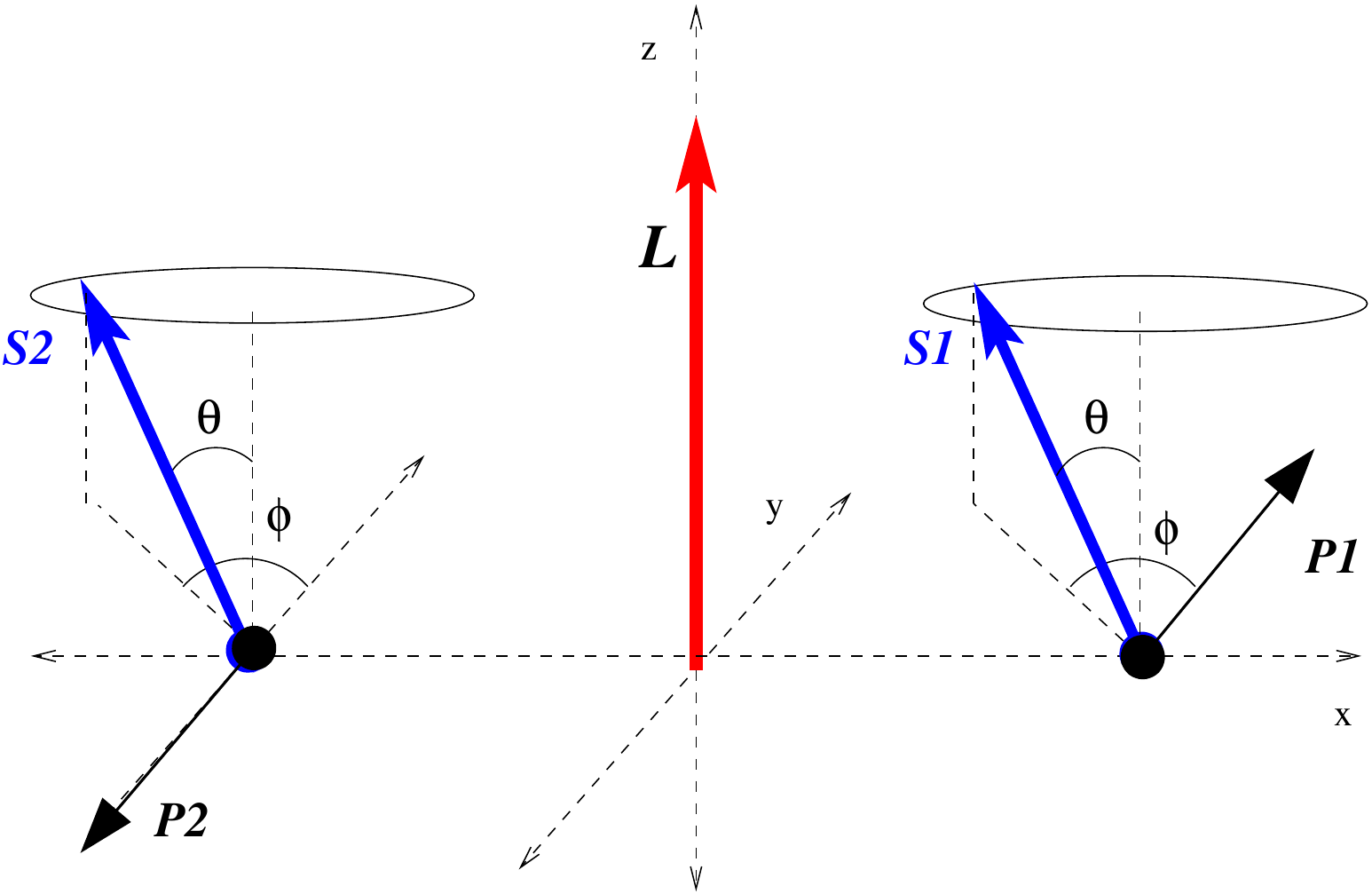}
  \caption{Configuration of the SP precessing binary.}
  \label{fig:SP}
\end{figure}

In order to single out the effects of precession we study
 BHBs consisting of two BHs with identical masses and spins
 (see Fig.~\ref{fig:SP}). Two such
configurations were first studied in~\cite{Campanelli:2006fy},
 soon after the breakthroughs in numerical relativity.
For a given spin-inclination angle, these
configurations maximize precession effects, at least in the PN regime,
because they maximize the total spin magnitude $S$. As seen in 
 Eq. (\ref{Sconstevol}) below, $\dot{\vec S}$ ($\vec S$ only changes
due to precession at  2PN order) is proportional to $S$.
In addition, the spin-orbit contribution to the radiation of angular
momentum [see Eq.~(3.28c) of Ref.~\cite{Kidder:1995zr}] is
maximized for these configurations as well, since, in the equal-mass case,
all terms in the formula for the radiation of angular
momentum are proportional to $S$.
Note that, when the two spins are fully aligned or fully counteraligned with the
orbital angular momentum, the BHB is nonprecessing~\cite{Campanelli:2006uy}.
Also note that these configurations \cite{Campanelli:2006fy} are symmetric under parity
[i.e.\ $(x,y,z)\to(-x,-y,-z)$]  and, consequently, there is no recoil
of the final remnants.

Our numerical simulations use the {\sc Cactus}/{\sc
EinsteinToolkit}~\cite{cactus_web,
einsteintoolkit} infrastructure and {\sc
Carpet}~\cite{Schnetter-etal-03b} mesh refinement driver, as well as
both private and open-source {\it thorns}.
We use the TwoPunctures thorn~\cite{Ansorg:2004ds} to generate initial
puncture data~\cite{Brandt97b} for the BHB simulations. These data are characterized by mass parameters $m_p$ (which
are not the horizon masses), as well as the momentum and spin,  of
each BH.  We evolve these BHB data sets using the {\sc
LazEv}~\cite{Zlochower:2005bj} implementation of the moving puncture
approach~\cite{Campanelli:2005dd, Baker:2005vv} with the conformal
function $W=\sqrt{\chi}=\exp(-2\phi)$ suggested by
Ref.~\cite{Marronetti:2007wz}.  For the runs presented here, we use
centered, eighth-order finite differencing in
space~\cite{Lousto:2007rj} and a fourth-order Runge - Kutta time
integrator. Note that we do not upwind the advection terms. 

We locate the apparent horizons using the {\sc AHFinderDirect}
thorn~\cite{Thornburg2003:AH-finding} and measure the horizon spin
using the isolated horizon (IH) algorithm detailed
in~\cite{Dreyer02a}.

The families of BHBs
considered here are equal-mass, equal-spins BHBs that are further
characterized by three parameters, the initial
orbital frequency, the polar inclination of the individual BH spins,
and the azimuthal orientation of the spins. Due to the \hangup
effect, we choose larger initial orbital frequencies for the BHBs
with small polar angles. For each polar angle $\theta$ we evolve a set
of six azimuthal angles $\phi=0^\circ, 30^\circ,\cdots,150^\circ$
(except for $\theta=0^\circ$ and $\theta=180^\circ$). We
choose initial polar angles 
$\theta=0^\circ, 48.2^\circ, 70.5^\circ, 90^\circ, 109.5^\circ,
131.8^\circ, 146.4^\circ, 180^\circ$ (the polar angles were chosen to
cover the sphere uniformly in $\cos\theta$, but with higher density
near $\theta=180^\circ$). We denote these configurations by
SPTHXXPHYY below, where XX is the polar angle (in degrees) and
YY is the azimuthal angle. We also performed a high-resolution run
of the SPTH0 configuration, which we denoted by SPTH0H below. For all
the SP configurations, the dimensionless spin of each BH was
$\alpha=0.8$. We chose this because it is relatively large, but can
also be evolved relatively efficiently (spins of 0.85 require
substantially more resources due to gauge effects).

In addition to the SP runs, we also evolve a set of equal-mass runs,
with one BH spin aligned with the orbital angular momentum, and the
other counter aligned. The two spins have the same magnitude. This
type of configuration has been studied before in, e.g.,
\cite{Herrmann:2007ac, Rezzolla:2007xa}.
Here we are interested in measuring the radiated energy and final
remnant spin. In previous modeling formulas, the effect of spin for
this configuration was usually ignored~\cite{Barausse:2012qz, Barausse:2009uz}.
We denote these
configuration below by UD0.XX, where 0.XX gives the magnitude of the
spin. We performed the UD simulations using 3 different resolutions
to monitor the accuracy of the results. We denote these resolutions
C, M, and F for coarse, medium, and fine, respectively. Here the
coarse resolution corresponds to the resolution used for all SP
configurations. The medium and fine results had 1.2 times and $1.2^2$
times the number of gridpoints in each direction. As seen in
Fig~\ref{fig:up_E_and_A}, for a spin of $\alpha=0.8$, we can expect a relative
error of $0.04\%$ in the final remnant spin and $0.1\%$ in the final
remnant mass by using only the coarse resolution. Note also that the
error increases dramatically when the spin is increased to
$\alpha=0.85$.

For the computation of the radiated angular momentum components, we
use formulas based on ``flux-linkages''~\cite{Winicour_AMGR} and
explicitly written in terms of $\psi_4$  in
\cite{Campanelli:1998jv, Lousto:2007mh}.

From the waveform extracted at a given radius $R$ (i.e.\ the
extraction sphere), we calculate
the rate with which angular momentum is radiated out of the extraction
sphere to infinity, which
we denote with the symbol
$$
  \frac{d}{dt}\left( \overrightarrow
{\delta J}_{\rm rad}(t) \frac{}{}\right),
$$ and
the net
angular momentum radiated through the extraction sphere from the
initial slice to time $t$, which is given by
$$
\overrightarrow
{\delta J}_{\rm rad}(t)
=
\int_0^{t} \frac{d}{d\tau}\left( \overrightarrow
{\delta J}_{\rm rad}(\tau) \frac{}{}\right)\, d\tau.
$$
We then define the total angular momentum at time $t$ to be
$$
  \vec J(t) = J_{\rm ADM} - \overrightarrow
{\delta J}_{\rm rad}(t),
$$
which is the total angular momentum contained on slice $t$ within the
extraction sphere.
Because of the finite propagation
speed of the radiation, when comparing local and radiative techniques
for the angular momentum, we translate the radiative quantities in
time
 by $R$,
which is a crude, but accurate enough, measure of the propagation
time.

For the current paper, it is important to measure the direction of the
orbital angular momentum. To do this we use a few different
techniques. First, we define
$\vec L$ via
\begin{eqnarray}
  \vec L(t) = \vec J(t) - \vec S(t), \nonumber \\
  \hat L(t) = \frac{\vec L(t)}{|\vec L(t)|},
\label{eq:LfromJ}
\end{eqnarray}
Here $\vec S(t) = \vec S_1(t) + \vec S_2 (t)$ is the sum of the spins
of the individual BHs, as measured
by the IH formalism
 (its direction is inferred from the zeros of the
approximate Killing field~\cite{Campanelli:2006fy}).

We can also define a purely coordinate based measure of the orbital
angular momentum direction using the coordinate trajectory $\vec r(t)$, given by
\begin{equation} \hat L_{\rm coord} = \frac{\vec{r}(t) \times
\dot{\vec{r}}(t)}{|\vec{r}(t) \times \dot{\vec{r}}(t)|}.
\end{equation} Note that while this equation resembles the Newtonian
definition of angular momentum, we apply it to fully nonlinear
numerical trajectories, thus including relativistic corrections. One
could also use an alternative definition in terms of the quasilocal linear
momentum $\vec{p}$, as  was proposed in \cite{Krishnan:2007pu}.

In the text below, when needed for clarity, we refer to $\hat{L}(t)$ as the
{\it radiation angular momentum} and $\hat{L}_{\rm coord}(t)$ as the
{\it coordinate angular momentum}.

We also compare $\hat{L}(t)$ and $\hat{L}_{\rm coord}(t)$
with the purely radiation-based measures of
the preferred asymptotic frames  of O'Shaughnessy et
al.~\cite{O'Shaughnessy:2012vm} and Boyle~\cite{Boyle:2013nka}.
Briefly, these two are based on the symmetric matrix $\langle
f|\J{(a}\J{b)}|f\rangle$ [${}_{(a\ b)}$ denotes symmetrization over the
two indices] and vector $\Im(\langle f|\J{a}| \partial_t
f\rangle)$, where $\J{a}$ is the angular momentum operator and $f$
is a waveform-related function, e.g.\ the strain $h$, the Bondi News
$N$, or the Newman-Penrose scalar $\psi_4$, and $\Im(z)$ denotes the
imaginary part of $z$. The angular momentum
operator for spin-weighted functions is given by~\cite{Ruiz:2007yx}
\footnote{We use the conventions
of~\cite{Boyle:2013nka}  to define $\J{}$ which differ from
the conventions of \cite{Ruiz:2007yx}
by a factor of $-i$.}.
\begin{eqnarray}
  \J{x} &=& i \sin \phi \partial_\theta + \cos\phi(i\cot\theta
\partial_\phi + s\csc\theta),\\
  \J{y} &=& -i\cos\phi \partial_\theta +\sin\phi(i\cot\theta \partial_\phi
+ s \csc \theta),\\
  \J{z} &=& -i \partial_\phi,
\end{eqnarray}
where $s$ is the spin-weight. Note that the radiated angular momentum
is given by~\cite{Lousto:2007mh, Ruiz:2007yx}
$$
  \frac{d}{dt}\left( \overrightarrow
{\delta J}_{\rm rad}(t) \frac{}{}\right)=
\lim_{r\to\infty}
  \frac{r^2}{16 \pi} \Im{\langle h|\J{a}| \partial_t
h\rangle}.$$
 O'Shaughnessy et al.
define the preferred frame to be one with $z$ axis aligned with the
eigenvector corresponding to the largest eigenvalue of  $\langle
f|\J{(a}\J{b)}|f\rangle$ and Boyle defines it to be the frame aligned
with the vector $\langle f|\J{(a}\J{b)}|f\rangle^{-1} (\Im{\langle
f|\J{a}|\partial_t f\rangle})$. For this work we explore using the
choices $f=\psi_4$, the Bondi News $N$, and the strain $h$ (note that
Boyle defines $f=h$ and with this choice is able to recover the
post-Newtonian $\vec L$). We denote the directions defined by
O'Shaughnessy et al. by $\vec O_{X}$, where $X$ is $\psi$,
$N$, or $h$, corresponding to using $f=\psi_4$, $f=N$, and $f=h$,
respectively. Similarly, we denote the preferred direction of Boyle
by $\vec B_{X}$, where $X$ is $N$ or $h$, corresponding to $f=N$ and
$f=h$.

\section{Results}\label{sec:results}
\subsection{Results from the SP configurations}
Here we present results relating to the precession of the SP
configurations. Detailed tables with the initial data parameters,
radiated energy-momentum, and remnant properties can be found in
Appendix~\ref{app:SPResults}. 

For all the SP configurations, we find that, while $\vec J$ changes in magnitude by more than 30\%, the direction
of $\vec J$ is largely unchanged during the entire simulation.
 We
thus only observe {\it simple} precession in these equal-mass BHBs.
For convenience, we introduce the notation
$\{\vec A\cdot \vec B\} = \cos^{-1}(\hat A\cdot\hat B)$ for the angle
between two vectors.
 As
seen in Table~\ref{tab:Jprec} and Fig~\ref{fig:Jprec}, the net
rotation
of $\hat J$ (i.e.\ the maximum of $\ang{\vec J(t)}{\vec J(0)}$)
is under $5^\circ$. The table also shows how far
$\overrightarrow {\delta J}_{\rm rad}(t)$ deviates from $\hat J$
($\ang{\overrightarrow {\delta J}_{\rm rad}(t)}{\vec J(t)}_{\rm max}$)
for
selected runs.

\begin{figure}
  \includegraphics[width=\columnwidth]{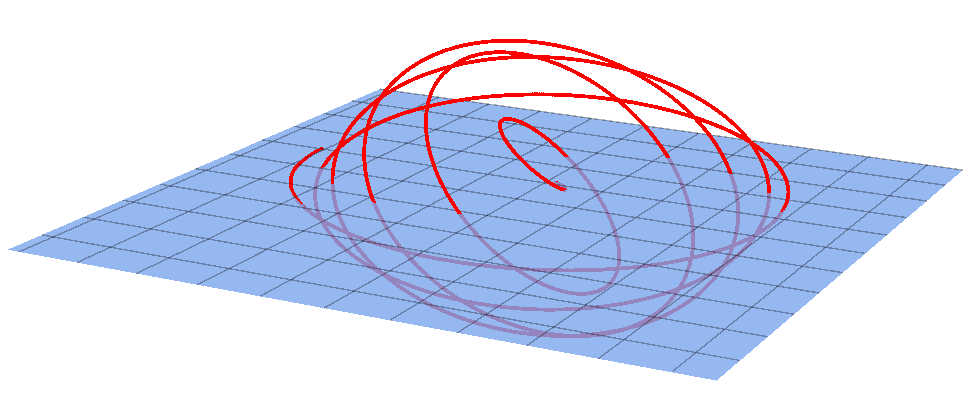}
  \caption{The coordinate trajectory for a typical SP configuration.}
 \label{fig:trackn}
\end{figure}

\begin{figure}
  \begin{tabular}{l|r}
  \includegraphics[width=0.49\columnwidth]{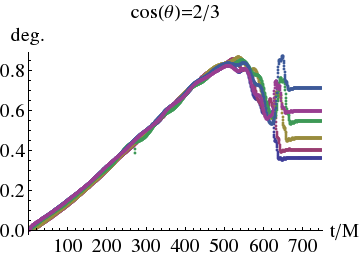} &
  \includegraphics[width=0.49\columnwidth]{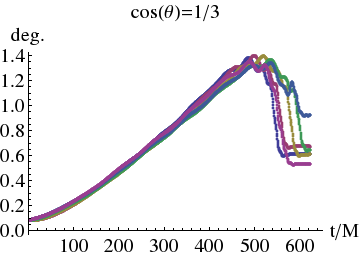} \\
\includegraphics[width=0.49\columnwidth]{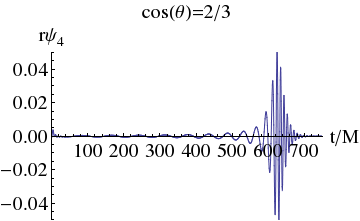} &
  \includegraphics[width=0.49\columnwidth]{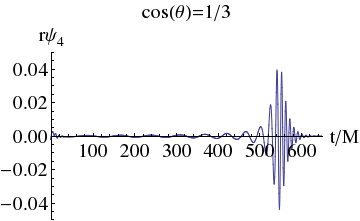}\\
\hline
  \includegraphics[width=0.49\columnwidth]{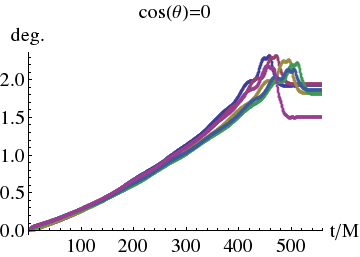} &
  \includegraphics[width=0.49\columnwidth]{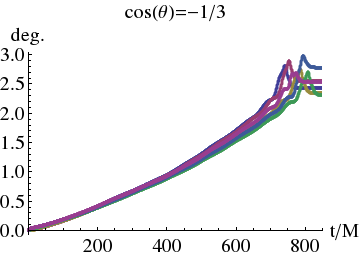} \\
  \includegraphics[width=0.49\columnwidth]{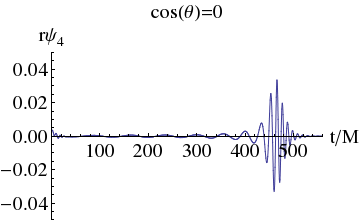} &
  \includegraphics[width=0.49\columnwidth]{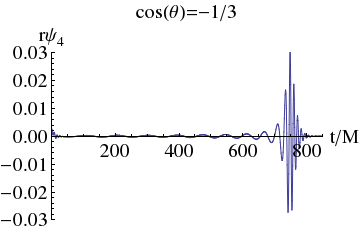}\\
\hline
  \includegraphics[width=0.49\columnwidth]{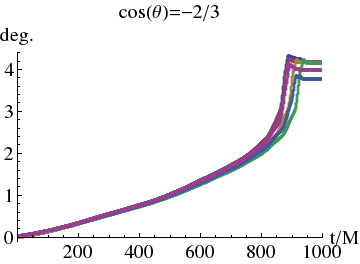}&
  \includegraphics[width=0.49\columnwidth]{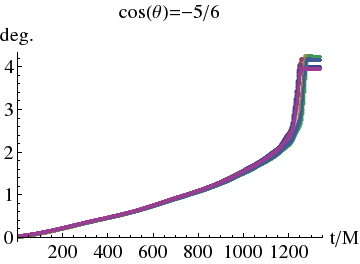}\\
  \includegraphics[width=0.49\columnwidth]{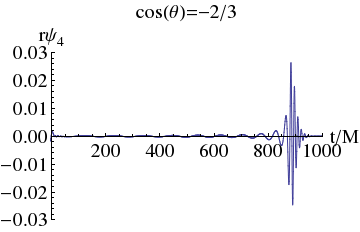} &
  \includegraphics[width=0.49\columnwidth]{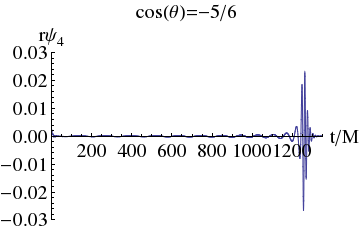}\\
  \end{tabular}
  \caption{The precession of $\hat J$ (compared to its initial direction)
  as a function of time for the SPTH48 (top left), SPTH70 (top right),
  SPTH90 (middle left), SPTH110 (middle right), SPTH132 (bottom left),
and SPTH146 (bottom right) configurations.
The lower plots show the $(\ell=2,m=2)$ mode of $\psi_4$ for the same
time scale. Note that
 the precession angle decreases during the large burst of
radiation at merger. Each azimuthal configuration (for a given
inclination angle) is shown in a different color to give an idea of
how the final angle varies with the initial azimuthal angle.
 }\label{fig:Jprec}
\end{figure}

\begin{figure}
   \includegraphics[width=.9\columnwidth]{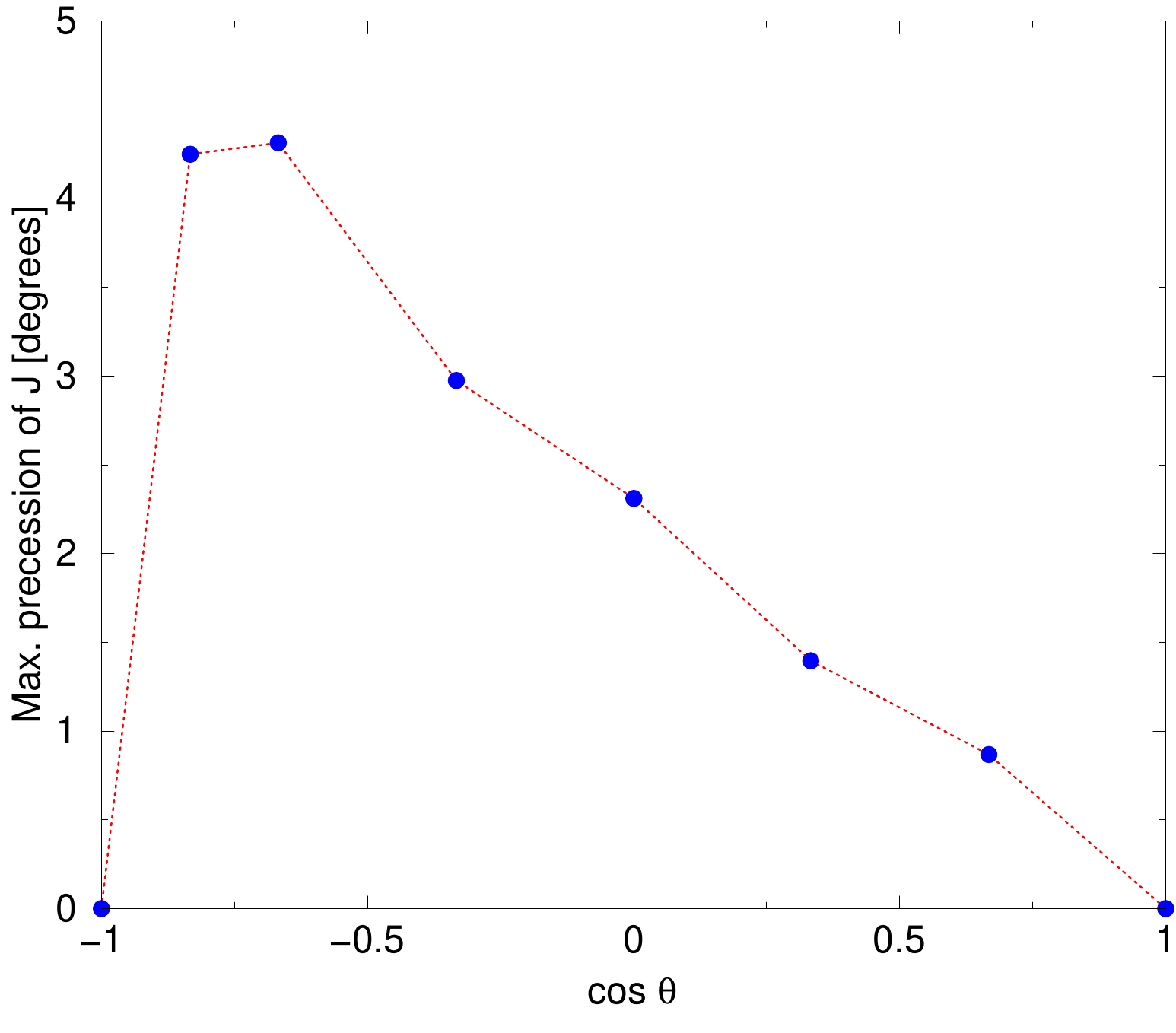}
   \caption{The maximum deviation of $\vec J$ from its initial
orientation versus $\theta$. Here the maximum is over time and over
all $\phi$ configurations for a given $\theta$ configuration.}
\label{fig:Jprecmax}
\end{figure}

 Figure~\ref{fig:trackn} shows a typical trajectory [$\vec r(t)
= \vec r_1(t) - \vec r_2 (t)$], while Fig.~\ref{fig:Jprec}, shows the
precession of $\vec J(t)$ for all SP configurations.  For a given
configuration, the maximum precession
occurs near merger, with $\vec J(t)$ moving back towards its initial
location during the plunge. This {\it antiprecession} effect is most
noticeable for smaller values of $\theta$.
The figure also shows the real part of the $(\ell=2,m=2)$ mode of
$\psi_4$ to provide a reference time scale. By comparing $\psi_4$ with $\vec
J(t)$, we can see that the {\it antiprecession} occurs roughly at
merger.
As shown in Fig.~\ref{fig:Jprecmax} and Table~\ref{tab:Jprec},
 the maximum deviation of $\vec J(t)$ from its initial
orientation is a strong function of $\theta$.
 Note how the
precession of $\vec J(t)$ increases as the polar angle $\theta$
increases until about $\theta=140^\circ$, where it then must rapidly
decrease to zero at $\theta=180^\circ$ (which is a non-precessing
configuration).
\begin{table}
  \caption{The change in the direction and magnitude of $\vec J$ for
configurations with different inclination ($\theta$) angles. Here $\ang{\vec A}{\vec B} =
\cos^{-1}(\hat A \cdot \hat B)$ measures the angle between two
vectors.
}
\label{tab:Jprec}
  \begin{ruledtabular}
  \begin{tabular}{lllllll}
$\cos\theta$   &  2/3 & 1/3 & 0 & -1/3 & -2/3 & -5/6\\
\hline
${\rm max}_{t \phi} \ang{\hat J(t)}{\hat J(0)} $ & $0.87^\circ$ & $1.40^\circ$
& $2.31^\circ$ &$ 2.97^\circ$ &
$4.31^\circ$ & $4.25^\circ$\\
${\rm max}_ {t \phi} \ang{\widehat{\delta J}(t)}{ \hat J(t)}$ &
$15.68^\circ$ & $21.53^\circ$ & $25.18^\circ$ & $27.39^\circ$ &
$27.12^\circ$ & $23.81^\circ$\\
  \end{tabular}
  \end{ruledtabular}
\end{table}

The reason why the maximum deviation of $\vec{J}$  from its initial
orientation
occurs when $\vec{S}$ is nearly
counteraligned with $\vec{L}$ is in part due to
the smaller magnitude of $\vec{J}$ for anti-aligned configurations, which
makes it easier to tilt $\vec J$ even though the anti-aligned configurations
radiate less than the aligned ones.
In the maximally spinning case, the largest precession of $\vec J$
should occur when the spins are even closer to
full counteralignment (see Sec.~\ref{subsec:hangup}).

\begin{figure}
  \includegraphics[width=\columnwidth]{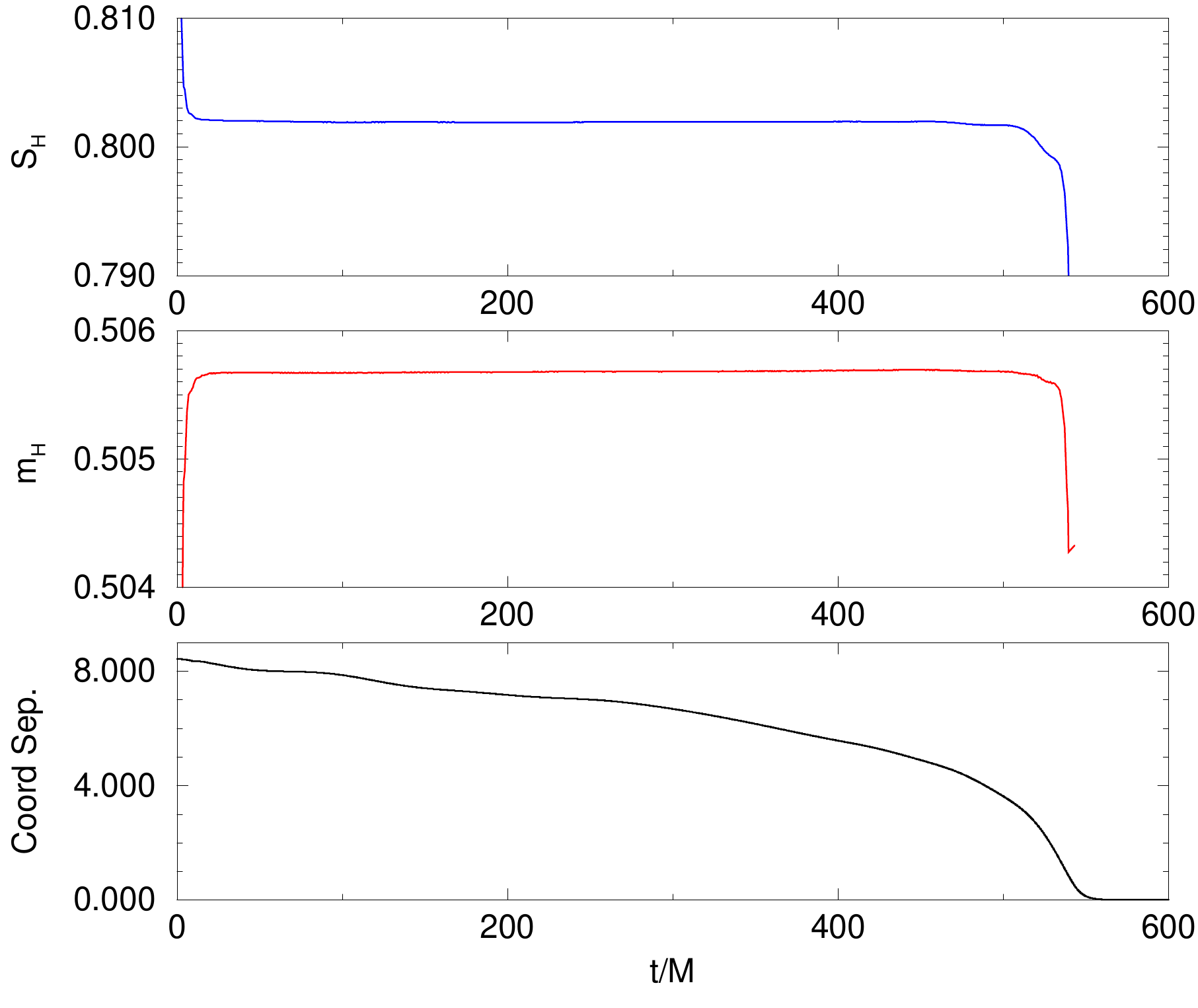}
  \caption{ The individual horizon spin angular momentum (not rescaled by the
horizon mass), horizon mass, and coordinate separation
of the two BHs for  a prototypical simulation (SPTH70PH0).
 Note how the spin magnitude (and horizon mass) is conserved over the entire
inspiral until just before merger. At merger, the isolated horizon
approximation, which measures the spins, is no longer valid.}
 \label{fig:spin_cons}
\end{figure}

As shown in Fig.~\ref{fig:spin_cons}, the magnitude of the individual
BH spins is conserved to a very high degree during the simulation. The
total angular momentum $\vec J$, however, is not. Indeed the fraction
of angular momentum lost to radiation during an inspiral approaches
$1$ as the initial separation approaches infinity.  We interpret this
to mean that orbital angular momentum is lost to radiation, while spin
angular momentum is largely conserved. Note that, for the
configurations studied here, the total spin must also be conserved in
magnitude because the two spins cannot rotate with respected to each
other, as that would break parity symmetry.

When $\vec{J}$ exhibits simple precession, as in our simulations,
even though its magnitude is always dominated by the highly precessing 
$\vec{L}$,
the spins of the individual BHs
are always oriented such that
the direction of $\vec J$ is largely unchanged.
Thus the direction of the spin of the merger
remnant $\vec{S}_\text{\rm rem}$ agrees
very closely with that of
$\vec{J}_\text{initial}$ (which was also observed
by Barausse et al.~\cite{Barausse:2009uz}).

\begin{figure}
  \includegraphics[width=0.49\columnwidth]{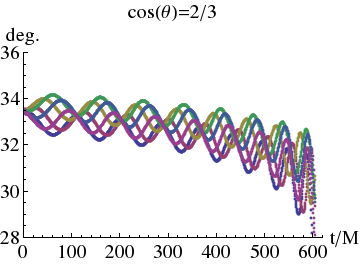}
  \includegraphics[width=0.49\columnwidth]{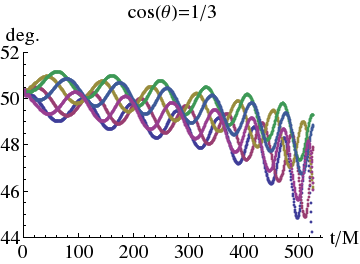}

  \includegraphics[width=0.49\columnwidth]{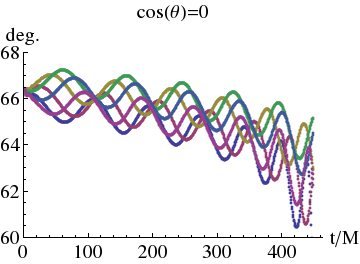}
  \includegraphics[width=0.49\columnwidth]{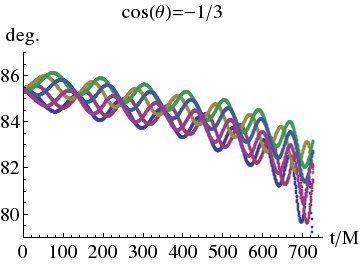}

  \includegraphics[width=0.49\columnwidth]{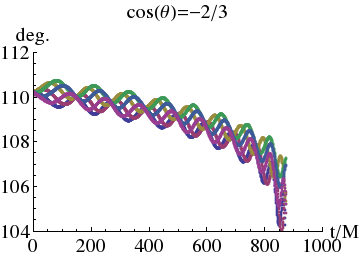}
  \includegraphics[width=0.49\columnwidth]{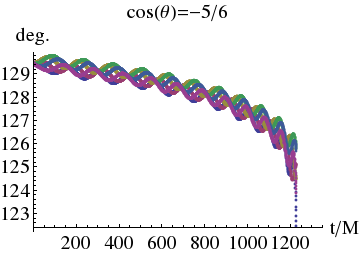}
  \caption{The angle between $\hat J$  and $\hat S_1=\hat S_2$
for the SPTH48 (top left), SPTH70 (top right),
  SPTH90 (middle left), SPTH110 (middle right),
SPTH132 (bottom left), and SPTH146 (bottom right)
configurations.
 }\label{fig:jsprec}
\end{figure}

\begin{figure}
  \begin{tabular}{l|r}

  \includegraphics[width=0.49\columnwidth]{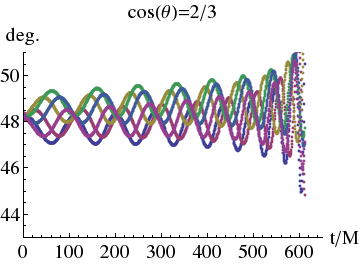} &
  \includegraphics[width=0.49\columnwidth]{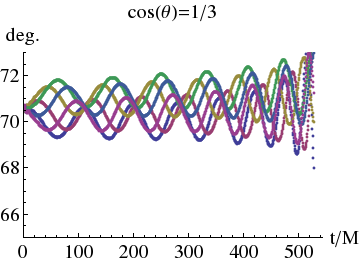} \\
  \includegraphics[width=0.49\columnwidth]{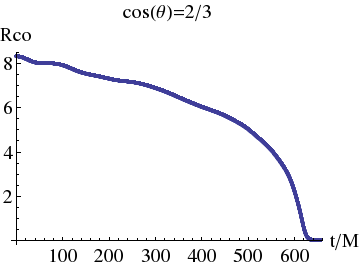} &
  \includegraphics[width=0.49\columnwidth]{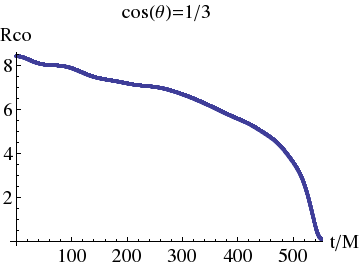}\\
\hline
  \includegraphics[width=0.49\columnwidth]{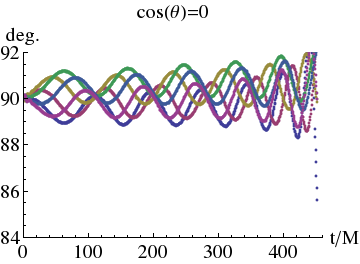}&
  \includegraphics[width=0.49\columnwidth]{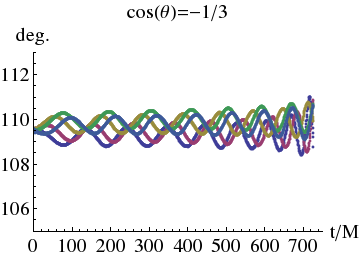}\\
  \includegraphics[width=0.49\columnwidth]{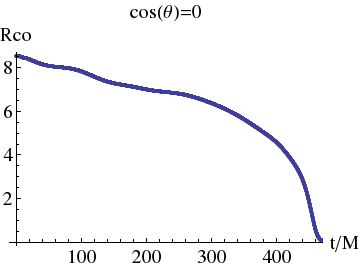} &
  \includegraphics[width=0.49\columnwidth]{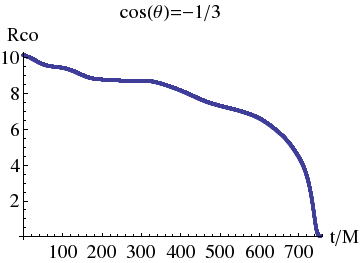}\\
\hline
  \includegraphics[width=0.49\columnwidth]{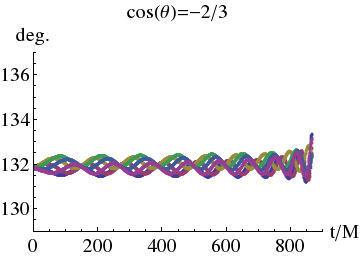} &
  \includegraphics[width=0.49\columnwidth]{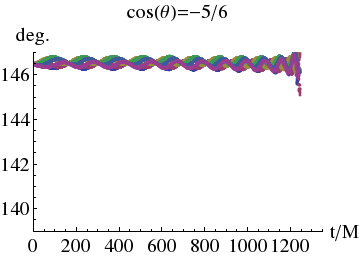} \\
  \includegraphics[width=0.49\columnwidth]{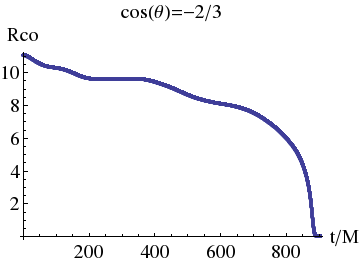} &
  \includegraphics[width=0.49\columnwidth]{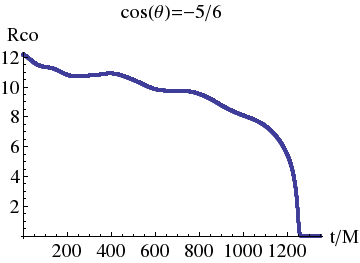}\\
  \end{tabular}
  \caption{The angle between $\hat{L}(t)$ (radiation) and $\hat S_1=\hat S_2$
 for the SPTH48 (top left), SPTH70 (top right),
SPTH90 (middle left), SPTH110 (middle right), SPTH132 (bottom left), and SPTH146 (bottom right) configurations. The lower
plots show the orbital separation versus the same time scale.
 }\label{fig:lsprec}
\end{figure}

\begin{figure}
  \includegraphics[width=0.49\columnwidth]{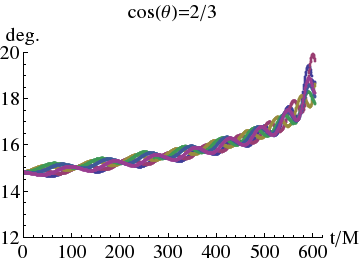}
  \includegraphics[width=0.49\columnwidth]{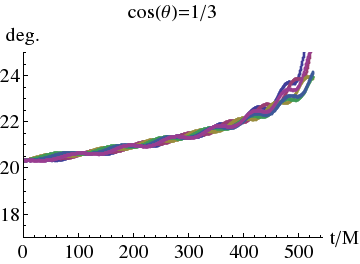}

  \includegraphics[width=0.49\columnwidth]{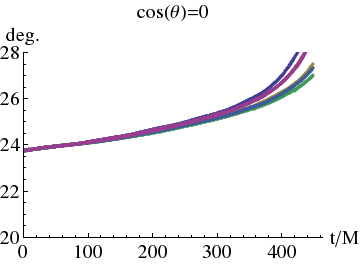}
  \includegraphics[width=0.49\columnwidth]{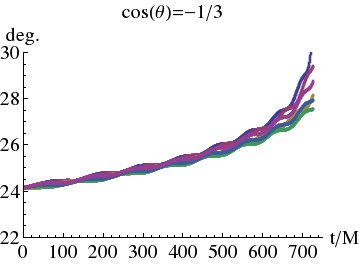}

  \includegraphics[width=0.49\columnwidth]{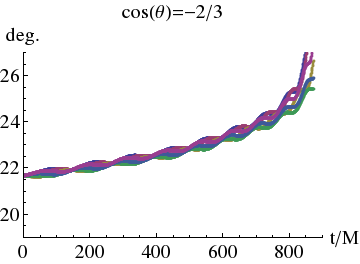}
  \includegraphics[width=0.49\columnwidth]{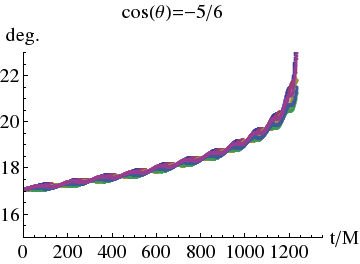}
  \caption{The angle between $\hat J$  and $\hat{L}(t)$
for the SPTH48 (top left), SPTH70 (top right),
  SPTH90 (middle left), SPTH110 (middle right), SPTH132 (bottom left), and SPTH146 (bottom right)
configurations.}
\label{fig:jlprec}
\end{figure}

Since the spin magnitude is conserved to a very high degree, and the
binary radiates angular momentum continuously, it is not surprising that
the
angle between the total spin $\vec S(t)$ and $\vec J(t)$ decreases
secularly, as shown in
Fig.~\ref{fig:jsprec}. From this, we can conclude that the component
of $\vec S$ along $\vec J$ is not conserved on secular timescales.
On the other hand, as seen in Fig.~\ref{fig:lsprec}, the
 angle between $\vec S(t)$ and $\vec L(t)$, as defined using
Eq.~(\ref{eq:LfromJ}), shows no noticeable secular trend. That is to
say, the average (over an orbital timescale) of $\hat L\cdot \hat S$
is nearly constant (in PN theory, it is the
orbit average of $\hat L\cdot \vec S$ that is approximately conserved
too).
From this, and the fact that $|\vec
S(t)|$ is conserved, we can conclude that, on
average, the component of $\vec S$ along $\vec L(t)$ is conserved.
Finally, in Fig.~\ref{fig:jlprec}, we show that the angle between
$\vec L(t)$ and $\vec J(t)$ increases on secular timescales.
\clearpage

\begin{figure}
  \includegraphics[width=0.6\columnwidth]{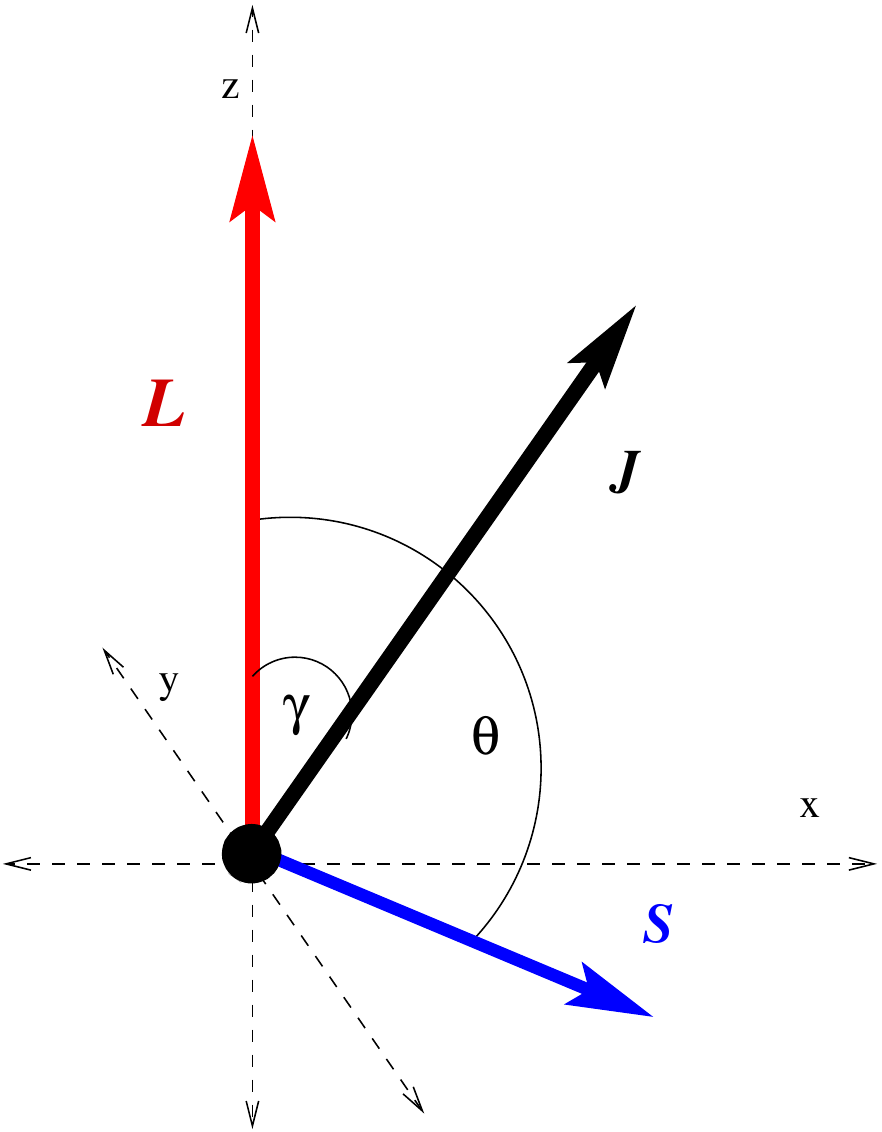}
  \caption{The orbital, spin, and net angular momentum. We choose
the polar axis to always lie along the orbital angular momentum. The
angles $\gamma$ and $\theta$ measure the inclination angles of $\vec
J$ and $\vec S$, respectively.}
  \label{fig:LJS}
\end{figure}

Given that we see an approximate conservation of the magnitude of
the total spin $\vec{S}$ (see Fig.~\ref{fig:spin_cons}),
as well as the conservation (see Fig.~\ref{fig:lsprec})
of the component $\vec{S}$ along
the angular momentum $\vec{L}$ (denoted by $S_\|=\vec{S}\cdot\hat{L}$) we can
deduce that the  angle between $\vec{J}$ and $\vec{L}$ should increase
on secular timescales, while the angle between
$\vec{J}$ and $\vec{S}$  should decrease.
The argument is as follows; since
$|\vec S|^2 = S_\perp^2 + S_\|^2$ and both $|\vec S|$ and
$S_\|$ are conserved, then $S_\perp$ must be conserved, as well (in
magnitude only). $|\vec J|$, on the other hand decreases on secular
timescales due to the radiation of angular momentum to infinity. Hence
(see Fig.~\ref{fig:LJS})
\begin{equation}
\cos^2\gamma=(\hat{J}\cdot\hat{L})^2=1-\frac{S_\perp^2}{J^2}.
\end{equation}
decreases on the same secular timescale. For the configurations
considered here, $\gamma < \pi/2$ (i.e.\ there is only simple
precession), hence $\gamma$ increases (as seen in Fig.~\ref{fig:jlprec}).

Likewise we can show that the constancy of the spin magnitude and
its projection along $\vec{L}$ leads to
the decrease of the angle between $\vec{J}$ and $\vec{S}$
seen in Fig.~\ref{fig:jsprec}.
Again, from $\vec{J}=\vec{L}+\vec{S}$, we have (see Fig.~\ref{fig:LJS})
\begin{equation}
\cos^2(\theta-\gamma)=(\hat{J}\cdot\hat{S})^2=1-
\frac{L^2}{J^2}\sin^2\theta,
\end{equation}
and since
\begin{equation}
\frac{J^2}{L^2}=1+\frac{2}{L}(\hat{L}\cdot\vec{S})+\frac{S^2}{L^2}
\end{equation}
will increase or decrease
with the radiation of $L$ depending on the sign of $(2\vec{L}+\vec{S})\cdot\vec{S}$
(and hence depending on the sign of $(\hat{L}\cdot\vec{S})$ for the simple
precession case).
Thus for our configurations, the angle
$(\theta-\gamma)$ has to decrease on the secular scale
of the orbital decay (as seen in Fig.~\ref{fig:jsprec}).

\begin{figure}
  \includegraphics[width=0.99\columnwidth]{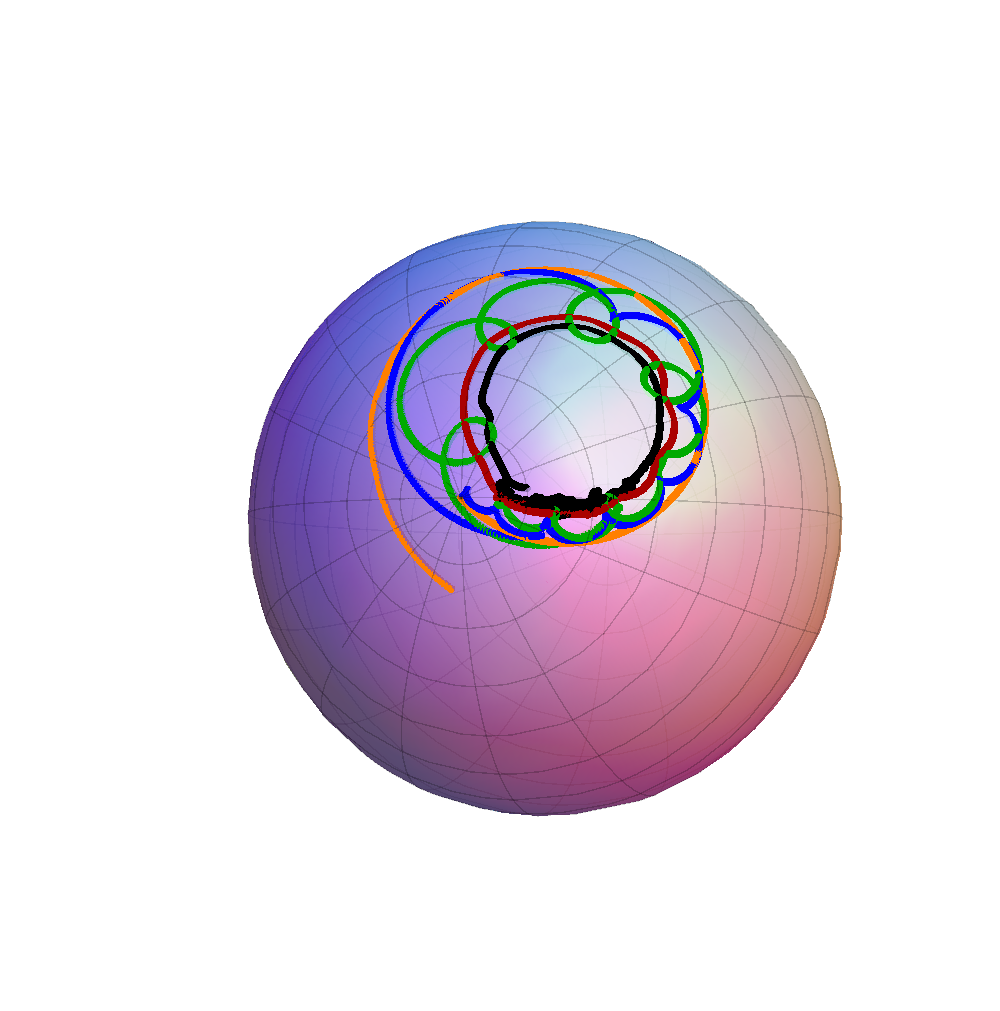}
  \includegraphics[width=0.99\columnwidth]{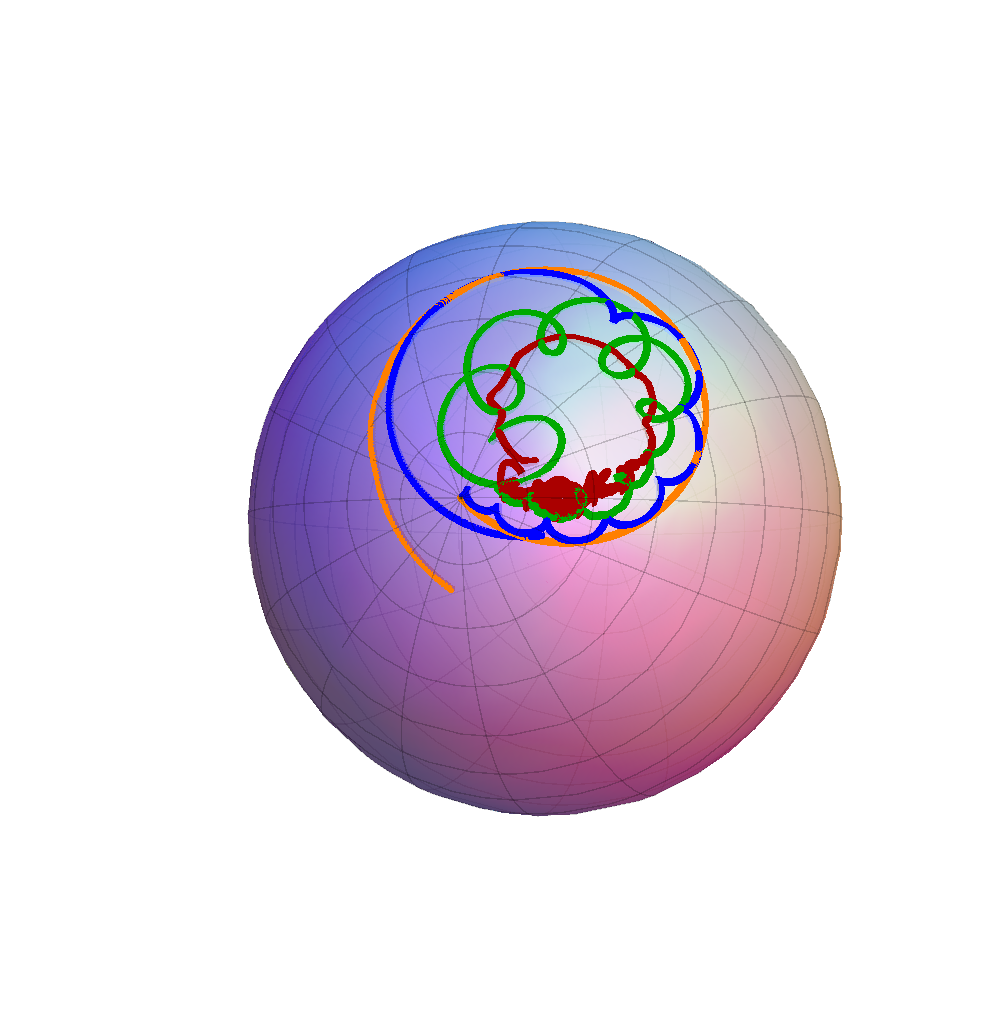}
  \caption{ A comparison of the directions of 
$\hat L_{\rm coord}$ (blue), $\hat L_{\rm rad}$ (orange),
with the principle eigenvector direction of
$\langle {\bf L L} \rangle$ (see Ref.~\cite{O'Shaughnessy:2012vm})
defined using $\psi_4$ (black), $N$ (red), and $h$ (green)
(top),
and 
with the vectors
$\langle {\bf L L} \rangle^{-1} \langle {\bf LT \rangle}$ (see
Ref.~\cite{Boyle:2013nka})
defined using $N$ (red) and $h$ (green) (bottom). Both plots are from the
SPTH90PH0 configuration.} \label{fig:OB_cmp}
\end{figure}

\begin{figure}
  \includegraphics[width=0.99\columnwidth]{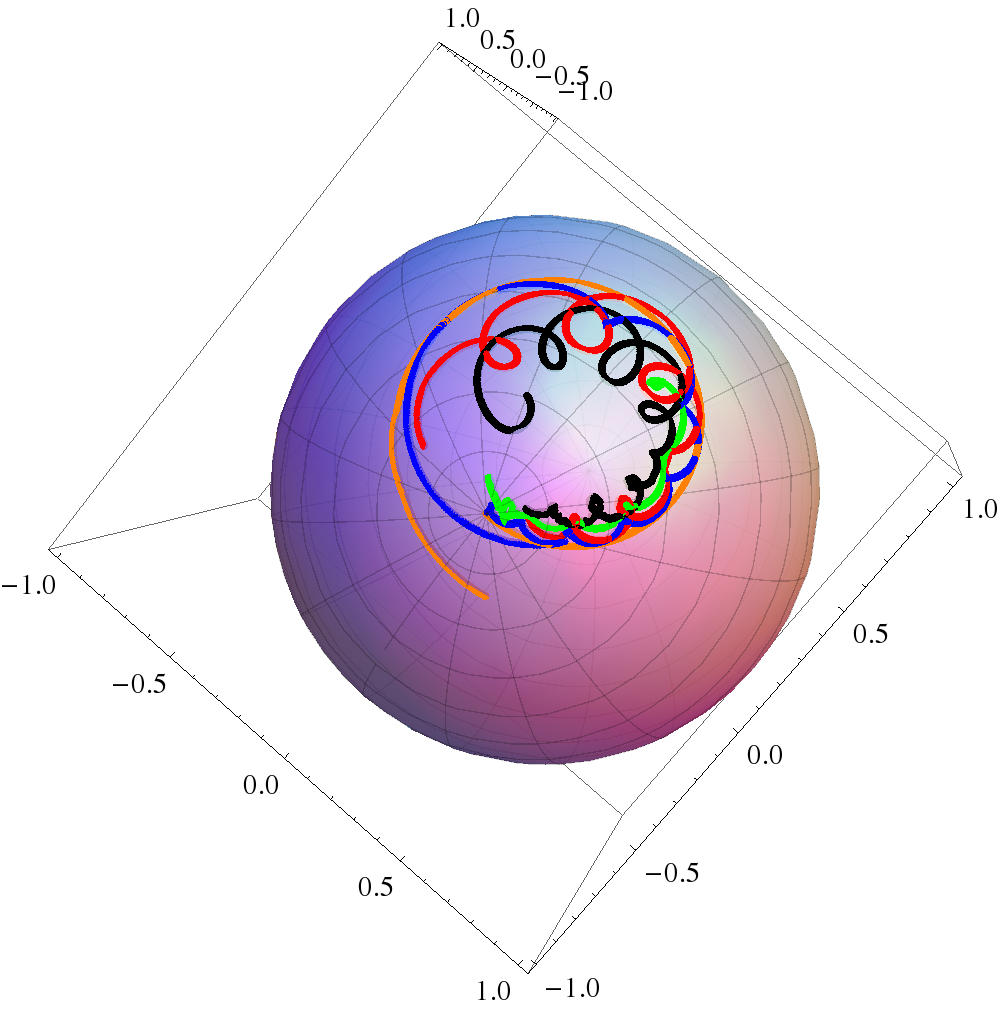}
  \caption{A plot comparing different measures of the direction of
orbital angular momentum for the SPTH90PH0 configuration. Plotted are
the trajectories of  $\hat B_h(t)$ (black), $\hat O_h(t)$ (red),
$\vec{\delta J}_{\rm rad}$ (green), $\hat{L}_{\rm
coord}$ (blue), and $\hat L(t)$ (orange).
$\hat L_{\rm coord}$ shows interesting nutation behavior not seen in
the other curves.}\label{fig:Lcmp}
\end{figure}

We now briefly return to the question of which approximation to
the direction of the orbital angular momentum
provides the best properties for subsequent modeling of  BHBs.
In Fig.~\ref{fig:OB_cmp} and~\ref{fig:Lcmp} we plot
the directions defined
by O'Shaughnessy et al., $\hat O_\psi$, $\hat O_N$, $\hat O_h$,
the directions defined by Boyle, $\hat B_h(t)$, $\hat B_N(t)$,
as well as $\hat L(t)$, and the purely coordinate-based $\hat L_{\rm
coord}(t)$. Based on these figures, one may suspect that $\hat L_{\rm
coord}(t)$ most faithfully follows the dynamics of the binary (i.e.\
nutation effects are present). However, for our purposes, we are
interested in determining the direction $\hat V$ such that
$\hat V\cdot \hat S$ is as constant as possible.

\begin{figure}
  \includegraphics[width=0.9\columnwidth]{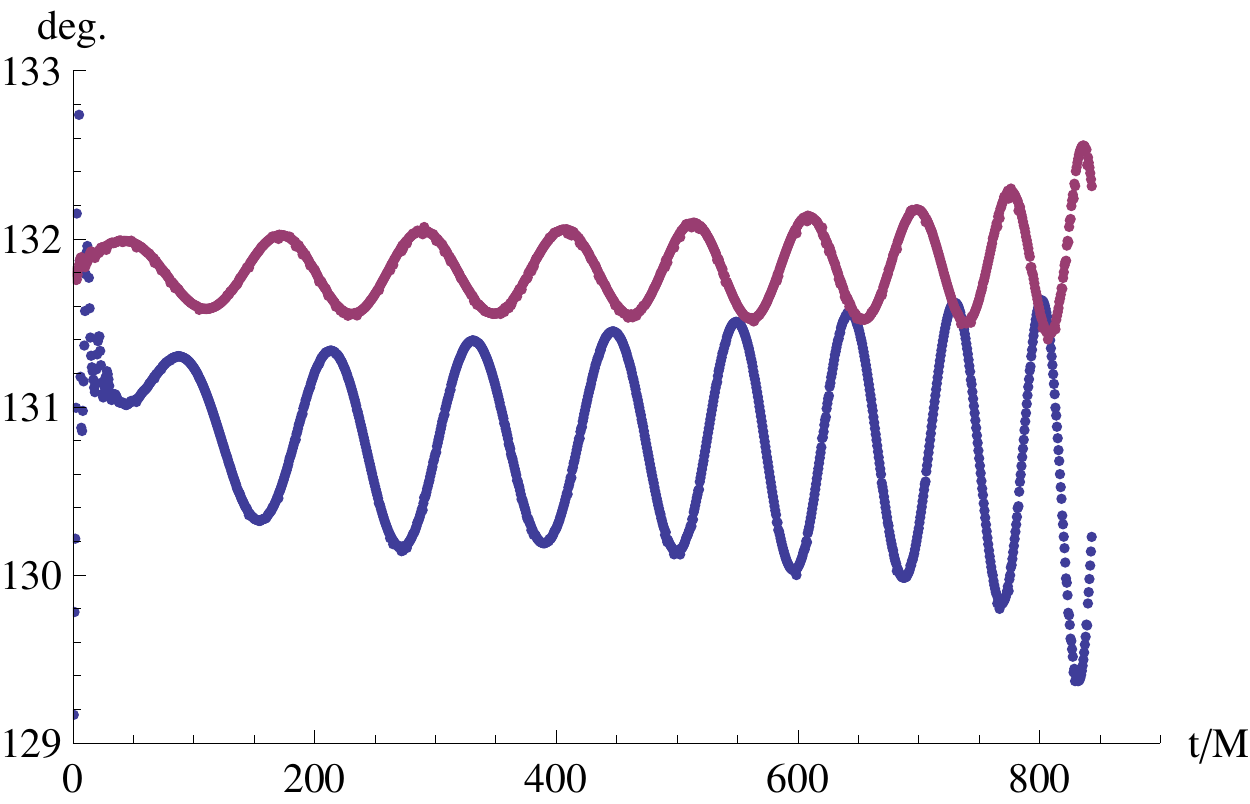}
  \caption{
A plot comparing the angles $\{\vec S(t),\vec L(t)\}$ (upper, red)
 and $\{\vec S(t), \vec L_{\rm coord}(t)\}$ (lower, blue) for the SPTH132PH0
configuration. Note how  $\{\vec S(t), \vec L_{\rm coord}(t)\}$ shows a
stronger secular trend.}\label{fig:LLcomp}
\end{figure}

\begin{figure}
  \includegraphics[width=0.9\columnwidth]{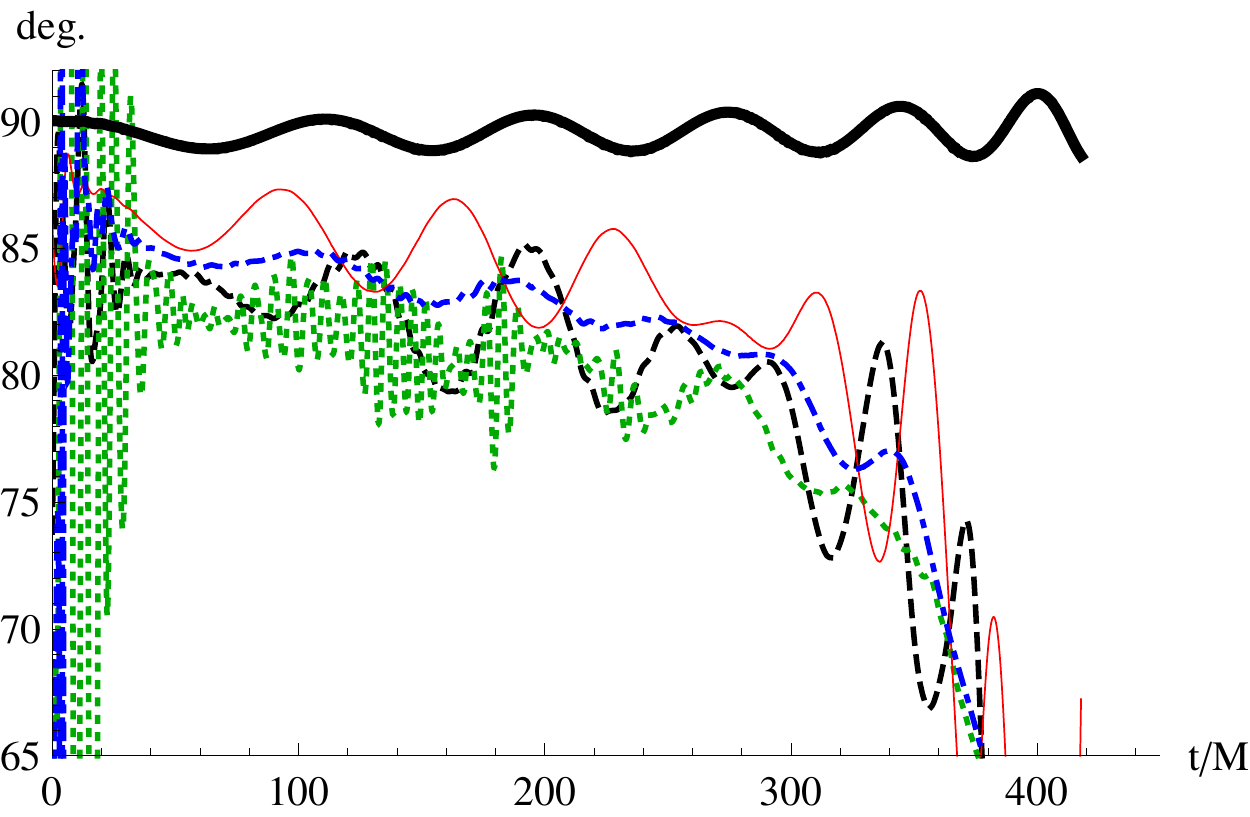}
\caption{
The angle between $\vec S(t)$ and $\vec B_h(t)$ (black, dashed),
$\vec B_n(t)$ (green, dotted), $\vec O_h(t)$ (red, solid), $\vec
O_n(t)$ (blue, dashed-dot), and
$\vec L(t)$ (black, solid) for the SPTH90PH0 configuration.  All angles show a strong secular trend
except the angle between $\vec S(t)$ and $\vec L(t)$.}
\label{fig:ls_def_comp}
\end{figure}

To this end, we compare the time dependence of the two angles
$\{\vec S(t), \vec L(t)\}$ and $\{\vec S(t), \vec L_{\rm coord}(t)\}$.
As shown in Fig.~\ref{fig:LLcomp}, in all cases we saw that
$\{\vec S(t), \vec L(t)\}$ shows a smaller secular change than
$\{\vec S(t), \vec L_{\rm coord}(t)\}$ (the effect was not always
significant). This raises the question of whether the more geometric
directions of O'Shaughnessy et al. and Boyle lead to even better
performance.  In Fig.~\ref{fig:ls_def_comp}, we plot the angles
 $\{\vec S(t), \vec O_N(t)\}$,
 $\{\vec S(t), \vec O_h(t)\}$,
 $\{\vec S(t), \vec B_N(t)\}$,
 $\{\vec S(t), \vec B_h(t)\}$,
and $\{\vec S(t), \vec L(t)\}$.
Here we find that the angle between $\vec S(t)$ and $\vec L(t)$ is
conserved to a much higher degree than any of the other angles.

\subsection{Results from the UD configurations}

\begin{figure}
 \includegraphics[width=0.9\columnwidth]{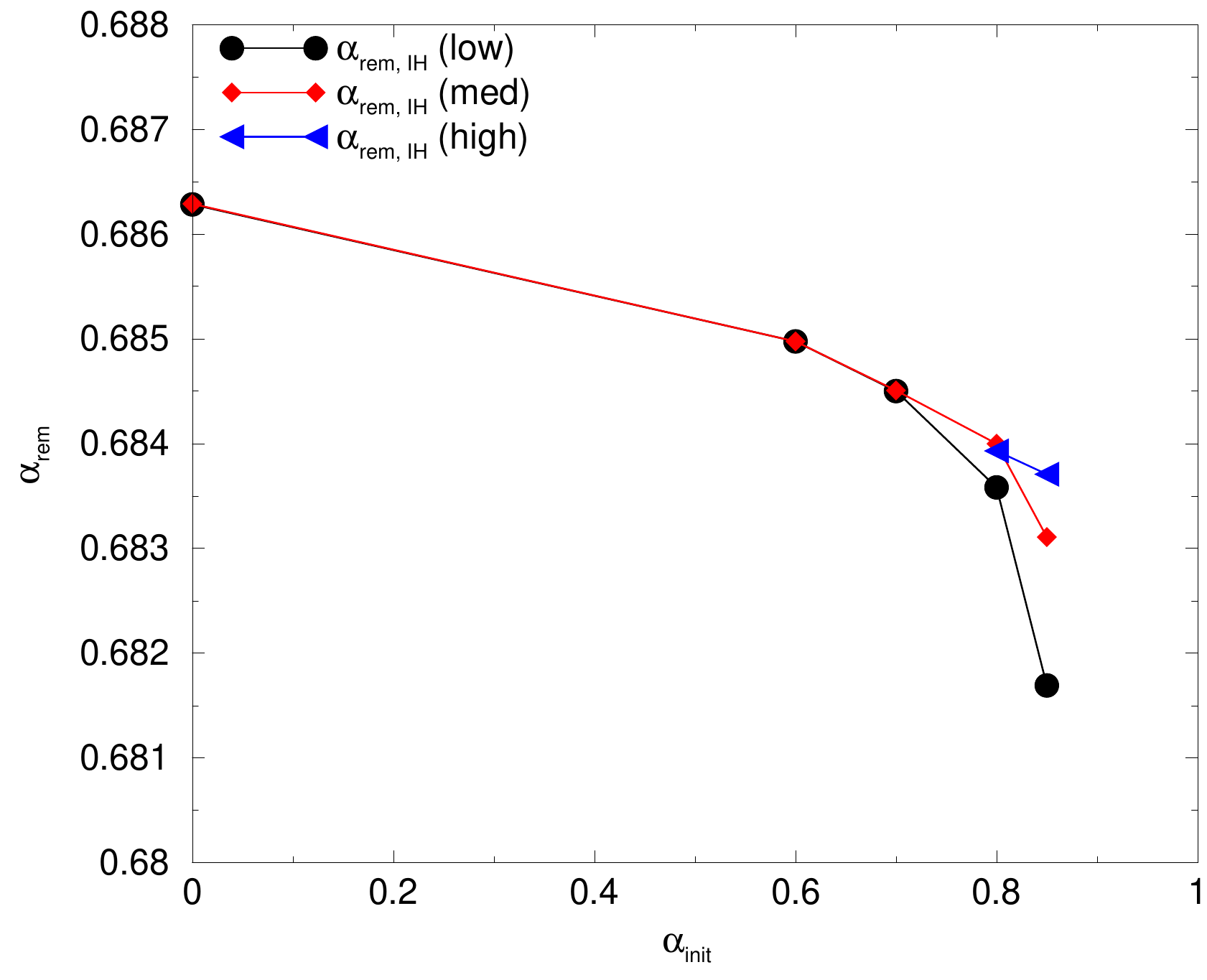}
 \includegraphics[width=0.9\columnwidth]{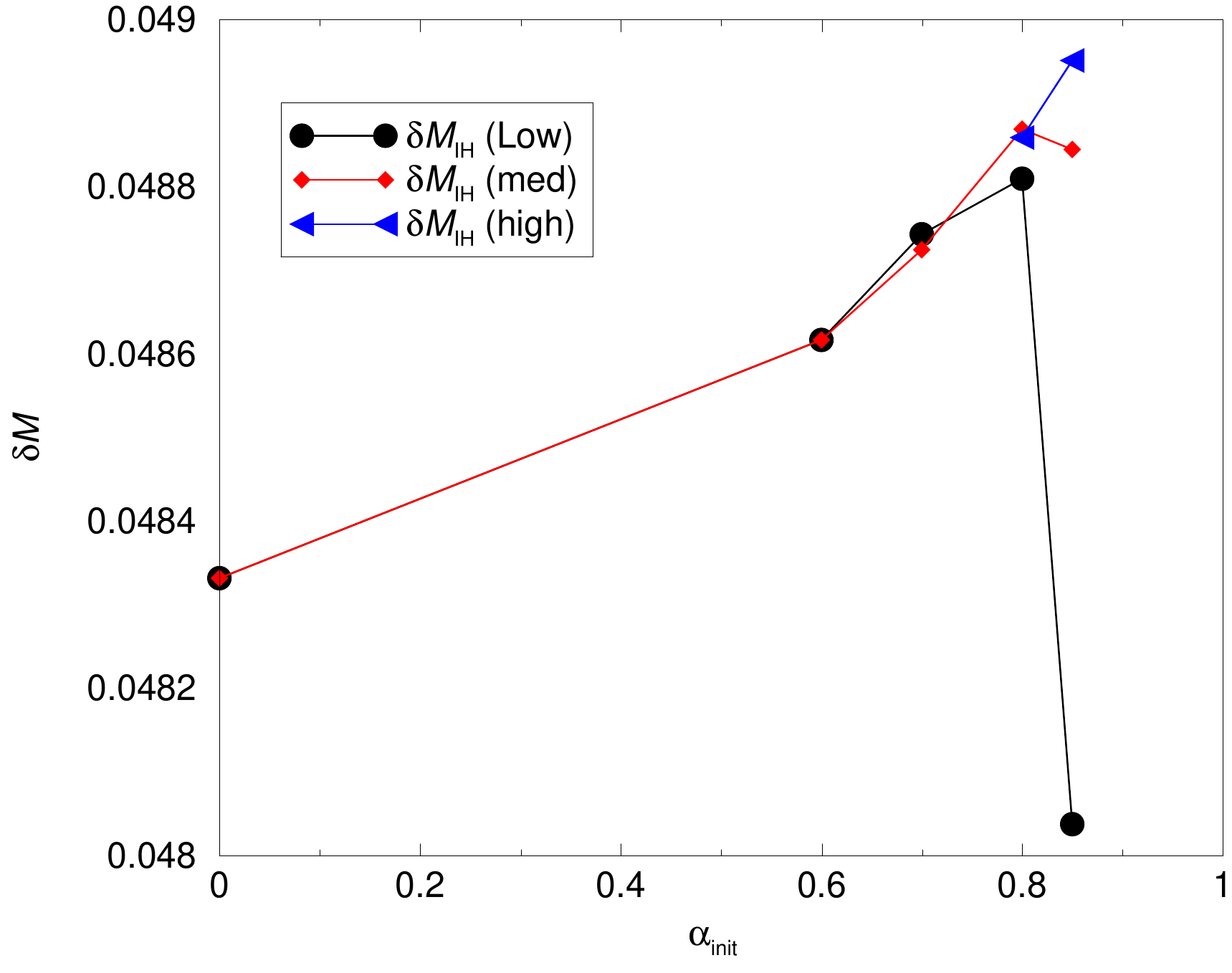}
  \caption{The total radiated energy ($\delta {\cal M}$) and final remnant
spin $\alpha_{\rm rem}$ versus initial spin for the UD
configurations. Note how trends for both the radiated mass and final
remnant spins need higher resolution runs at high spins.}
\label{fig:up_E_and_A}
\end{figure}

In order to model the dependence of the final mass (or radiated energy)
and spin of the merger remnant of BHBs as a function $\Delta_\|$,
we perform a series of additional runs of equal-mass binaries
with one BH spin aligned and the other counteraligned to the orbital
angular momentum.
These up-down (UD)
configurations are non precessing.
For these configurations, we are interested in calculating the final
remnant properties to high accuracy.
As shown in Fig.~\ref{fig:up_E_and_A}, the
calculation of the total radiated
mass and the final remnant spin require increasingly higher resolutions as the
initial spin magnitude increases.

Since the UD configuration with initial spins $\pm\alpha$ is
equivalent to the configuration with initial spins $\mp\alpha$, the
remnant spin and total mass cannot depend on the sign of $\alpha$ (and
hence cannot depend on the sign of $\Delta_\|$).
Consequently, we fit the remnant spin $\alpha_{\rm rem}$ and total radiated mass
to the forms
\begin{eqnarray}
  \alpha_{\rm rem}^2 = A_{0} + A_{2}
\left(\frac{\Delta_\|}{m^2}\right)^2 + A_{4}
\left(\frac{\Delta_\|}{m^2}\right)^4,\label{eq:udafit}\\
  \delta {\cal M} = E_{0} + E_{2} \left(\frac{\Delta_\|}{m^2}\right)^2 + E_{4}.
\left(\frac{\Delta_\|}{m^2}\right)^4,\label{eq:udefit}
\end{eqnarray}
We give the results from the fits in Table~\ref{tab:UDfit}.
\begin{table}
  \caption{The values for the parameters in
Eqs.~(\ref{eq:udafit})~and~(\ref{eq:udefit}) after fitting to the results
of the UD configurations. Here ``0'' indicates that the uncertainty in
the remaining fitting parameters was substantially
reduced when this coefficient was
set to zero.}\label{tab:UDfit}
  \begin{ruledtabular}
  \begin{tabular}{rl|rl}
     $A_{0}$ & $(0.470978\pm 0.000051)$ & $E_0$ & $ (0.048332\pm0.000002)$ \\
     $A_{2}$ & $-(0.00492\pm0.00010)$ &  $E_{2}$ & $ (0.000743 \pm 0.000018)$ \\
     $A_{4}$ & 0 & $E_{4}$ & $(0.000124\pm0.000028)$ \\

  \end{tabular}
  \end{ruledtabular}
\end{table}

\section{Modeling Precessing binaries and their Remnants}\label{sec:modeling}

In this section we  use analytic techniques to gain
insight into the observed dynamics of precessing BHBs
and the modeling of the merger remnant.

\subsection{Post-Newtonian Analysis}\label{subsec:PN}

As show in Fig.~\ref{fig:lsprec}, the angle between $\vec L$ and
$\vec S$ oscillates at roughly the orbital frequency, and the
amplitude of these oscillations increases as the binary inspirals. In
addition, the baseline of these oscillations is a function of the
azimuthal angle $\phi$. This behavior is predicted by PN theory, as we
will demonstrate here.

At 2PN order, the evolution of the
total spin for equal-mass binaries is given by~\cite{Racine:2008kj} 
\begin{equation}
\frac{d\vec{S}}{dt} =\frac{7}{2r^3}(\vec{L}\times\vec{S})
+ \frac{3}{r^3}(\hn\cdot\vec{S})(\hn\times\vec{S}), \label{Sconstevol}
\end{equation}
where $\vec L$ is the orbital angular momentum and $\hn$ is the
unit vector pointing along the line joining the two BHs.
At this order, the evolution is conservative, and hence
$\dot{\vec{J}}=\dot{\vec{L}}+\dot{\vec{S}}=0,$ where an overdot means total
time derivative.
Consequently, $
\vec{S}\cdot\dot{\vec{S}}=0$ and therefore 
\begin{equation}
\frac{d(\vec{J}\cdot\vec{S})}{dt}=
\frac{d(\vec{L}\cdot\vec{S})}{dt}=
-\frac{d(\vec{J}\cdot\vec{L})}{dt}=
\vec{L}\cdot\dot{\vec{S}}.
\end{equation}
While the magnitude of $\vec{S}$ remains constant the
magnitude of
$\vec{L}$ varies according to
\begin{equation}
\frac{d(\vec{S}\cdot\vec{S})}{dt}=0,\quad
\frac{d(\vec{L}\cdot\vec{L})}{dt}=-2\,\vec{L}\cdot\dot{\vec{S}}.
\end{equation}

The variation of the component of $\vec S$ along $\vec L$ is then
given by
\begin{eqnarray}\label{eq:L.S}
\frac{d}{dt}(\hL\cdot\vec{S})&=&
\hL\cdot\dot{\vec{S}}+\dot\hL\cdot\vec{S}\nonumber\\
&=&\frac{3}{r^3}(\hn\cdot\vec{S})\,(\hl\cdot\vec{S})
\left[1+(\hL\cdot\vec{S})/\ell\right]\nonumber\\
=\dot{S}_{\hL}&=&\frac{3}{r^3}(S_{\hn})\,(S_{\hl})
\left[1+(S_{\hL})/\ell\right]
\end{eqnarray}
where $\hn$ and $\hl$ are the unit vectors on the orbital plane along
the radial and tangential directions, and $\hL=\hn\times\hl$
and $\ell=\|\vec L\|.$

Similarly, the variation of the component of $\vec{S}$ perpendicular to
$\vec{L}$ is given by
\begin{eqnarray}\label{eq:LxS}
\frac{d}{dt}(\hL\times{\vec{S}})&=&
\dot{S}_{\hn}\hl-\dot{S}_{\hl}\hn+
{S}_{\hn}\dot\hl-{S}_{\hl}\dot\hn\nonumber\\
&=&\hL\times\dot{\vec{S}}+\dot\hL\times{\vec{S}}
\end{eqnarray}
where
\begin{eqnarray}
\dot\hL&=&\left(-\frac72+\frac{3S_{\hL}}{\ell}\right)
\frac{S_{\hn}}{r^3}\hl+\frac{7S_{\hl}}{2r^3}\hn,\nonumber\\
\dot\hl&=&-\left(-\frac72+\frac{3S_{\hL}}{\ell}\right)
\frac{S_{\hn}}{r^3}\hL-\frac{V_\lambda}{r}\hn,\nonumber\\
\dot\hn&=&\frac{V_\lambda}{r}\hl-\frac{7S_{\hl}}{2r^3}\hL,
\end{eqnarray}
where $V_\lambda$ is the tangential velocity of the binary.

The two transverse components of the total spin therefore obey
\begin{eqnarray}\label{eq:Sln}
\dot{S}_{\hn}&=&\Omega_-\,S_{\hl},\nonumber\\
-\dot{S}_{\hl}&=&\Omega_+\,S_{\hn},
\end{eqnarray}
where
\begin{equation}
\Omega_-=\frac{V_\lambda}{r}-\frac{7\ell}{2r^3}(1+S_{\hL}/\ell),
\end{equation}
and
\begin{equation}
\Omega_+=\Omega_-+\frac{3S_{\hL}}{r^3}(1+S_{\hL}/\ell).
\end{equation}

Equations (\ref{eq:L.S}) and (\ref{eq:Sln}) represent a system
of equations for the evolution
of the precessing spin which can solved in the small oscillations regime
to give
\begin{eqnarray}\label{eq:SSS}
S_{\hn}(t)&=&S_{\hn}(0)\cos(\Omega_\pm t)+
\frac{\Omega_-}{\Omega_\pm}S_{\hl}(0)
\sin(\Omega_\pm t),\\
S_{\hl}(t)&=&-\frac{\Omega_\pm}{\Omega_-}S_{\hn}(0)
\sin(\Omega_\pm t)+S_{\hl}(0)\cos(\Omega_\pm t),\\
S_{\hL}(t)&=&S_{\hL}(0)+\frac{3}{2\Omega_-r^3}
\left[S_{\hn}^2(t)-S_{\hn}^2(0)\right],
\end{eqnarray}
where $\Omega_\pm^2=\Omega_+\Omega_-,$
\begin{equation}
S_{\hn}(0)=S\sin\theta_0\cos\phi_0,
\end{equation}
\begin{equation}
S_{\hl}(0)=-S\sin\theta_0\sin\phi_0,
\end{equation}
and the angles $\phi_0$ and $\theta_0$ give the initial azimuthal
and polar orientations of $\vec S$ (where the polar axis is
aligned with $\hL$).
These PN expressions
thus reproduce the amplitude and frequency of the 
oscillations observed in
Fig.~\ref{fig:lsprec}.
Taking the orbit average we get,
\begin{eqnarray}
&&\langle S_{\hL}(t)\rangle\, =S_{\hL}(0)+\nonumber\\
&&\frac{3|\vec{S}|^2}{4r^3\Omega_\pm^2}\sin^2(\theta_0)
\left[\Omega_-\cos^2(\phi_0)-\Omega_+\sin^2(\phi_0)\right].
\end{eqnarray}
Since $S$ is conserved, we find that the orbit-averaged
angle $\cos^{-1}(\hat S\cdot \hat L)$ changes by less than
three degrees from infinite separation down to separations where the
orbital frequency approaches 0.1/M (where
the PN approximation breaks down) and that the average angle depends
of $\phi_0$ (i.e.\ the initial azimuthal orientation).

\subsection{The Hangup  effect in precessing binaries}\label{subsec:hangup}

To estimate which configuration  maximizes the amount of
angular momentum radiated perpendicular to $\vec J$, we look
for configurations that maximize the angle  $\gamma$ between the orbital
angular momentum $\vec L$ and $\vec J$
near merger (at merger, there is an intense burst of angular momentum
radiated).

Since $\vec J = \vec L + \vec S$,
 $\cos \gamma$ is given
by
\begin{equation}\label{eq:gamma}
\cos\gamma=\frac{\vec L\cdot\vec J}{|\vec L||\vec J|}=
\frac{1+(S/L)\cos\theta}{\sqrt{1+2(S/L)\cos\theta+(S/L)^2}},
\end{equation}
where $\theta$ is the angle between $\vec S$ and $\vec L$ (see
Fig.~\ref{fig:LJS}).

While $S$ is preserved during evolutions of our
configurations (see Fig.~\ref{fig:spin_cons}),
to estimate $L$ at merger we use the hangup configuration
results (as this proves to be the dominant effect for
precessing binaries \cite{Lousto:2013vpa}).
Our ansatz is
\begin{equation}
L^{\rm hangup}=J^{\rm hangup}_{\rm rem}-S\cos\theta
\end{equation}
where $L_{\rm merger} = L^{\rm hangup}$ and $J^{\rm hangup}_{\rm rem}$ is given by \cite{Hemberger:2013hsa}
\begin{eqnarray}
J^{\rm hangup}_{\rm rem}/M^2_{\rm rem}=&&0.686402+0.30660\,s-0.02684\,s^2\nonumber\\
&&-0.00980\,s^3-0.00499\,s^4,\\
M_{\rm rem}/M_{\rm init}=
&&\left(1-0.00258+\frac{0.07730}{s-1.6939}\right),
\end{eqnarray}
$M_{\rm init}=M_1+M_2$ (the horizons masses $M_1$ and $M_2$ are equal
in this case),
 $s = 2 S/M_{\rm init}^2 \cos\theta$
is the component of the dimensionless spin along the orbital angular
momentum (note and $s=0.8\cos\theta$ and 
$S/M^2_{\rm rem}=0.4\,M^2/M^2_{\rm rem}$ 
for our configurations).
By minimizing $\cos\gamma$ with respect to
 $\cos\theta$, we find that
\begin{equation}
\gamma\approx34^\circ\quad{\rm and}\quad\theta\approx110^\circ,
\end{equation}
Extrapolating the above expressions to a pair of maximally spinning black holes
gives,
\begin{equation}
\gamma_{max}\approx42^\circ\quad{\rm and}\quad\theta_{max}\approx117^\circ.
\end{equation}
Thus we find by using straightforward geometric arguments that the
largest precession of the total angular momentum will occur with spins
partially anti-aligned with the orbital angular momentum, which
agrees qualitatively
with our results
in Fig.~\ref{fig:Jprecmax}.

\subsection{Symmetries}\label{subsec:symmetries}
In this section, we use $\vec J$ to denote the final remnant
spin angular momentum and $\vec S$ to denote the spin of the precursor
binary. Here, the component $V_\perp$ of any vector $\vec V$ is
understood to be $V_\perp = \vec V \cdot \hat n$, where $\hat n$ is 
a vector in the orbital plane (e.g.\ the direction from BH1 to BH2).
We always define these directions such that, under exchange of
labels BH1$\leftrightarrow$BH2, $\hat n$ changes sign.

In order to develop a new phenomenological formula for the
final masses and spins of merged binary black holes, we
consider a Taylor expansion of those
quantities~\cite{Boyle:2007sz,Boyle:2007ru} in terms of
spin components $S_\|$, $S_\perp$, $\Delta_\|$, $\Delta_\perp$.
This choice of variables allows us to relate the expansion of
the final mass (or energy radiated) and the final spin with
the expansion we developed for modeling the final recoil velocity in
\cite{Lousto:2012gt}. There we argued that the leading-order
dependence of the recoil could be modeled by the variable 
$\vec \Delta$ rather than the alternative
$\vec{\delta S} = \vec S_2 - \vec S_1$ 
(see Fig. 2 in Ref.~\cite{Lousto:2012gt}).
In the equal-mass limit, $\vec \Delta = 2 \vec{\delta S}$ and the two
variables are equivalent. However, for $q<1$, we showed that the
spin-dependence of the recoil is fit with fewer terms if we expand in
terms of $\vec \Delta$.
Here too, we may consider alternative variables to $\vec{S}$, for
example, $\vec{S}_0=m(\vec{S}_2/m_2+\vec{S}_1/m_1)$ or
the $\vec{S}_{\rm eff}=\vec{S}/4+3\vec{S}_0/4$ which arise from the
leading-order spin terms in the PN Hamiltonian \cite{Damour:2001tu}.
While the infinite series expansion for all of these choices of spin
variables are equivalent, a low-order truncated series in a preferred 
 set of variables
may have a substantially smaller error than the other choices.
Determining the best choice of the variables will involve
a detailed analysis of a large set
of new {\it unequal} mass binaries that we will postpone
for a forthcoming paper \cite{LZ}. Note that the alternative variables
we considered share the symmetries of $\vec{S}$, hence the expansion
we give below can be easily adapted to them.

A Taylor expansion of a function with $v$ independent variables
of a given order of expansion $o$ has
$n$ terms, where $n$ is given by
\cite{2005mmp..book.....A}
\begin{equation}\label{eq:terms}
n=\frac{(o+v-1)!}{o!\,(v-1)!}.
\end{equation}
Our models will depend on five variables, $\delta m$, $S_\|$,
$S_\perp$, $\Delta_\|$, and $\Delta_\perp$. Hence $n=1, 5, 15, 35,
70,\cdots$ for $0^{\rm th}$ order through $4^{\rm th}$ order
expansions, respectively.
However, only certain combinations of variables are allowed according to
the symmetry properties of the object we want to build up.
 In order to take into account the correct combinations of variables
for each component of the final mass and spin of the remnant black hole
at a given order, we consider
the symmetry properties summarized in Table \ref{table:symmetries}
(note that this corrects typos in Table I of
Ref~\cite{Lousto:2012gt}). While the number of terms in
 $J_\|$ and $J_\perp$ is 
less than Eq.~(\ref{eq:terms}), the sum of the terms in both agrees
with  Eq.~(\ref{eq:terms}), as seen in Table~\ref{table:numberterms}.
This is due to the fact that an arbitrary product of powers of
$S_\|$, $S_\perp$, $\Delta_\|$, and $\Delta_\perp$ either has the
has the symmetries  of $J_\|$ or $J_\perp$, or this product times
$\delta m$ has the symmetries  of $J_\|$ or $J_\perp$.
\begin{table}
\caption{The number of terms at a given order of expansion (with
respect to $\vec S$ or $\vec \Delta$) for $J_\|$ and $J_\perp$ and a
subtotal including all terms in both $J_\|$ and $J_\perp$ up to
the given order. Note that the subtotals agree with
formula (\ref{eq:terms}) indicating that all possible terms are used
in expanding $J_\|$ and $J_\perp$.
Here 1 indicates terms that are retained in the
equal-mass limit (or proportional to even powers of $\delta m$)
and  $\delta m$
indicates terms proportional to $\delta m$ (and other odd powers). The
number of terms in an expansion of $M_{\rm rem}$ agrees with the
corresponding number of terms in $J_\|$ and is provided for
convenience. }
\label{table:numberterms}
\begin{tabular}{|c|c|c|c|c|c|c|c|c|c|c|c|}
\hline\hline
Order & 0th & 0th & 1st & 1st & 2nd & 2nd & 3rd & 3rd & 4th & 4th & total\\
\hline
term & 1 & $\delta m$ & 1 & $\delta m$ & 1 & $\delta m$  & 1 & $\delta m$ & 1 & $\delta m$ & Sum \\
\hline\hline
$J_\|$ & 1 & 0 & 1 & 1 & 4 & 2 & 5 & 5 & 11 & 8 & 38\\
\hline
$J_\perp$ & 0 & 0 & 1 & 1 & 2 & 2 & 5 & 5 & 8 & 8 & 32\\
\hline
subtotal & -- & 1 & -- & 5 & -- & 15 & -- & 35 & -- & 70 & --\\
\hline\hline
$M_{\rm rem}$ & 1 & 0 & 1 & 1 & 4 & 2 & 5 & 5 & 11 & 8 & 38\\
\hline\hline
\end{tabular}
\end{table}

The possible terms to a given expansion order in spin (i.e., products
of $S$ and $\Delta$) are summarized in
Tables~\ref{table:Jp}-\ref{table:Jzm}.
In our phenomenological description the terms in Tables~\ref{table:Jp}
and~\ref{table:Jzm} are all multiplied by fitting coefficients.
Note that the coefficients of these terms can depend on higher
powers of $\delta m$ (i.e., higher even powers in $\delta m$ for those
terms proportional to $\delta m^0$, and
odd powers for terms proportional to $\delta m$).

\begin{table}
\caption{Symmetry properties of key quantities under parity (P) and
exchange of labels (X) [i.e., BH1$\leftrightarrow$ BH2].}
\label{table:symmetries}
\begin{tabular}{|l|c|c|}
\hline\hline
Symmetry & P & X \\
\hline\hline
$J_\|/m^2$ & + & + \\
\hline
$J_\perp/m^2$ & -- & -- \\
\hline
$m/M$ & + & + \\
\hline
$S_\perp/M^2=(S_1+S_2)_\perp/M^2$ & -- & -- \\
\hline
$S_\|/M^2=(S_1+S_2)_\|/M^2$ & + & + \\
\hline
$\Delta_\perp/M^2=(S_2/m_2-S_1/m_1)_\perp/M$ & -- & + \\
\hline
$\Delta_\|/M^2=(S_2/m_2-S_1/m_1)_\|/M$ & + & -- \\
\hline
$\hn=\hat{r}_1-\hat{r}_2$ & + & -- \\
\hline
$\delta m=(m_1-m_2)/M$ & + & -- \\
\hline\hline
\end{tabular}
\end{table}

\begin{widetext}

\begin{table}
\caption{Parameter dependence at each order of expansion
 for the final spin component perpendicular to the reference
 $\vec{L}$ direction.
 Here 1 indicates terms present in the
equal-mass limit (or proportional to even powers of $\delta m$)
and  $\delta m$
indicates terms proportional to $\delta m$ (and other odd powers).}
\label{table:Jp}
\begin{tabular}{|c|l|}
\hline\hline
$J_\perp$ & 0th order\\
\hline
1 & 0\\
\hline
$\delta m$ & 0\\
\hline\hline
$J_\perp$ & 1st order\\
\hline
1 & $S_\perp$\\
\hline
$\delta m$ & $\Delta_\perp$\\
\hline\hline
$J_\perp$ & 2nd order\\
\hline
1 & $S_\perp.S_\|+\Delta_\|.\Delta_\perp$\\
\hline
$\delta m$ & $S_\perp.\Delta_\|+\Delta_\perp.S_\|$\\
\hline\hline
$J_\perp$ & 3rd order\\
\hline
1 & $\Delta_\|.\Delta_\perp.S_\|+S_\perp.S_\|^2+S_\perp.\Delta_\|^2+S_\perp^3+S_\perp.\Delta_\perp^2$\\
\hline
$\delta m$ & $\Delta_\perp.\Delta_\|^2+\Delta_\perp.S_\|^2+S_\perp.\Delta_\|.S_\|+\Delta_\perp.S_\perp^2+\Delta_\perp^3$\\ 
\hline\hline
$J_\perp$ & 4th order\\
\hline
1 & $\Delta_\perp.\Delta_\|^3+S_\perp.S_\|^3+S_\perp.S_\|.\Delta_\|^2+\Delta_\perp.\Delta_\|.S_\|^2+S_\perp^3.S_\|+\Delta_\perp^3.\Delta_\|+S_\perp^2.\Delta_\perp.\Delta_\|+S_\perp.\Delta_\perp^2.S_\|$\\
\hline
$\delta m$ &
$S_\perp.\Delta_\|^3+\Delta_\perp.S_\|^3+\Delta_\perp.S_\|.\Delta_\|^2+S_\perp.\Delta_\|.S_\|^2+S_\perp^3.\Delta_\|+\Delta_\perp^3.S_\|+S_\perp^2.\Delta_\perp.S_\|+\Delta_\perp^2.S_\perp.\Delta_\|$\\ 
\hline\hline
\end{tabular}
\end{table}

\begin{table}
\caption{Parameter dependence at each order of expansion
for the final spin component along the reference
$\vec{L}$ direction and similarly for the remnant mass $M_{\rm rem}$.
Here 1 indicates terms present in the
equal-mass limit (or proportional to even powers of $\delta m$) 
and  $\delta m$ 
indicates terms proportional to $\delta m$ (and to other odd powers).}
\label{table:Jzm}
\begin{tabular}{|c|l|}
\hline\hline
$J_\|$ or $M_{\rm rem}$ & 0th order\\
\hline
1 & $L(S=0)$ or $M(S=0)$ \\
\hline
$\delta m$ & 0\\ 
\hline\hline
$J_\|$ or $M_{\rm rem}$ & 1st order\\
\hline
1 & $S_\|$\\ 
\hline
$\delta m$ & $\Delta_\|$\\
\hline\hline
$J_\|$ or $M_{\rm rem}$ & 2nd order\\
\hline
1 & $\Delta_\|^2+S_\|^2+\Delta_\perp^2+S_\perp^2$\\
\hline
$\delta m$ & $\Delta_\|.S_\|+\Delta_\perp.S_\perp$\\
\hline\hline
$J_\|$ or $M_{\rm rem}$ & 3rd order\\
\hline
1 & $S_\|.\Delta_\|^2+S_\|.S_\perp^2+\Delta_\perp.S_\perp.\Delta_\|+S_\|.\Delta_\perp^2+S_\|^3$\\ 
\hline
$\delta m$ & $\Delta_\perp.S_\perp.S_\|+\Delta_\|.S_\|^2+\Delta_\|.\Delta_\perp^2+\Delta_\|^3+\Delta_\|.S_\perp^2$\\
\hline\hline
$J_\|$ or $M_{\rm rem}$ & 4th order\\
\hline
1 & $\Delta_\perp.\Delta_\|.S_\perp.S_\|+\Delta_\perp^4+\Delta_\|^4+S_\perp^4+S_\|^4+\Delta_\perp^2.\Delta_\|^2+\Delta_\perp^2.S_\perp^2+\Delta_\perp^2.S_\|^2+\Delta_\|^2.S_\perp^2+\Delta_\|^2.S_\|^2+S_\perp^2.S_\|^2$\\
\hline
$\delta m$ & $S_\perp.\Delta_\perp^3+\Delta_\|.S_\|^3+\Delta_\|.S_\|.\Delta_\perp^2+S_\perp.\Delta_\perp.S_\|^2+\Delta_\|^3.S_\|+S_\perp^3.\Delta_\perp+\Delta_\|^2.S_\perp.\Delta_\perp+\Delta_\|.S_\perp^2.S_\|$\\
\hline\hline
\end{tabular}
\end{table}

\end{widetext}

For the discussion below, we will denote terms in $J$ independent of 
$\delta m$ (or dependent on even powers of $\delta m$)
 by $J(1)$ and terms proportional
to odd powers by $J(\delta m)$.
The expansion terms reported in Tables~\ref{table:Jp}
and~\ref{table:Jzm} are related to each other (which is a consequence
of the symmetries in Table~\ref{table:symmetries}).
For example, all terms for
the components of $\vec J$,
with the exception of the even-order spin dependent terms of $J_\|(1)$,
 can be obtained
from the terms in $J_\perp(1)$ under the following
transformations
\begin{eqnarray}
  J_\perp(1) \leftrightarrow J_\perp(\delta m) \quad{\rm under}\quad
[(S_\perp,\Delta_\perp)\leftrightarrow (\Delta_\perp,S_\perp)], \label{eq:sim1}\\
  J_\perp(\delta m) \leftrightarrow J_\|(\delta m) \quad{\rm
under}\quad [(\Delta_\perp,\Delta_\|)\leftrightarrow
(\Delta_\|,\Delta_\perp)]. \label{eq:sim2}
\end{eqnarray}
and for the odd-power in the spin variables only
\begin{equation}
J_\perp(1)\leftrightarrow J_\|(1)\quad{\rm under}\quad
[(S_\perp,S_\|)\leftrightarrow (S_\|, S_\perp)].\label{eq:sim3}
\end{equation}
The even-power (in the spin) terms in $J_\|(1)$ can
be obtained from $J_\perp(1)$, but the process is more complex.
Basically, one needs to take all terms in
$J_\|(1)$ that are at least linear in $S_\perp$ and then divide by
$S_\perp$. This means that there is a direct relationship
between the even-order terms in $J_\|(1)$ and the next
higher-order term in $J_\perp(1)$. This relationship also holds for
the odd-order terms and $\delta m$ terms, but 
Eqs.~(\ref{eq:sim1})-(\ref{eq:sim3})
are much more straightforward.
It remains to be seen if the corresponding coefficients can be
obtained (at least approximated) using these transformations.

\subsubsection{Modeling the equal-mass case}\label{sec:equal}

Using the above properties, and limiting our expansion to
fourth-order in the spin variables, we find that there are 
16 terms that
contribute to the perpendicular component of the final angular
momentum in the
equal-mass case (i.e.\ $J_\perp(1)$).
We can regroup all these terms
in the following symbolic form
\begin{eqnarray}\label{eq:regroup}
J_\perp=&&S_\perp.(1+\Delta_\perp^2+\cdots).(1+\Delta_\|^2+\cdots).\nonumber\\
&&.(1+S_\perp^2+\cdots).(1+S_\|+S_\|^2+S_\|^3+\cdots)+\nonumber\\
&&+\Delta_\perp.\Delta_\|.(1+\Delta_\perp^2+\cdots).(1+\Delta_\|^2+\cdots).\nonumber\\
&&.(1+S_\perp^2+\cdots).(1+S_\|+S_\|^2+\cdots).
\end{eqnarray}
Similar expansions can be generated for $J_\perp (\delta m)$ and
$J_\|(\delta m)$ using properties (\ref{eq:sim1}) and (\ref{eq:sim2}).

Because the perpendicular components of $\vec S$ and $\vec \Delta$ are
constructed by taking an inner product with unit vectors in the
plane, these components vary sinusoidally under rotation of the
directions of $\vec S$ and $\vec \Delta$. Thus $S_\perp = |S_\perp|
\cos\varphi$, etc., where we use the notation $|S_\perp|$ to denote
the magnitude of the projection of $\vec S$ onto the orbital plane.
Indeed, each term linear in the subindex $\perp$ would lead to
a $\cos(\varphi +\varphi_{[]})$ dependence ($\varphi_{[]}$ denotes
a constant that, in principle, is different for each term).
 In
our expansion above, there are ten terms directly proportional to 
$\cos(\varphi+\varphi_{[]})$ 
\begin{eqnarray}\label{eq:cosphi}
J_\perp^{\cos\varphi}&=&\left[S_\perp.(1+\Delta_\|^2+\cdots).(1+S_\|+S_\|^2+S_\|^3+\cdots)\right.\nonumber\\
	&&\left.+\Delta_\perp.\Delta_\|.(1+\Delta_\|^2+\cdots)\right.\nonumber\\
&&\left.\times(1+S_\|+S_\|^2+\cdots)\right].
\end{eqnarray}

In the same fashion we can regroup the 16 terms contributing up to
the fourth order expansion of $J_\|$ (or $M_{\rm rem}$)
\begin{eqnarray}\label{eq:regroupJz}
J_\|=&&(1+\Delta_\perp^2+\Delta_\perp^4+\cdots).(1+\Delta_\|^2+\Delta_\|^4+\cdots).\nonumber\\
&&.(1+S_\perp^2+S_\perp^4+\cdots).(1+S_\|+S_\|^2+S_\|^3+S_\|^4+\cdots)\nonumber\\
&&+\Delta_\perp.\Delta_\|.S_\perp.(1+\Delta_\perp^2+\cdots).(1+\Delta_\|^2+\cdots).\nonumber\\
&&.(1+S_\perp^2+\cdots).(1+S_\|+\cdots).
\end{eqnarray}

Nine terms are independent of $\perp$ components of $\vec S$ or $\vec
\Delta$ and are thus independent of $\cos(\varphi)$. The next set of
terms are quadratic in the subindex $\perp$. We may expect that these
quadratic terms lead a $\cos^2\varphi$ dependence. However, they
actually lead to both terms proportional to  $\cos^2\varphi$ and terms
independent of $\cos \varphi$. To see why, consider the term $S_\perp^2$.
This should really be written as $(\vec S\cdot \hat n_1)(\vec S\cdot
\hat n_2) = \vec S_\perp \cos\phi \vec S_\perp\cos(\phi-\phi_0)$,
where $\phi_0$ is the angle between $\hat n_1$ and $\hat n_2$. We then
have $(\vec S\cdot \hat n_1) (\vec S\cdot \hat n_2) =
S_\perp^2 \cos^2\varphi + S_\perp^2 (1-\cos\phi_0)/2$, where
$\varphi = \phi - \phi_0/2$ and $\phi_0$ is independent of
orientation. These terms are 
\begin{eqnarray}\label{eq:Jzcosphi}
J_\|^{1+\cos^2\varphi}=(1+\Delta_\|^2+\Delta_\|^4+\cdots)\nonumber\\
\times(1+S_\|+S_\|^2+S_\|^3+S_\|^4+\cdots)\nonumber\\
+\left[(\Delta_\perp^2+S_\perp^2)(1+\Delta_\|^2+\cdots)
(1+S_\|+S_\|^2+\cdots)\right.\nonumber\\
+\left.\Delta_\perp.\Delta_\|.S_\perp(1+S_\|+\cdots)
\right].
\end{eqnarray}
With an analogous expression for the final mass of the merged black
hole $M_{\rm rem}$.

Note that we actually fit to the magnitude of the remnant spin
$m^2 \alpha_{\rm rem}^2 = \left(J_\perp^{\cos\varphi}\right)^2 +
\left(J_\|^{1+\cos^2\varphi}\right)^2$ which contains terms proportional
to $(\cos\varphi)^0$, $(\cos\varphi)^2$, $\cdots$.

\section{Fitting Full Numerical Results}\label{sec:fitting}

We fit  the total radiated mass and final remnant spin as a
function of initial configuration for  the SP
and UD configurations described above,  the \hangup kick configurations (we
will denote these configurations with HK)
described in~\cite{Lousto:2011kp, Lousto:2012su},
and the N configurations of~\cite{Lousto:2012gt}.
The HK configurations are equal-mass, equal-spin-magnitude
configurations where the components of the spins along $\vec L$ are 
equal, and the components in the orbital plane are equal in magnitude
but opposite in direction. We evolved two types of HK configurations:
one with a dimensionless spin of 0.7 and the other with a
dimensionless spin of 0.9. We denote the latter by HK9 in the tables
below. The N configurations have equal-mass
BHs with one spinning and the other nonspinning.
The SP configurations  have $\vec S\neq0$ but
$\vec \Delta=0$, while the HK configurations have $\vec S_\perp =0$ and
$\Delta_\|=0$. For both families, $S_\|\neq0$, while for the UD
configurations, $S_\|=0$.
 In addition, we use
the results of the L, K, and S configurations of ~\cite{Lousto:2012gt} 
to verify the accuracy of our model. 
The K configurations are equal-mass, equal-spin magnitude
configurations where the components of the spin along $\vec L$ are
equal in magnitude and opposite in direction and the components in the
plane are equal in both magnitude and direction. The L configurations
are again equal-mass, equal-spin magnitude configurations, but here,
initially,
one BH spin is aligned with $\vec L$ and the other lies in the orbital
plane. Finally, the S configurations are equal-mass, equal spin
magnitude configurations, where, initially, the components of the spin
along $\vec L$ are equal in magnitude and opposite in direction and
the components in the plane are also equal in magnitude and opposite
in direction. For the N, L, K, and S configurations the dimensionless
spin of the BHs (when not zero) was set to 0.8.

Here we are interested in the total mass loss of the binary, starting from
an infinite separation 
  $$\delta {\cal M} = \frac{M_{1}^\infty + M_{2}^\infty- M_{\rm
rem}}{M_{1}^\infty + M_{2}^\infty} $$
(where $M_{1,2}^\infty$ are the masses of two BHs at infinite separation)
and the final remnant spin $\alpha_{\rm rem}$. We approximate $\delta
{\cal M}$  by
$$
  \delta {\cal M} \approx \frac{M_1 + M_2 - M_{\rm rem}}{M_1 + M_2},$$ where $M_{\rm rem}$ is the
remnant BH's mass and $M_1$ and $M_2$ are the initial masses of the two BHs in
the binary (which is a good approximation for their masses when the
binary was infinitely separated due to the near constancy of the BH
mass during the inspiral, as seen in, e.g., Fig~\ref{fig:spin_cons}).

The measured values of $\delta {\cal M}$ and $\alpha_{\rm rem}$ for the SP
and UD configurations are given in Table~\ref{tab:RemComp}. We found
that the values of $\delta {\cal M}$ and $\alpha_{\rm rem}$ obtained by the
IH formalism were significantly more accurate than the radiation-based measures of
these quantities (likely due to the resolution we used
in the wave zone and to our placement of the extraction spheres). Hence, for all fits, we use the IH results.

\begin{figure}
  \includegraphics[width=0.9\columnwidth]{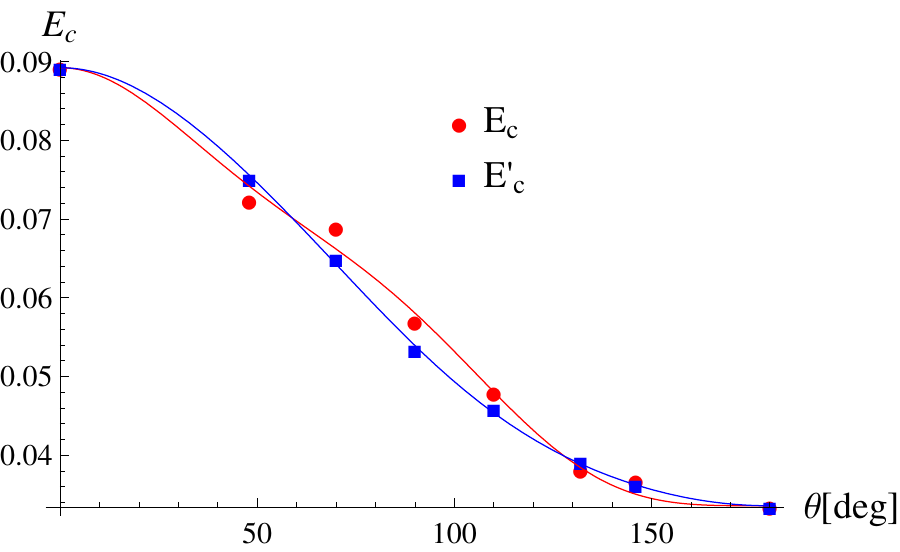}
  \includegraphics[width=0.9\columnwidth]{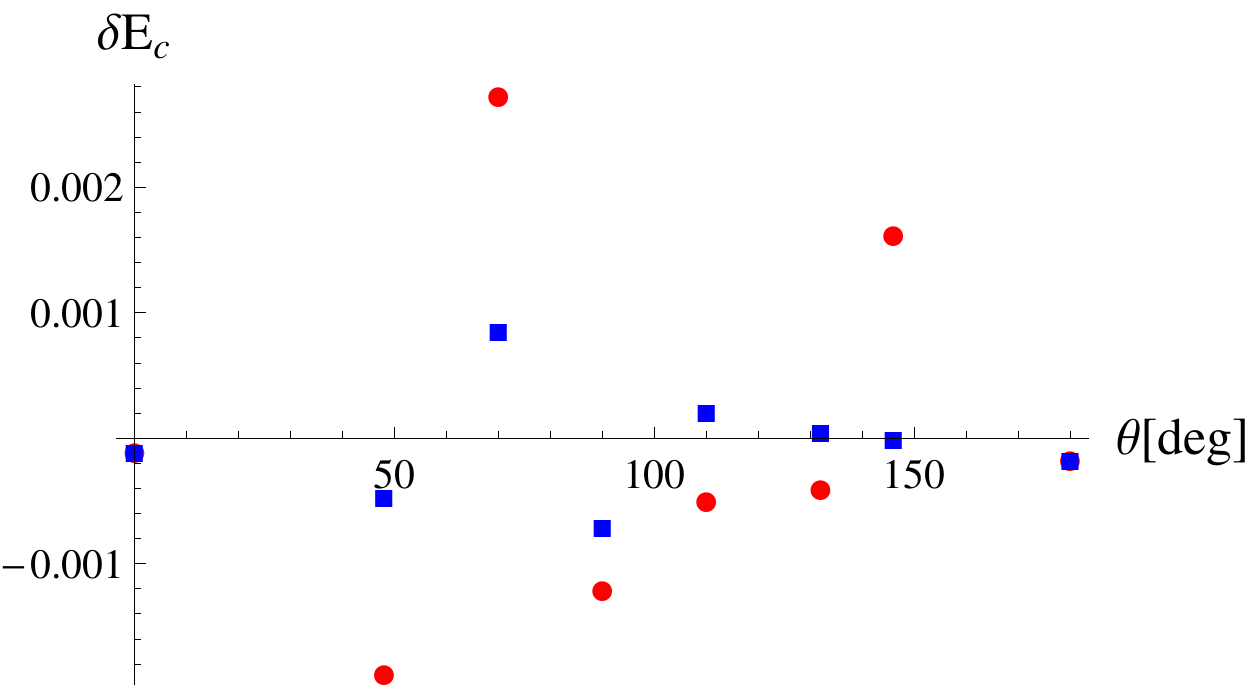}
  \caption{Fits of $E_{c}$ and $E_c'$ versus $\theta$ for the SP
configurations.  The upper panel shows the data and fit, while the
lower shows the residuals. Note that the scatter in $E_c'$ is
substantially better than for $E_c$.} \label{fig:FitE}
\end{figure}

\begin{figure}
  \includegraphics[width=0.9\columnwidth]{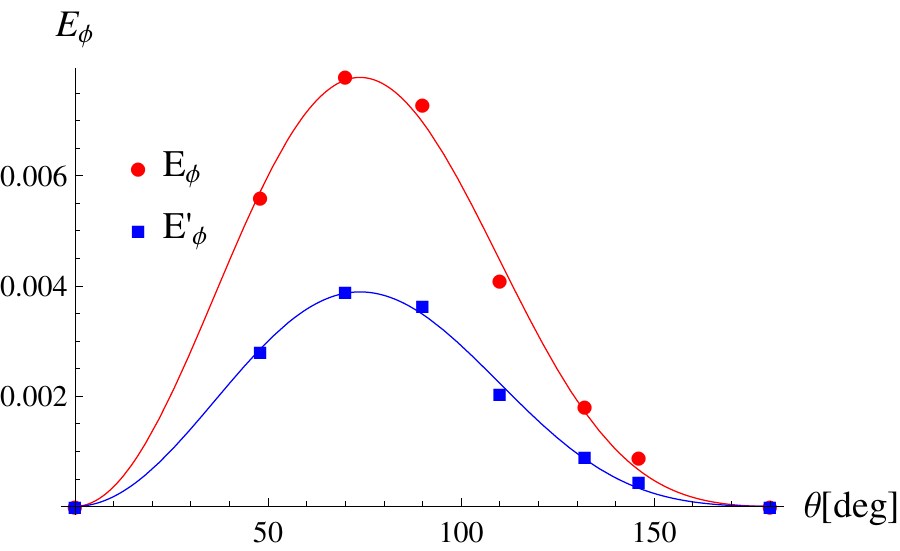}
  \includegraphics[width=0.9\columnwidth]{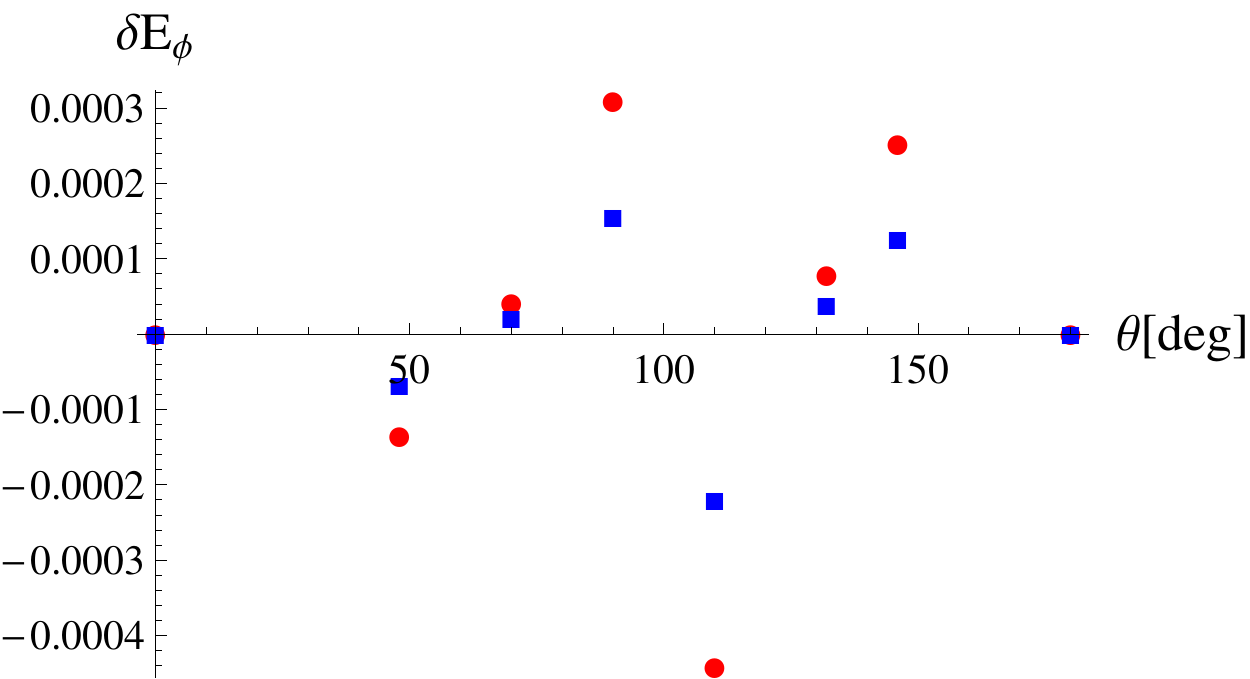}
  \caption{Fits of $E_\phi$ and $E_\phi'$
 (note that $E_\phi'=E_\phi/2$ identically)
versus $\theta$ for the SP configurations.  The upper panel shows the
data and fits, while the
lower shows the residuals.
}
\label{fig:FitEV}
\end{figure}

\begin{figure}
  \includegraphics[width=0.9\columnwidth]{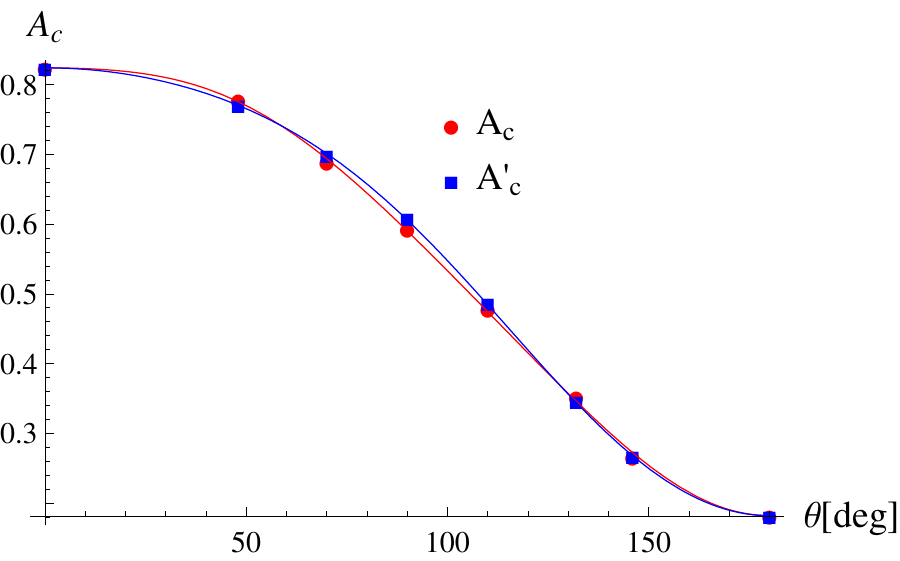}
  \includegraphics[width=0.9\columnwidth]{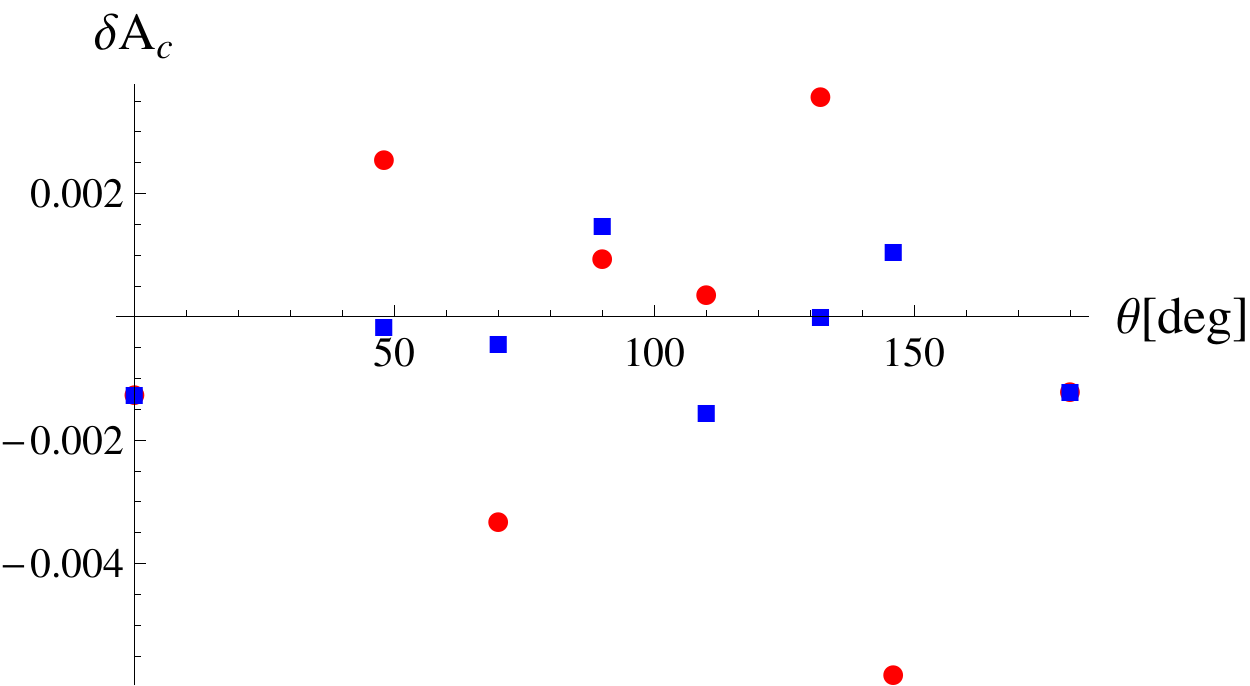}
  \caption{Fits of $A_{c}$ and, $A_c'$
versus $\theta$ for the SP configurations.  The upper panel shows the
data and fit, while the
lower shows the residuals.}
\label{fig:FitA}
\end{figure}

\begin{figure}
  \includegraphics[width=0.9\columnwidth]{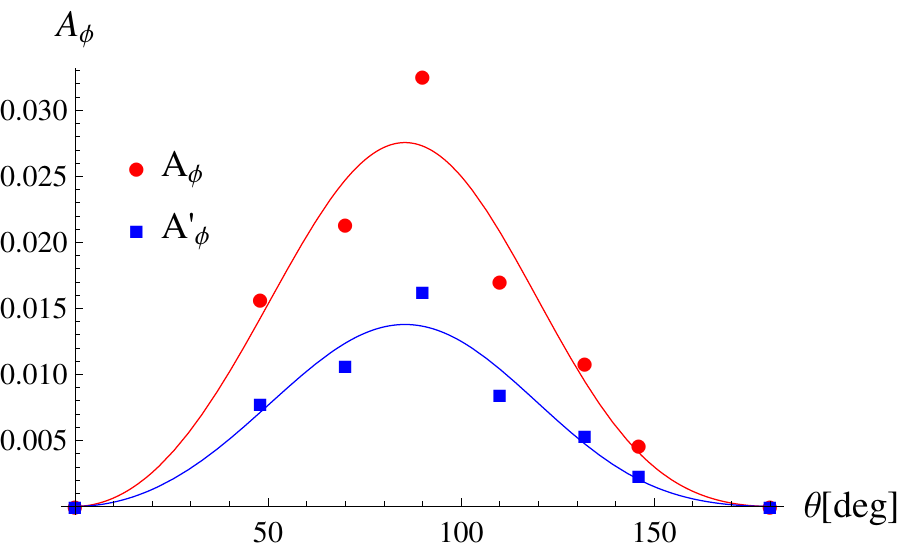}
  \includegraphics[width=0.9\columnwidth]{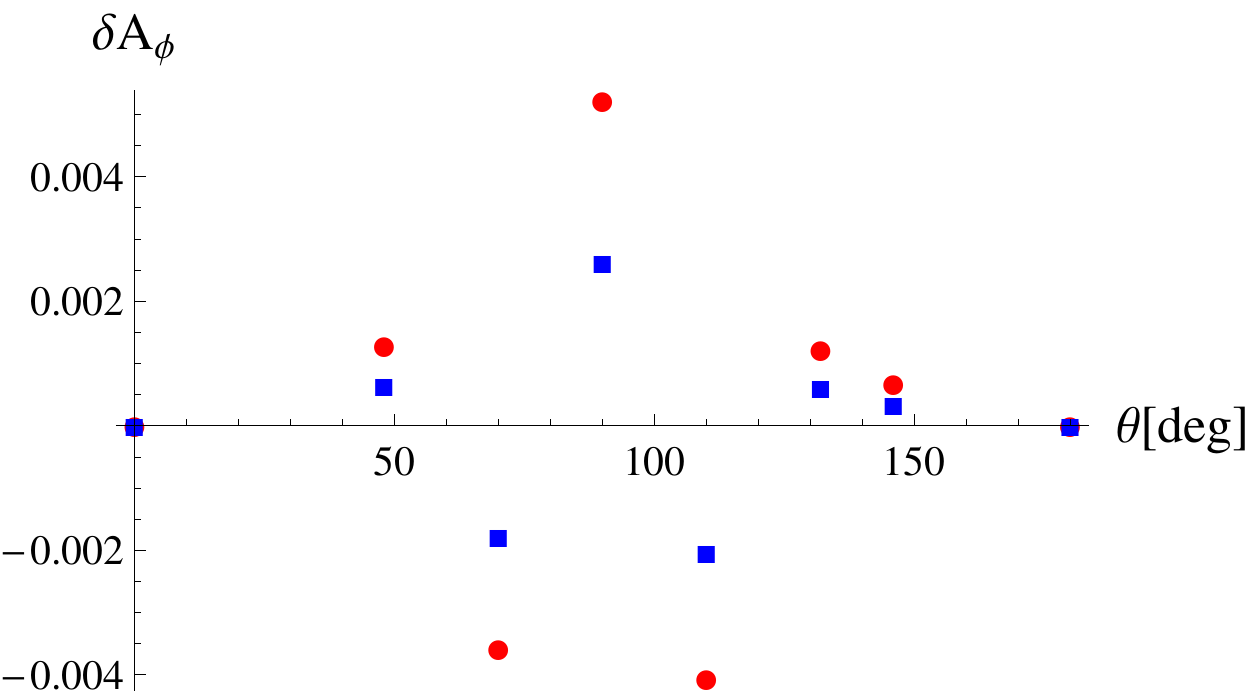}
  \caption{Fits of $A_{\phi}$ and $A_\phi'$
versus $\theta$ for the SP configurations
(note that $A_\phi'= A_\phi/2$ identically).  The upper panel shows the
data and fits, while the
lower shows the residuals.
Note that the fits to $A_\phi$  and $A_\phi'$ are rather poor.}
\label{fig:FitAV}
\end{figure}

Our fitting procedure is as follows. For each set of simulations with
the same (initial) inclination angle  $\theta$, we
fit $\delta {\cal M}$ and $\alpha_{\rm rem}^2$ to
\begin{eqnarray}
\delta {\cal M} = E_c + E_\phi \cos(\varphi-\varphi_e)^2,\label{eq:E}\\
\delta {\cal M} = E_c' + E_\phi' \cos (2 \varphi-2 \varphi_e'),\label{eq:E'}\\
\alpha_{\rm rem}^2 = A_c + A_\phi \cos(\varphi-\varphi_a)^2,\label{eq:A}\\
\alpha_{\rm rem}^2 = A_c' + A_\phi' \cos(2 \varphi-2 \varphi_a'),\label{eq:A'}
\end{eqnarray}
where $E_c$, $E_\phi$, $\varphi_e$,  $A_c$, $A_\phi$,
and $\varphi_a$ are fitting constants. While the expansion
$C_1 + C_2 \cos^2 \phi$ is mathematically equivalent to the
expansion $C_1' + C_2' \cos 2 \phi $ (i.e.\ $C_1' = C_1 + C_2/2$ and
$C_2'
= C_2/2$), we find that the constant term $C_1'$ is better modeled
as a function of $\theta$, as shown in
Figs.~\ref{fig:FitE}, \ref{fig:FitEV}, \ref{fig:FitA}, and \ref{fig:FitAV}. Fits to $A_\phi$ show
substantial scatter when compared to other quantities. The most likely
explanation for the scatter is that there were not enough $\phi$
configurations to accurately extract $A_\phi$.

As an aside, we note that we actually measure the angle $\phi$ using the
techniques of \cite{Lousto:2008dn}. Briefly, we measure the spin
direction of each BH near merger in a frame rotated in such a way
that the late-time trajectories overlap. The angle $\phi$ is then
the orientation of the spin of $BH1$ in this frame.

We then fit $E_c$, $E_c'$, $A_c$, and $A_c'$ as functions of $\theta$,
as follows. First we note that at $\theta=0$ and $\theta=180^\circ$,
we should obtain the associated hangup energies and final remnant
spins. In addition, because the resulting formulas should be an even
function of $\theta$ (i.e., $\pm\theta$ are equivalent), we fit
these functions to the form
$f(\theta) = \sin^2\theta\left(c_0 + c_1 \cos\theta + c_2 \cos^2\theta
+ \cdots\right)$.
However, since  $S_\perp\propto\sin\theta$ and $S_\|\propto\cos\theta$,
our actual fitting functions are
\begin{eqnarray}
  E_c = E_{\rm hu} +
    \left(\frac{S_\perp}{m^2}\right)^2\left(e_1  + e_2
\left(\frac{S_\|}{m^2}\right) +
 e_3 \left(\frac{S_\|}{m^2}\right)^2 \right),\nonumber \\
  A_c = A_{\rm hu}^2 +
 \left(\frac{S_\perp}{m^2}\right)^2\left(a_1  + a_2
\left(\frac{S_\|}{m^2}\right) +
 a_3 \left(\frac{S_\|}{m^2}\right)^2 \right),
\label{eq:constantfit}
\end{eqnarray}
where $E_{\rm hu}$ is the radiated energy for a pure \hangup
configuration with the same $S_\|$ and $A_{\rm hu}$ is the remnant spin
for the same \hangup configuration.
We use the formula provided in~\cite{Hemberger:2013hsa} for
both $E_{\rm hu}(\theta)$ and $A_{\rm hu}(\theta)$.
The fitting constants (with errors) are given in
Table~\ref{tab:sp_fits}.
In addition, we fit $E_\phi$ and $A_\phi$ to the forms
\begin{eqnarray}
  E_\phi = \left(\frac{S_\perp}{m^2}\right)^2\left({\cal E}_1  +
{\cal E}_2
\left(\frac{S_\|}{m^2}\right) +
 {\cal E}_3 \left(\frac{S_\|}{m^2}\right)^2 \right),\nonumber \\
  A_\phi = \left(\frac{S_\perp}{m^2}\right)^2\left({\cal A}_1  + {\cal
A}_2
\left(\frac{S_\|}{m^2}\right) +
 {\cal A}_3 \left(\frac{S_\|}{m^2}\right)^2 \right),
\label{eq:constantfitv}
\end{eqnarray}
These fitting constants (with errors) for the SP configurations are given in
Table~\ref{tab:sp_fits}, as well.
Note that we do not include terms proportional to $\Delta_\perp^4$. In
principle these terms should appear, but based the results from the
{\it hangup kick} effect~\cite{Lousto:2012su}, where we found the subleading terms
in $\Delta_\perp$ do not contribute significantly to the recoil, we
omit these terms here, as well. We test the validity of this
assumption later below by comparing the predictions of our formula
with the results from over 100 additional configurations.

\begin{widetext}

\begin{table}
\caption{Fitting parameters for the radiated energy and final remnant
spin for each TH family of SP configurations. The $\varphi_e$ and
$\varphi_a$ parameters are not shown (prime and unprimed).}
\begin{ruledtabular}
\begin{tabular}{l|llll}
 Family &  $E_c$ & $E_\phi$ & $E_c'$ & $E_\phi'$\\
SPTH48 & $0.07223\pm0.00066$ &$0.00560\pm0.00089$ &$0.07503\pm0.00040$ & $0.00280\pm0.00045$ \\
SPTH70 & $0.06874\pm0.00091$ &$0.00779\pm0.00107$ &$0.06484\pm0.00040$ &$0.00389\pm0.00054$\\
SPTH90 & $0.05687\pm0.00035$ &$0.00729\pm0.00060$ &$0.05323\pm0.00014$
&$0.00364\pm0.00030$ \\
SPTH110 & $0.04782\pm0.00037$ &$0.00410\pm0.00055$
&$0.04577\pm0.00018$ &$0.00205\pm0.00027$\\
SPTH132 & $0.03808\pm0.00006$ &$0.00182\pm0.00008$
&$0.03898\pm0.00003$ &$0.00091\pm0.00004$ \\
SPTH146 & $0.03660\pm0.00005$ &$0.00089\pm0.00008$
&$0.03615\pm0.00003$ &$0.00045\pm0.00004$\\
\hline
 Family &  $A_c$ & $A_\phi$ & $A_c'$ & $A_\phi'$\\
SPTH48 & $0.77775\pm0.00451$ &$0.01563\pm0.00602$ &$0.76994\pm0.00262$ &$0.00781\pm0.00301$\\
SPTH70 & $0.68821\pm0.00290$ &$0.02132\pm0.00291$ &$0.69887\pm0.00220$ &$0.01066\pm0.00146$\\
SPTH90 &$0.59141\pm0.00642$ &$0.03252\pm0.01126$ &$0.60767\pm0.00261$
&$0.01626\pm0.00563$\\
SPTH110 & $0.47790\pm0.00233$ &$0.01700\pm0.00338$
&$0.48640\pm0.00111$ &$0.00850\pm0.00169$\\
SPTH132 & $0.35056\pm0.00164$ &$0.01079\pm0.00209$
&$0.34516\pm0.00088$ &$0.00539\pm0.00105$\\
SPTH146 & $0.26443\pm0.00055$ &$0.00462\pm0.00081$
&$0.26674\pm0.00030$ &$0.00231\pm0.00041$\\
\hline
\end{tabular}
\end{ruledtabular}
\end{table}
\end{widetext}

\begin{table}
\caption{Fitting parameters for coefficient $E_c$ and $A_c$ as a function
of $S_\perp$ and $S_\|$ for the SP configurations.}
\label{tab:sp_fits}
\begin{ruledtabular}
\begin{tabular}{ll|ll}
$e_1$ & $0.0613\pm0.0080$ & $e_1'$ & $0.0356\pm0.0025$\\
$e_2$ & $0.0764\pm0.0388$ & $e_2'$ & $0.0960\pm 0.0122$\\
$e_3$ & $-0.3842\pm0.2133$ & $e_3'$ & $0.1217\pm 0.0671$\\
${\cal E}_1$ & $0.0434\pm0.0013$ &  ${\cal E}_1'$ & $0.0217\pm0.0006$ \\
${\cal E}_2$ & $0.0839\pm0.0061$ &  ${\cal E}_2'$ & $0.0419\pm0.0030$ \\
${\cal E}_3$ & $-0.0214\pm0.0334$ &  ${\cal E}_3'$ & $-0.0107\pm0.0167$ \\
$a_1$ & $ 0.7422\pm0.0168 $ & $a_1'$ & $ 0.8401 \pm 0.0061 $ \\
$a_2$ & $ -0.2588\pm0.0812 $ & $a_2'$ & $-0.3277 \pm 0.0294$ \\
$a_3$ & $ 1.3068\pm 0.4461$ & $a_3'$ & $-0.6088 \pm0.1616 $\\
${\cal A}_1$ & $0.1699\pm0.0158$ &  ${\cal A}_1'$ & $0.0849\pm0.0079$ \\
${\cal A}_2$ & $0.1000\pm0.0764$ &  ${\cal A}_2'$ & $0.0500\pm0.0382$ \\
${\cal A}_3$ & $-0.5023\pm0.4197$ &  ${\cal A}_3'$ & $-0.2512\pm0.2099$ \\
\end{tabular}
\end{ruledtabular}
\end{table}

The new formula for $A_c$ agrees quite will with the ansatz
in~\cite{Lousto:2013vpa}. In~\cite{Lousto:2013vpa}, we conjectured,
based on most of the above runs, that the final remnant spin could be
obtained from
$$
  \alpha_{\rm rem}^2 = \alpha_{\rm hu}^2 + \frac{S_\perp^2}{m^2}.
$$
The idea was that the in-plane spin was preserved and therefore
contributed entirely to the final remnant spin. The other components
of the spin arising from the hangup effect.
 Here we find, $$\alpha_{\rm rem}^2 = \alpha_{\rm hu}^2 +  0.84
\frac{S_\perp^2}{m^2}+ 0.085 \frac{S_\perp^2}{m^2}\cos(2\phi) + \mbox{(small corrections)}.$$ We thus see that, depending on orientation
between $3.8\%$ and $13.1\%$  (with an average of 8.4\%)
of the in-plane spin is actually lost to radiation.

We use an essentially identical procedure to fit the mass loss and
remnant spin from the \hangup kick (HK) configurations. The main difference
here is that $\sin\theta$ is proportional to $\Delta_\perp$, rather
than to $S_\perp$ ($\cos \theta$ is still proportional to $S_\|$).

The fitting function for the HK runs has the form
\begin{eqnarray}
  E_c &=& E_{\rm hu} +
    \left(\frac{\Delta_\perp}{m^2}\right)^2\left(\epsilon_1  + \epsilon_2
\left(\frac{S_\|}{m^2}\right) +
 \epsilon_3 \left(\frac{S_\|}{m^2}\right)^2 \right),\nonumber \\
  E_\phi&=&  \left(\frac{\Delta_\perp}{m^2}\right)^2\left(\tilde E_1  +
\tilde E_2
\left(\frac{S_\|}{m^2}\right) +
 \tilde E_3 \left(\frac{S_\|}{m^2}\right)^2 \right),\nonumber \\
  A_c &=& A_{\rm hu}^2 +
 \left(\frac{\Delta_\perp}{m^2}\right)^2\left(\zeta_1  + \zeta_2
\left(\frac{S_\|}{m^2}\right) +
 \zeta_3 \left(\frac{S_\|}{m^2}\right)^2 \right), \nonumber \\
 A_\phi &=&
 \left(\frac{\Delta_\perp}{m^2}\right)^2\left(\tilde A_1  + \tilde A_2
\left(\frac{S_\|}{m^2}\right) +
 \tilde A_3 \left(\frac{S_\|}{m^2}\right)^2 \right).
\end{eqnarray}
The fitted constants are provided in Table~\ref{tab:hu_fits}. Note here how the
correction to $A_c$ (above the hangup correction) is consistent with
zero.

\begin{table}
\caption{Fitting parameters for coefficient $E_c$, $E_\phi$, $A_c$,
and $A_\phi$ as functions
of $\Delta_\perp$ and $S_\|$ for the HK configurations. Note how the
$\zeta_i$ coefficients
are
consistent with zero.}
\label{tab:hu_fits}
\begin{ruledtabular}
\begin{tabular}{ll|ll}
$\epsilon_1$ & $ 0.0081 \pm 0.0062$ & $\epsilon_1'$ &  $
0.0043\pm0.0012 $ \\
$\epsilon_2$ &  $0.0075 \pm 0.0110$ & $\epsilon_2'$ & $ 0.0050\pm
0.0021$\\
$\epsilon_3$ & $-0.0984 \pm0.1394 $ & $\epsilon_3'$ & $-0.0090 \pm
0.0263$ \\
$\tilde E_1$ & $ 0.0051\pm 0.0004$ & $\tilde E_1'$ & $ 0.0025\pm
0.0002$ \\
$\tilde E_2$ & $ 0.0119\pm 0.0007$ & $\tilde E_2'$ & $ 0.0060\pm
0.0004$ \\
$\tilde E_3$ & $ 0.0316\pm0.0092 $ & $\tilde E_3'$ & $ 0.0158\pm0.0046 $ \\
$\zeta_1$ & $-0.0028 \pm0.0121 $ & $\zeta_1'$ & $ -0.0209\pm 0.0070$ \\
$\zeta_2$ & $-0.0007 \pm0.0213 $ & $\zeta_2'$ & $ -0.0381\pm0.0124 $\\
$\zeta_3$ & $-0.1420 \pm 0.2690$ & $\zeta_3'$ & $ 0.0429\pm 0.1568$\\
$\tilde A_1$ & $0.0234\pm0.0017$ & $\tilde A_1'$ & $0.0117\pm 0.0008$\\
$\tilde A_2$ & $0.0421\pm0.0030$ & $\tilde A_2'$ & $0.0210\pm 0.0015$\\
$\tilde A_3$ & $0.0561\pm0.0373$ & $\tilde A_3'$ & $0.0281\pm0.0187 $\\
\end{tabular}
\end{ruledtabular}
\end{table}

Finally, using the N configurations (only one of the black holes is spinning)
of Ref.~\cite{Lousto:2012gt}, we can
attempt to fit terms in the radiated mass and final spin proportional
to $\Delta_\|$. 
In order to do this, we must make an assumption regarding how the
$S_\perp^2$ terms and $\Delta_\perp^2$ terms interact. That is, each
one has an oscillatory behavior of the form $\cos^2(\phi-c)$, where
$c$ is some constant phase. Here, we assume that the two phases
are equal. This will introduce an error into our fits. However, as we
show below, this error is of the same order of magnitude as the error
in the $S_\perp^2$ and $\Delta_\perp^2$ terms themselves.

Thus, to fit the $\Delta_\|$ terms with the N configuration,
we take the coefficients in
Tables~\ref{tab:sp_fits}, \ref{tab:hu_fits}, and \ref{tab:UDfit}
 as exact and then fit
the energy radiated and final remnant spins to the forms
\begin{eqnarray}
  E_c &=& E_{c}^{\Delta_\|=0} +
\left(\frac{\Delta_\|}{m^2}\right)^2\left(E_{2} + k_2 S_\| + k_3
S_\|^2\right)\nonumber\\
  &&+ k_4 \Delta_\| \Delta_\perp S_\perp + E_{4} \Delta_\|^4 \label{eq:Ec}\\
  E_\phi &=& E_{\phi}^{\Delta_\|=0} +
  m_4 \Delta_\| \Delta_\perp S_\perp\\
  A_c &=& A_{c}^{\Delta_\|=0} +
\left(\frac{\Delta_\|}{m^2}\right)^2\left(A_{2} + n_2 S_\| + n_3
S_\|^2\right)\nonumber \\
 &&+ n_4 \Delta_\| \Delta_\perp S_\perp\\
  A_\phi &=& A_{\phi}^{\Delta_\|=0} +
 p_4 \Delta_\| \Delta_\perp S_\perp.\label{eq:Aphi}
\end{eqnarray}
Note here that $E_{2}$, $E_{4}$, and $A_{2}$ were taken to be the
values given in Table~\ref{tab:UDfit}.
Here, we found that most coefficients were consistent with zero (i.e.\
the confidence interval contained zero). We
removed those coefficients from the fitting functions and then
fit again.  Our results are summarized in Table~\ref{tab:N_fits}.
\begin{table}
\caption{Fitting parameters for coefficient $E_c$, $E_\phi$, $A_c$,
and $A_\phi$ as functions
of $\Delta_\|$ and $S_\|$ for the N configurations}
\label{tab:N_fits}
\begin{ruledtabular}
\begin{tabular}{ll|ll}
$k_2$ & $0$ & $k_2'$ & 0 \\
$k_3$ & $0$ & $k_3'$ & 0 \\
$k_4$ & $0.045\pm0.026$ & $k_4'$ & $0.0174\pm0.0086$ \\
$m_4$ & 0 & $m_4'$ & 0 \\
$n_2$ & 0 & $n_2'$ & $-0.0308\pm0.0086$ \\
$n_3$ & 0 & $n_3'$ &0 \\
$n_4$ & 0 & $n_4'$ &  0\\
$p_4$ & 0 & $p_4'$ & 0\\
\end{tabular}
\end{ruledtabular}
\end{table}

Generalizing these results from the N configurations introduces a
problem because there is a sign ambiguity in $ \Delta_\| \Delta_\perp
S_\perp$. To understand the source of this
ambiguity, we need to
consider how this term arises. 
In general, these terms come from interaction terms of the form
\begin{equation}
  (\vec \Delta\cdot \hat L)(\vec \Delta\cdot \hat e_\Delta)(\vec S\cdot
\hat e_S).
\end{equation}
If we align our coordinate system such that $\hat e_\Delta$ is aligned
with the $x$ axis and $\hat L$ is aligned with the $z$ axis, then
$$
   (\vec \Delta\cdot \hat L)(\vec \Delta\cdot \hat e_\Delta)(\vec
S\cdot
\hat e_S) = \Delta_\| |\Delta_\perp| |S_\perp| \cos\phi\ \cos(\phi -
\phi_S),$$
where $\phi_S$ is the angle between $\hat e_\Delta$ and $\hat e_S$.
This then becomes
\begin{eqnarray}
  (1/2)\left(\cos(2\phi-\phi_S) + \cos \phi_S\right)(\Delta_\| |\Delta_\perp| |S_\perp|).
\end{eqnarray}
That is, there are terms proportional to $\cos(2\phi - c)$ and a
{\it constant} term proportional the $\cos \phi_S$. Under exchange of
labels ($1\leftrightarrow2$)
$\vec \Delta$ rotates by $180^\circ$ and, consequently,
$\cos \phi_S$ changes sign.
We absorb $\cos \phi_S$ into the fitting constant, but have no way of
encoding the information about its sign. Later, we explore the
effects of simply dropping these ambiguous terms from our model.

\begin{widetext}

\begin{table}
\caption{The absolute and relative differences (${\rm AE}$ and ${\rm
RE}$) for
$E_c'$, $E_\phi'$, $A_c'$, $A_\phi'$,
 as measured directly by fitting to the
simulation data and
the predictions of Eqs.~(\ref{eq:Ec})-(\ref{eq:Aphi}).}
\label{tab:residuals}
\begin{ruledtabular}
\begin{tabular}{l|ll|ll|ll|ll|}
  Family & ${\rm AE}(E_c')$ & ${\rm RE}(E_c')\ \%$ &  ${\rm AE}(E_\phi')$ &
${\rm RE}(E_\phi')\ \%$ & ${\rm AE}(A_c')$ & ${\rm RE}(A_c')\ \%$ &  ${\rm
AE}(A_\phi')$ &
${\rm RE}(A_\phi')\ \%$\\
\hline
SPTH0 & 0.00011 & 0.122 & 0 & 0 & 0.00124 & 0.151 & 0 & 0\\
SPTH48 & 0.00047 & 0.625 & 0.0000674 & 2.407 & 0.01121 & 1.455 &
-0.00064 & -8.167\\
SPTH70 & -0.00085 & -1.314 & -0.0000209 & -0.536 & 0.00762 & 1.090 &
0.00179 & 16.783\\
SPTH90 & 0.00071 & 1.331 & -0.0001547 & -4.245 & -0.00149 & -0.245 &
-0.00261 & -16.030\\
SPTH110 & -0.00021 & -0.458 & 0.0002206 & 10.755 & -0.00233 & -0.478 &
0.00204 & 23.968\\
SPTH132 & -0.00005 & -0.124 & -0.0000390 & -4.294 & -0.00277 & -0.802
& -0.00061 & -11.253\\
SPTH146 & $10^{-5}$ & 0.013 & -0.0001264 & -28.355 & -0.00224 & -0.840
& -0.00033 & -14.447\\
SPTH180 & 0.00018 & 0.531 & 0 & 0 & 0.00121 & 0.669 & 0 & 0\\
HK22.5 & 0.00034 & 0.435 & -0.0000474 & -9.474 & -0.00200 & -0.263 &
-0.00023 & -12.523\\
HK45 & -0.00017 & -0.240 & $10^{-6}$ & 0.071 & 0.00097 & 0.142 &
0.00014 & 3.158\\
HK60 & 0.00003 & 0.048 & 0.0000170 & 1.116 & -0.00016 & -0.026 &
-0.00005 & -0.814\\
HK120 & 0.00006 & 0.152 & -0.0000301 & -3.891 & -0.00039 & -0.119 &
-0.00001 & -0.368\\
HK135 & -0.00007 & -0.187 & 0.0000327 & 6.846 & 0.00044 & 0.161 &
0.00002 & 0.816\\
HK9TH15 & -0.00045 & -0.458 & -0.0000770 & -14.759 & 0.01036 & 1.230
& -0.00090 & -38.694\\
HK9TH30 & -0.00042 & -0.475 & -0.0000148 & -0.989 & 0.01273 & 1.597 &
0.00263 & 114.180\\
HK9TH60L & -0.00023 & -0.332 & $-5.6*10^{-6}$ & -0.196 & -0.01295 &
-1.944 & -0.01431 & -56.678\\
UD00 & -0.00011 & -0.231 & 0 & 0 & 0.00016 & 0.033 & 0 & 0\\
UD60 & -0.00011 & -0.233 & 0 & 0 & 0.00017 & 0.036 & 0 & 0\\
UD70 & -0.00011 & -0.226 & 0 & 0 & 0.00017 & 0.036 & 0 & 0\\
UD80 & -0.00011 & -0.226 & 0 & 0 & 0.00021 & 0.045 & 0 & 0\\
UD85 & -0.00013 & -0.257 & 0 & 0 & 0.0001 & 0.021 & 0 & 0\\
NTH15 & -0.0001 & -0.154 & 0.0000228 & 22.374 & 0.00009 & 0.013 &
0.00003 & 6.638\\
NTH30 & -0.00003 & -0.044 & 0.0001060 & 30.942 & 0.00015 & 0.024 &
0.00009 & 6.264\\
NTH45 & 0.00006 & 0.104 & 0.0002808 & 49.797 & 0.00031 & 0.051 &
-0.00004 & -1.196\\
NTH60 & 0.00017 & 0.302 & 0.0003417 & 41.263 & 0.00039 & 0.067 &
0.00092 & 27.088\\
NTH120 & 0.00024 & 0.529 & 0.0002547 & 49.014 & -0.00001 & -0.002 &
0.00065 & 22.629\\
NTH135 & -0.00019 & -0.453 & 0.0000261 & 8.508 & 0.00064 & 0.170 &
-0.00010 & -6.273\\
NTH165 & 0.00010 & 0.258 & 0.0000393 & 105.030 & 0.00012 & 0.038 &
0.00017 & 93.699\\
S & 0.00010 & 0.205 & -0.0000629 & -7.169 & -0.00108 & -0.233 &
0.00023 & 6.555\\
L & 0.00008 & 0.118 & 0.0005827 & 44.456 & 0.00600 & 0.898 & 0.00151 &
32.595\\
KTH45 & -0.00076 & -1.450 & 0.0005348 & 44.254 & 0.00567 & 1.066 &
0.00037 & 5.665\\
KTH135 & -0.00069 & -1.371 & -0.0001528 & -23.034 & 0.00364 & 0.752 &
-0.00035 & -14.783\\
\end{tabular}
\end{ruledtabular}
\end{table}
\end{widetext}

Our final formula in the equal-mass case is given by
\begin{eqnarray}
  E_c = E_{\rm hu} &+
    \left(\frac{S_\perp}{m^2}\right)^2\left(e_1  + e_2
\left(\frac{S_\|}{m^2}\right) +
 e_3 \left(\frac{S_\|}{m^2}\right)^2 \right)+\nonumber\\
    &\left(\frac{\Delta_\perp}{m^2}\right)^2\left(\epsilon_1  + \epsilon_2
\left(\frac{S_\|}{m^2}\right) +
 \epsilon_3 \left(\frac{S_\|}{m^2}\right)^2 \right)+\nonumber \\
&\left(\frac{\Delta_\|}{m^2}\right)^2\left(E_{2} + k_2 S_\| + k_3
S_\|^2\right)+\nonumber\\
  &k_4 \Delta_\| \Delta_\perp S_\perp + E_{4} \Delta_\|^4
\end{eqnarray}
\begin{eqnarray}
  E_\phi = &\left(\frac{S_\perp}{m^2}\right)^2\left({\cal E}_1  +
{\cal E}_2
\left(\frac{S_\|}{m^2}\right) +
 {\cal E}_3 \left(\frac{S_\|}{m^2}\right)^2 \right)+\nonumber\\
  &\left(\frac{\Delta_\perp}{m^2}\right)^2\left(\tilde E_1  +
\tilde E_2
\left(\frac{S_\|}{m^2}\right) +
 \tilde E_3 \left(\frac{S_\|}{m^2}\right)^2 \right) +\nonumber\\
&m_4 \Delta_\| \Delta_\perp S_\perp\label{eq:empiricalE}
\end{eqnarray}

\begin{eqnarray}
A_c = A_{\rm hu}^2 &+
 \left(\frac{S_\perp}{m^2}\right)^2\left(a_1  + a_2
\left(\frac{S_\|}{m^2}\right) +
 a_3 \left(\frac{S_\|}{m^2}\right)^2 \right) +\nonumber\\
& \left(\frac{\Delta_\perp}{m^2}\right)^2\left(\zeta_1  + \zeta_2
\left(\frac{S_\|}{m^2}\right) +
 \zeta_3 \left(\frac{S_\|}{m^2}\right)^2 \right)+\nonumber\\
 &\left(\frac{\Delta_\|}{m^2}\right)^2\left(A_{2} + n_2 S_\| + n_3
S_\|^2\right)
\end{eqnarray}
\begin{eqnarray}
  A_\phi = &\left(\frac{S_\perp}{m^2}\right)^2\left({\cal A}_1  + {\cal
A}_2
\left(\frac{S_\|}{m^2}\right) +
 {\cal A}_3 \left(\frac{S_\|}{m^2}\right)^2\right) +\nonumber\\
&\left(\frac{\Delta_\perp}{m^2}\right)^2\left(\tilde A_1  + \tilde A_2
\left(\frac{S_\|}{m^2}\right) +
 \tilde A_3 \left(\frac{S_\|}{m^2}\right)^2 \right) + \nonumber\\
 &p_4 \Delta_\| \Delta_\perp S_\perp,\label{eq:empiricala}
\end{eqnarray}
where $\delta {\cal M} = E_c + E_\phi \cos(2\phi)$ and $\alpha_{\rm rem}^2 =
A_c + A_\phi \cos(2\phi)$ and the coefficients are given by the primed
quantities in Tables~\ref{tab:UDfit}, \ref{tab:sp_fits}, 
\ref{tab:hu_fits},
and \ref{tab:N_fits}, or  $\delta {\cal M} = E_c + E_\phi \cos^2(\phi)$ and
$\alpha_{\rm rem}^2 =
A_c + A_\phi \cos^2(\phi)$ and the coefficients are given by the unprimed
quantities. As the angle $\phi$ is not known in general, we
recommend using these formulas in statistical studies where one may
assume a uniform (or perhaps biased) distribution in $\phi$.

In terms of parameters of the binary, Eqs.~(\ref{eq:empiricalE})-(\ref{eq:empiricala}) can be
re-written as
\begin{eqnarray}
  \delta {\cal M} &=& \delta {\cal M}_{\rm hu} +\nonumber\\
  &&  \left(\frac{S_\perp}{m^2}\right)^2\left(e_1  + e_2
\left(\frac{S_\|}{m^2}\right) +
 e_3 \left(\frac{S_\|}{m^2}\right)^2 \right)+\nonumber\\
&&    \left(\frac{\Delta_\perp}{m^2}\right)^2\left(\epsilon_1  + \epsilon_2
\left(\frac{S_\|}{m^2}\right) +
 \epsilon_3 \left(\frac{S_\|}{m^2}\right)^2 \right)+\nonumber \\
&&\left(\frac{\Delta_\|}{m^2}\right)^2\left(E_{2} + k_2 \frac{S_\|}{m^2} + k_3
\left(\frac{S_\|}{m^2}\right)^2\right)+\nonumber\\
&&  k_4 \Delta_\| \Delta_\perp S_\perp + E_{4} \Delta_\|^4 +\nonumber\\
 && \left(\frac{S_\perp\cdot \hat n}{m^2}\right)^2\left({\cal E}_1  +
{\cal E}_2
\left(\frac{S_\|}{m^2}\right) +
 {\cal E}_3 \left(\frac{S_\|}{m^2}\right)^2 \right)+\nonumber\\
&&  \left(\frac{\Delta_\perp\cdot \hat l}{m^2}\right)^2\left(\tilde E_1  +
\tilde E_2
\left(\frac{S_\|}{m^2}\right) +
 \tilde E_3 \left(\frac{S_\|}{m^2}\right)^2 \right) +\nonumber\\
&&m_4 \Delta_\| \Delta_\perp S_\perp \label{eq:full_empirical_E}
\end{eqnarray}
\begin{eqnarray}
\alpha_{\rm rem}^2 &=& A_{\rm hu}^2 +\nonumber\\
 && \left(\frac{S_\perp}{m^2}\right)^2\left(a_1  + a_2
\left(\frac{S_\|}{m^2}\right) +
 a_3 \left(\frac{S_\|}{m^2}\right)^2 \right) +\nonumber\\
&& \left(\frac{\Delta_\perp}{m^2}\right)^2\left(\zeta_1  + \zeta_2
\left(\frac{S_\|}{m^2}\right) +
 \zeta_3 \left(\frac{S_\|}{m^2}\right)^2 \right)+\nonumber\\
&& \left(\frac{\Delta_\|}{m^2}\right)^2\left(A_{2} + n_2
\left(\frac{S_\|}{m^2}\right) + n_3
\left(\frac{S_\|}{m^2}\right)^2\right) +\nonumber\\
&&  \left(\frac{S_\perp\cdot \hat n}{m^2}\right)^2\left({\cal A}_1  + {\cal
A}_2
\left(\frac{S_\|}{m^2}\right) +
 {\cal A}_3 \left(\frac{S_\|}{m^2}\right)^2\right) +\nonumber\\
&& \left(\frac{\Delta_\perp\cdot \hat l}{m^2}\right)^2\left(\tilde A_1  + \tilde A_2
\left(\frac{S_\|}{m^2}\right) +
 \tilde A_3 \left(\frac{S_\|}{m^2}\right)^2 \right) + \nonumber\\
 && p_4 \Delta_\| \Delta_\perp S_\perp,\label{eq:full_empirical_A},
\end{eqnarray}
where $\hat n$ and $\hat l$ are unit vectors in the orbital plane. In
general, we do not know the orientation of these two vectors. In
our error analysis of the N configurations above, we assumed the two
directions are equal. This introduces an error into our formula which
we quantify by comparing the predicted and measured values of
$\delta {\cal M}$ and $\alpha_{\rm rem}$. In practice, we suggest
taking a fiducial orientation for $\hat n =\hat l$ and then considering
distributions of all possible azimuthal orientations of the initial spin.
The resulting distribution of remnants will be independent of the
fiducial choice of $\hat n$.
Note that
Eqs.~(\ref{eq:full_empirical_E})~and~(\ref{eq:full_empirical_A})
use the form $\delta {\cal M}  = E_c + E_\phi \cos^2\phi$, etc.. To use the
primed coefficients, replace terms like $(\vec V_\perp \cdot p)^2$
with $2[(\vec V_\perp \cdot p)^2 - \vec V_\perp^2]$.
The latter is more accurate only when dropping the $E_\phi$ and
$A_\phi$ terms, otherwise, the two expressions are equivalent.

For reference, we also give the  formulas of
\cite{Hemberger:2013hsa}
 for $\delta {\cal M}_{\rm hu}$
and $A_{\rm hu}$ in terms of $S_\|$ here
\begin{eqnarray}
  \delta {\cal M}_{\rm hu} &=& 0.0025829 -
   \frac{0.0773079}{2
\frac{S_\|}{m^2}  - 1.693959},\\
 A_{\rm hu} &=& 0.68640260 +  
0.3066014 \left(\frac{2 S_\|}{m^2}\right) -\nonumber\\
&&0.0268433 \left(\frac{2 S_\|}{m^2}\right)^2 - 
 0.0098019 \left(\frac{2 S_\|}{m^2}\right)^3 -\nonumber\\
&& 0.0049935
\left(\frac{2 S_\|}{m^2}\right)^4.
\end{eqnarray}

\begin{figure}[!h]
  \includegraphics[width=\columnwidth]{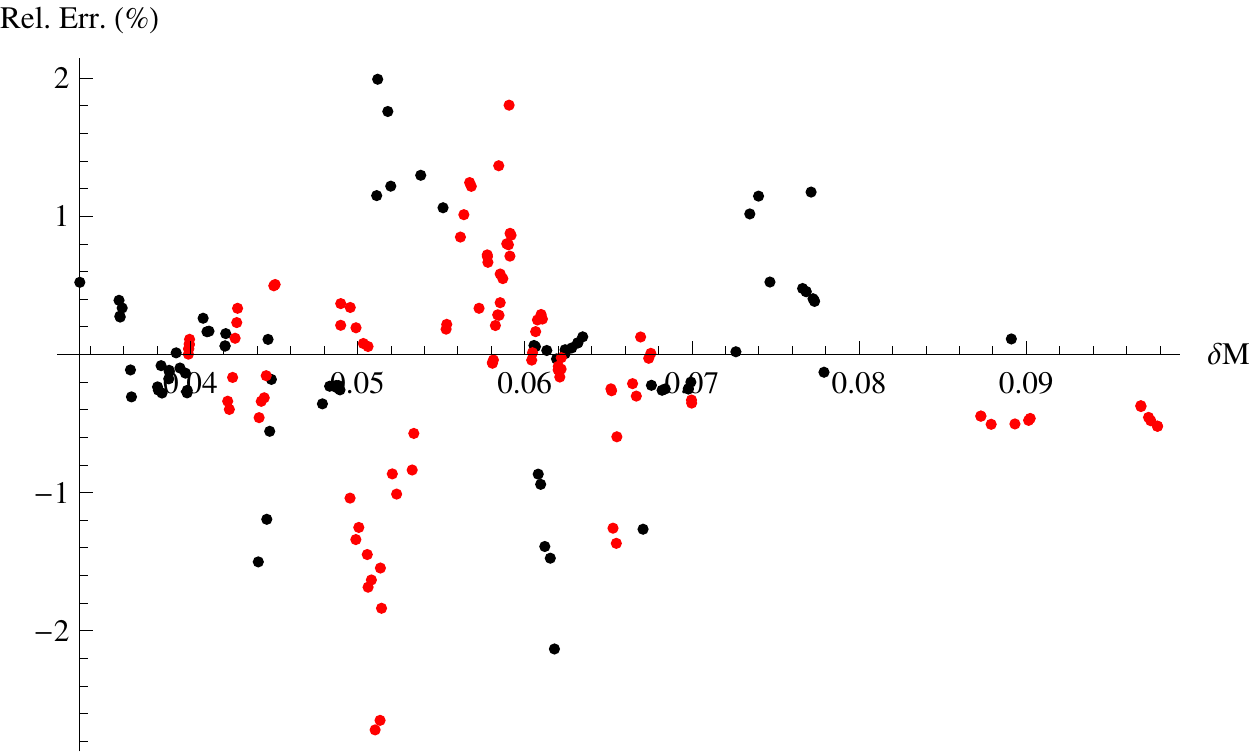}
  \includegraphics[width=\columnwidth]{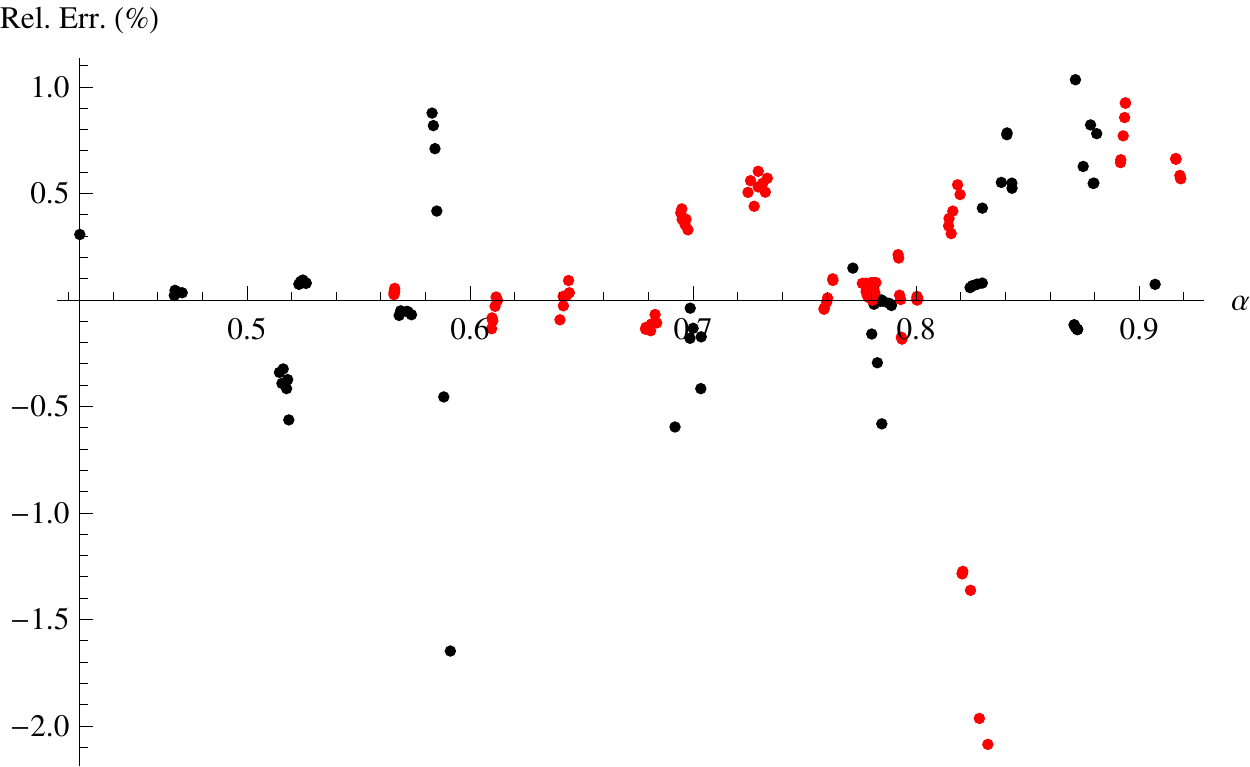}
  \caption{The relative errors in the predicted values of 
$\delta {\cal M}$ and $\alpha_{\rm rem}$  in percent for all 175 runs used in the
fits and analysis in this paper. Here the black dots correspond to
simulations used in the fits and the red dots to independent
simulations. Because we drop terms of the form $\Delta_\| \Delta_\perp
S_\perp$, the N configurations are independent runs here.}  \label{fig:all_resid}
\end{figure}

\begin{figure}[!h]
  \includegraphics[width=\columnwidth]{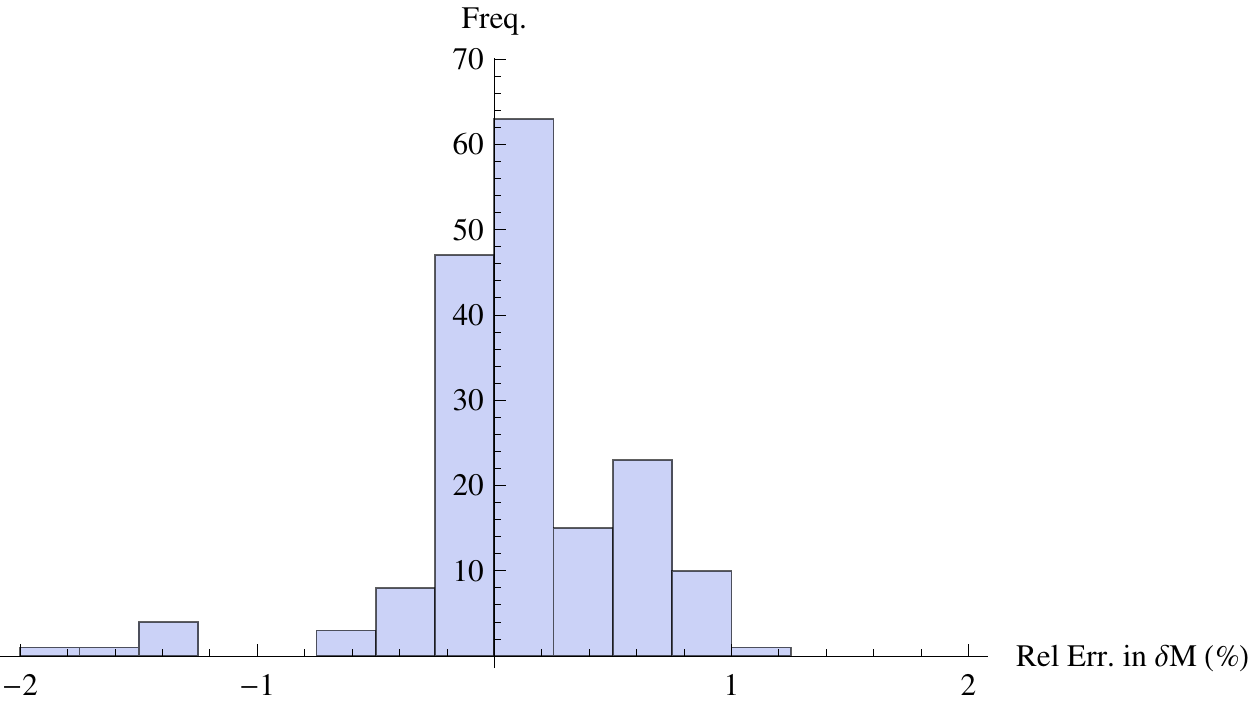}
  \includegraphics[width=\columnwidth]{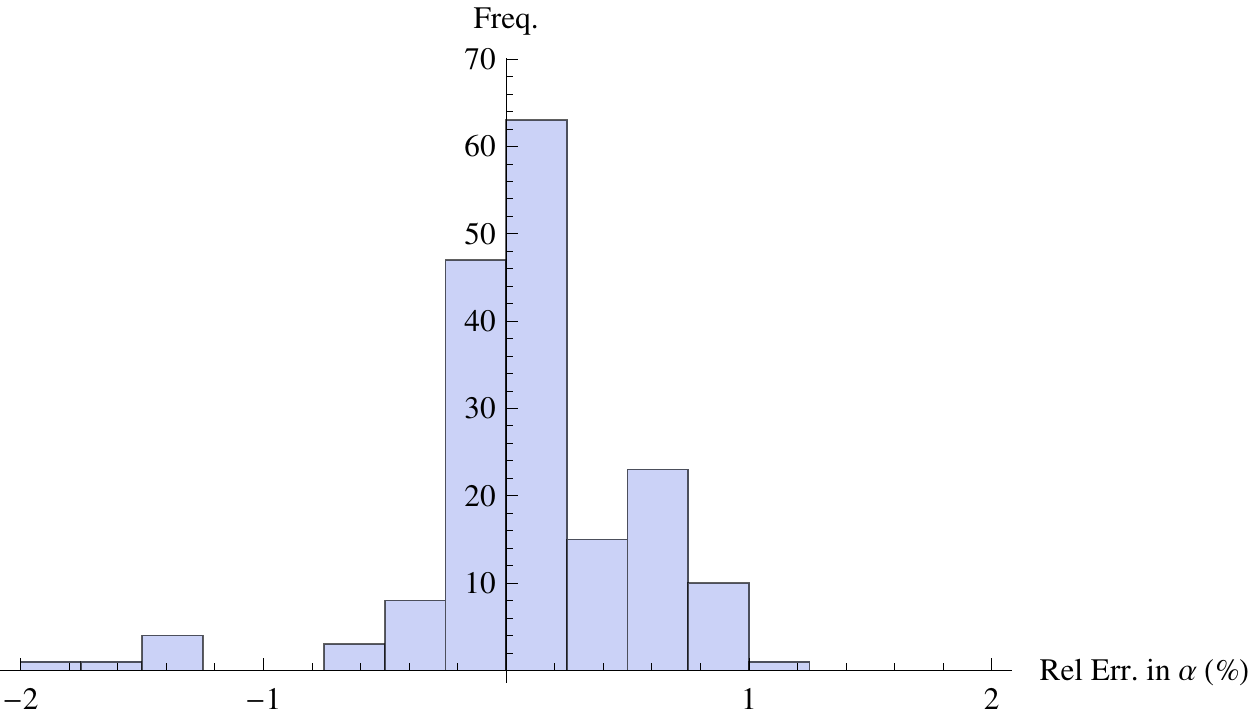}
  \caption{Histograms of the relative errors in
Fig.~\ref{fig:all_resid}. The $x$ axis is relative error in percent
and the bin width is $0.25\%$.}  \label{fig:resid_histo}
\end{figure}

To measure the error in our empirical formula, we provide two types
of residuals. First, for a given family with constant $\theta$ we
compare the measured $E_c'$, $E_\phi'$, $A_c'$, and $A_\phi'$ of
that family with the predictions of 
Eq.~(\ref{eq:empiricalE})-(\ref{eq:empiricala}), where we use the
average value of $S_\|$, $S_\perp$, $\Delta_\|$, and $\Delta_\perp$ of
that family. Second, we measure the residuals of each member of all
families using the known values of $\phi$ (which were obtained from
the original fits). When calculating these residuals, we also consider
the K, L, and S families of~\cite{Lousto:2012gt}. We show the first type of
residual in Table~\ref{tab:residuals} and we show the relative errors
in $\delta {\cal M}$ and $\alpha_{\rm rem}$ for all 175 configurations in
Fig.~\ref{fig:all_resid} and Fig.~\ref{fig:resid_histo}. However,
because of a sign ambiguity in the term $ \Delta_\|
\Delta_\perp S_\perp$, we do not include these terms when calculating
the residuals in Figs.~\ref{fig:all_resid} and \ref{fig:resid_histo}.
This also means that the N configuration do not contribute to the
fitting formulas and can therefore be used to verify the accuracy of
the formulas for more generic configurations.
We find that the relative errors are within $\pm2.5\%$ and
that the
distribution of errors in $\delta {\cal M}$  is wider but more symmetric then
the distribution of errors in $\alpha_{\rm rem}$.
As an aside, we note that the large relative errors in the
coefficients in $A_\phi$ and 
$E_\phi$ only have a small effect on the net error in the predictions.
The errors in the coefficients seen in
Table~\ref{tab:residuals} are a consequence of the smallness of $E_\phi$ and
$A_\phi$ relative to $E_c$ and $A_c$ (and because we would ideally
choose  twice as many $\phi$ configurations to accurately fit a $\cos 2\phi$
dependence).

\begin{figure}
\includegraphics[width=\columnwidth]{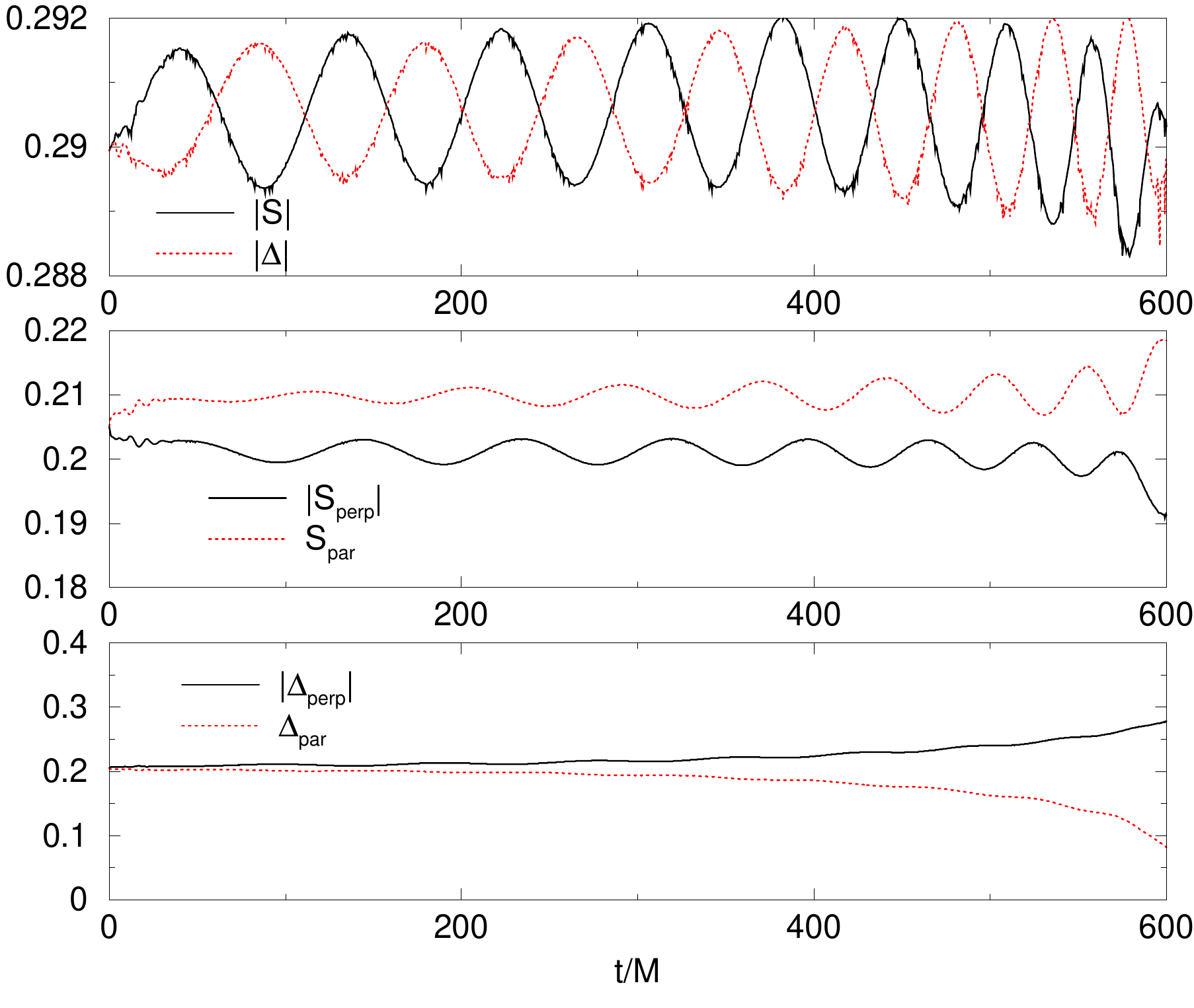}
\caption{
The magnitudes of $\vec S$ and $\vec \Delta$ (top panel),
  the magnitudes of the components of $\vec S$ (middle panel), and
the components of $\vec \Delta$ (lower panel) for a typical L
configuration (one spin initially in the plane, one spin perpendicular
to the plane). The magnitude of $\vec S$ and $\vec \Delta$ are both
conserved to a high degree (note the different scales). The
components of $\vec S$ are also conserved, but not to the same degree
as $|\vec S|$ itself. Finally, the components of $\vec \Delta$ are not
conserved.}\label{fig:L_Spin_Dir}
\end{figure}

\begin{figure}
\includegraphics[width=\columnwidth]{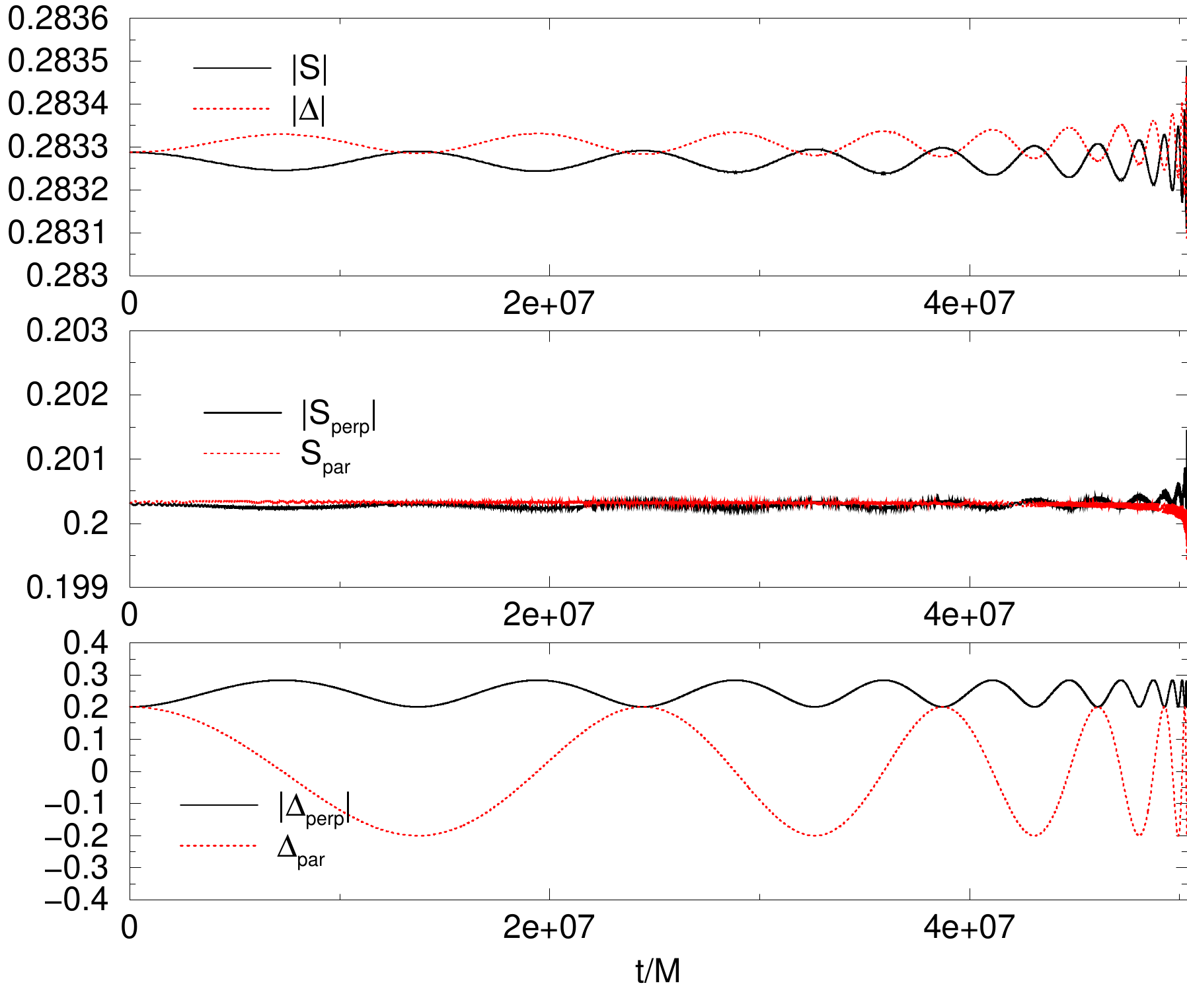}
\caption{The magnitudes of $\vec S$ and $\vec \Delta$ (top panel),
  the magnitudes of the components of $\vec S$ (middle panel), and
the components of $\vec \Delta$ (lower panel) for a 3.5PN evolution of a
typical L
configuration (one spin initially in the plane, one spin perpendicular
to the plane) starting from an initial quasicircular orbit at
a separation of $\sim 158M$. 
The magnitude of $\vec S$ and $\vec \Delta$ are both
conserved to a high degree (note the different scales). The
components of $\vec S$ are also conserved, but not to the same degree
as $|\vec S|$ itself. Finally, the components of $\vec \Delta$ are not
conserved.}\label{fig:PN_Spin_Dir}
\end{figure}

Our formulas are based on the vectors $\vec S$ and $\vec \Delta$. As
shown in Fig.~\ref{fig:L_Spin_Dir}, $S_\|$ and the magnitude of
$\vec S_\perp$ are conserved (approximately) throughout
a simulation. This is consistent with the predictions of PN theory
(see Fig.~\ref{fig:PN_Spin_Dir}).  However, while $|\vec \Delta|$ is
conserved (which is a consequence of $|\vec{\Delta}/2|^2 =
|\vec{S}_1|^2+|\vec{S}_2|^2 - |\vec{S}|^2$ for equal mass binaries),
 the components $\Delta_\|$ and $\Delta_\perp$ are not
(in general). Due to symmetry, the components of $\vec \Delta$ are
conserved for the SP, HK (due to $\pi$-symmetry the spins cannot rotate
with respect to each other), and N configurations (where
 $\vec \Delta = 2 \vec S$). The question then arises, where should
one evaluate $\Delta_\|$ and $\Delta_\perp$. For our runs, this issue
only arises for the K, L, and S configurations. Fortunately, this
represents 24 separate simulations. Thus we were able to explore this
question. We found that using the initial values of 
$\Delta_\|$ and $\Delta_\perp$ (which was done in
Table~\ref{tab:residuals} and Fig.~\ref{fig:all_resid})
reproduced the
measured $\delta {\cal M}$ and $\alpha_{\rm rem}$ with sufficient
accuracy (i.e.\ within $2.5\%$).


Finally, we note that while Eq.~(\ref{eq:full_empirical_A}) 
only provides the magnitude
of $\alpha_{\rm rem}$, we can obtain its direction (to within a few
degrees) from the direction
of $\vec J$ at any point along the binaries evolution.

\section{discussion}\label{sec:discussion}

\begin{figure}
  \includegraphics[width=0.9\columnwidth]{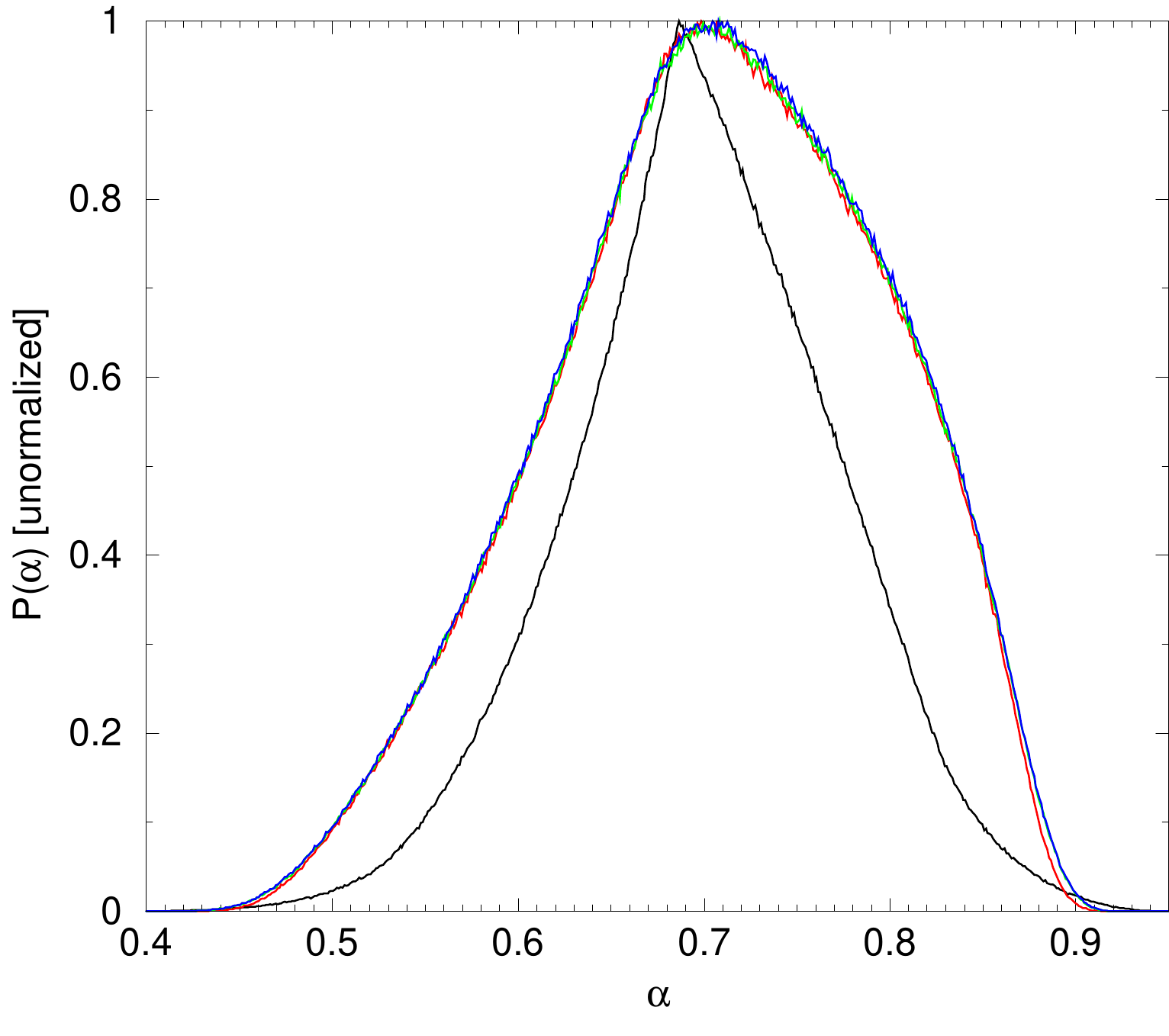}
  \caption{The approach to a {\it universal} spin distribution after a
few mergers when the BH spins are assumed to be uniform on the 
entire sphere $\theta < 180^\circ$. The wide black curve corresponds to
the first generation. The wide red curve corresponds to the second
generation. The remaining generations are green (3rd), and blue
(4th).}\label{fig:th180_dist}
\end{figure}

\begin{figure}
  \includegraphics[width=0.9\columnwidth]{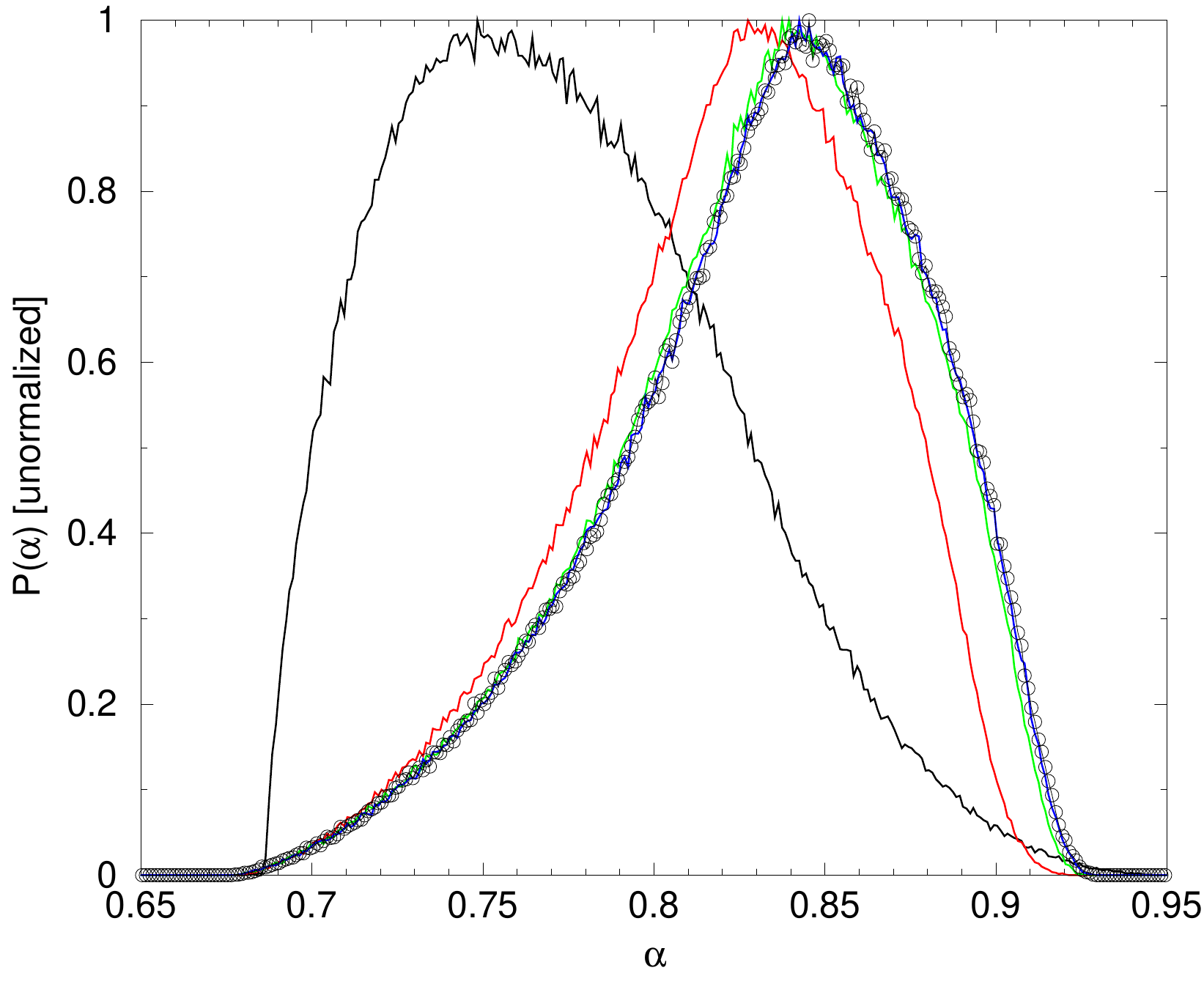}
  \caption{The approach to a {\it universal} spin distribution after a
few mergers when the BH spins are assumed to be uniform on the section
of the sphere $\theta < 90^\circ$. The wide black curve corresponds to
the first generation. The wide red curve corresponds to the second
generation. The remaining generations are green (3rd), blue (4th), and
black circles (5th).}\label{fig:th90_dist}
\end{figure}

\begin{figure}
  \includegraphics[width=0.9\columnwidth]{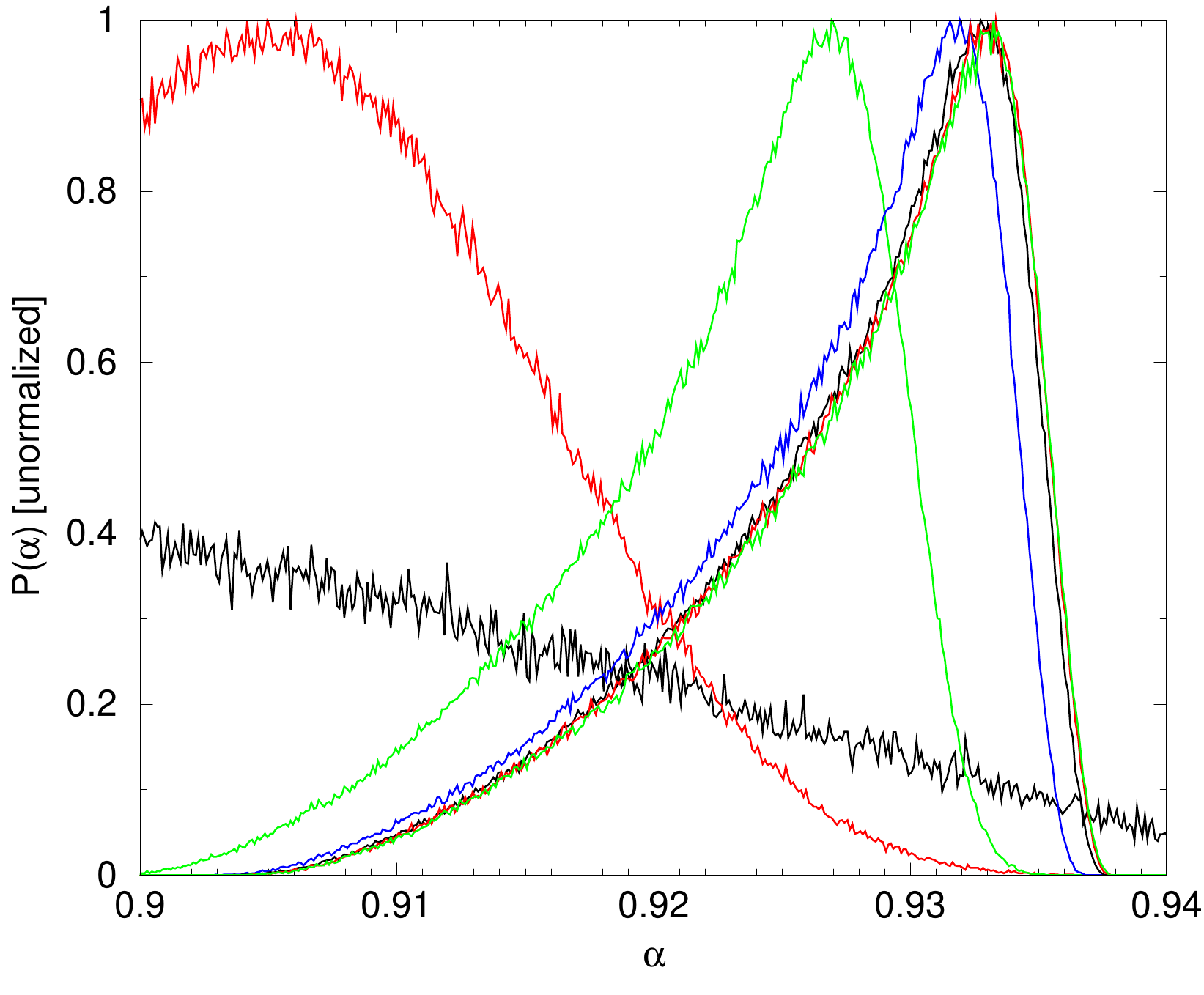}
  \caption{The approach to a {\it universal} spin distribution after a
few mergers when the BH spins are assumed to be uniform on the section
of the sphere $\theta < 30^\circ$. The wide black curve corresponds to
the first generation. The wide red curve corresponds to the second
generation. The remaining generations are green (3rd), blue (4th), followed by the
narrow black (5th), red (6th), and green(7th)
curves.}\label{fig:th30_dist}
\end{figure}

\begin{figure}

  \includegraphics[width=0.99\columnwidth]{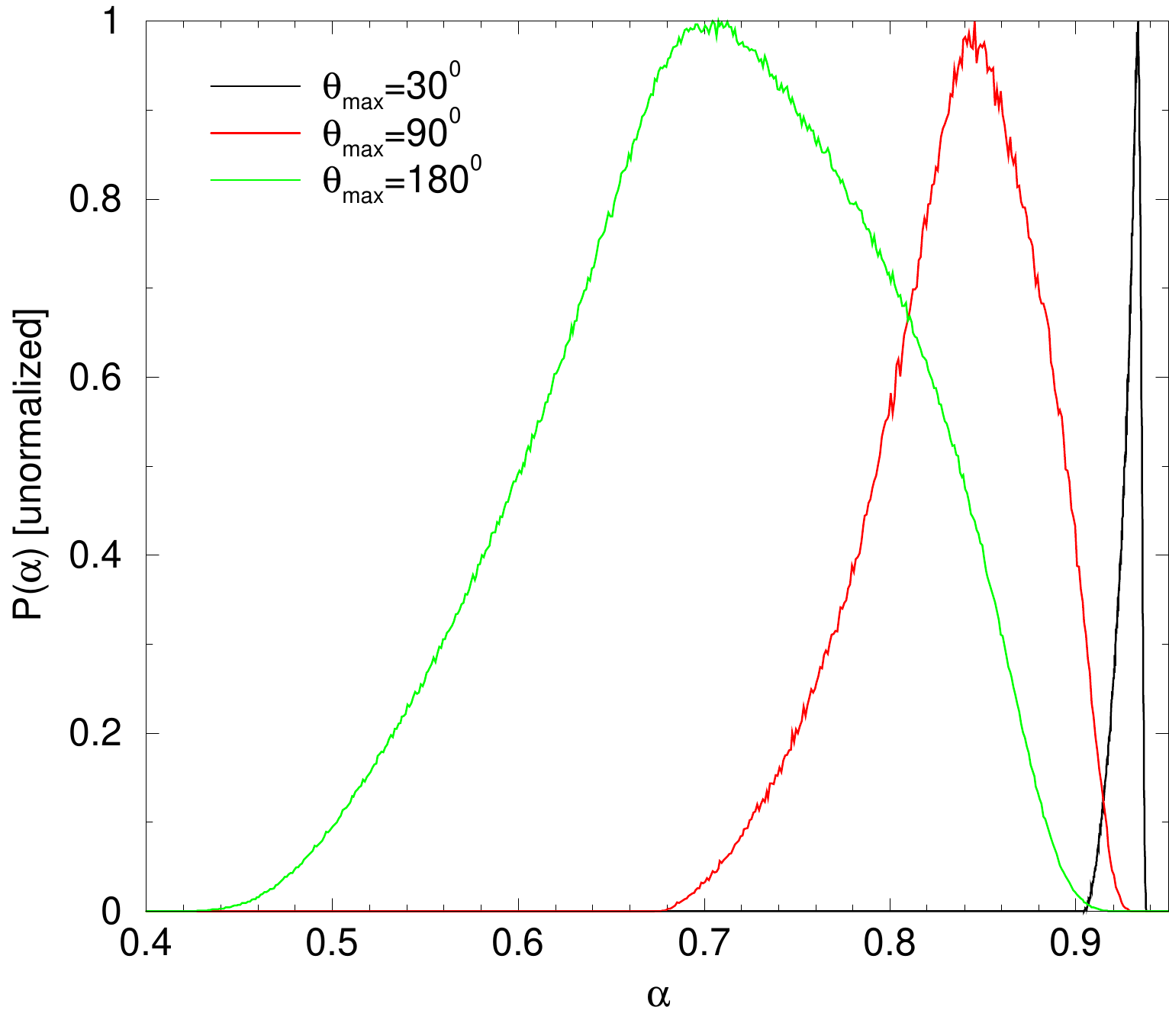}
  \includegraphics[width=0.99\columnwidth]{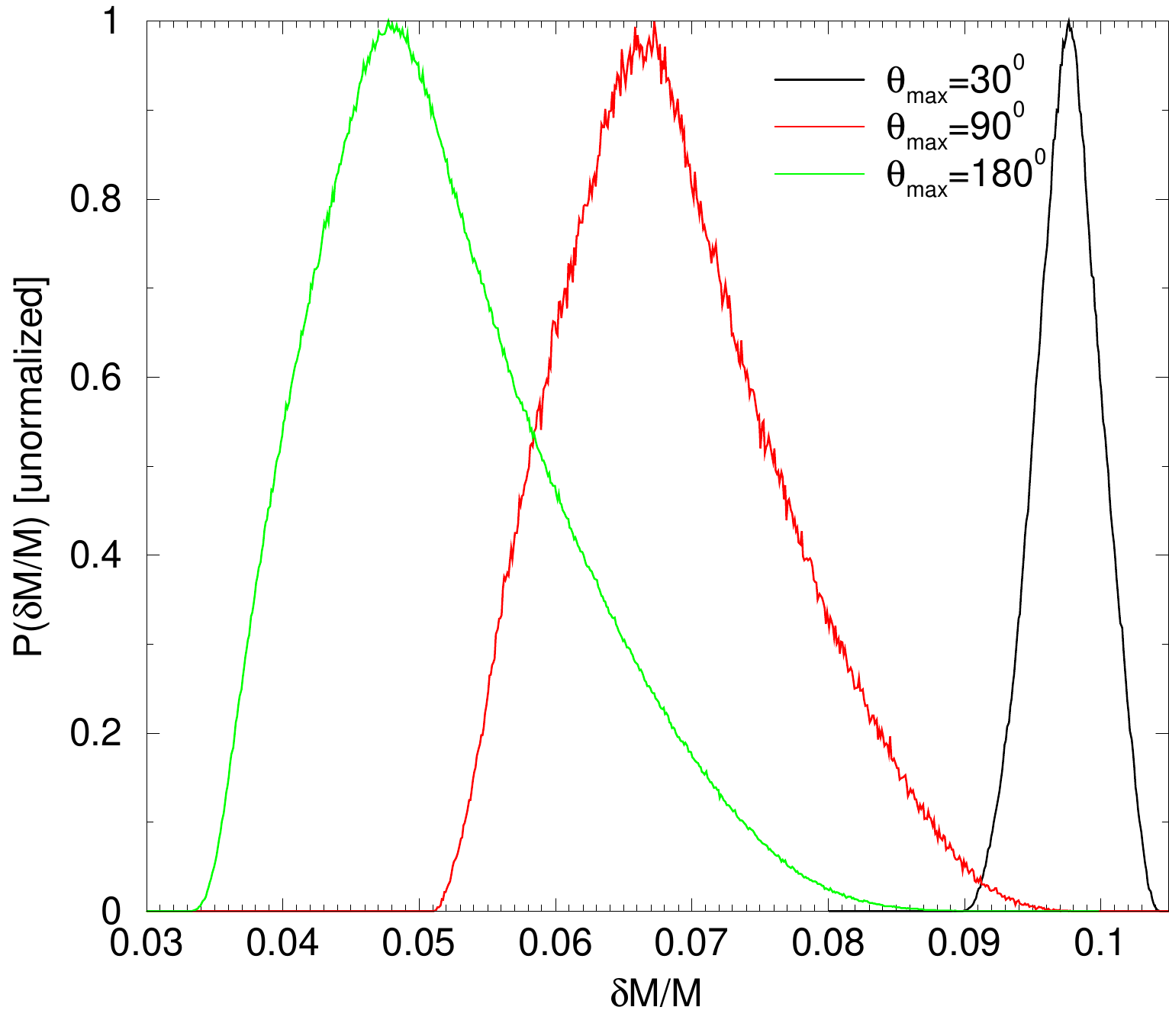}
  \caption{The final distributions of spin magnitudes and radiated mass
  for the remnants of BHB mergers. The angular distributions are
assumed to be uniform on the following segments of the sphere $\theta
< 180^\circ$, $\theta < 90^\circ$, and $\theta < 30^\circ$.
}
\label{fig:EandAdist}
\end{figure}

As an application to cosmological growth of black holes,
it is interesting to determine the spin distribution of BHs after many
generations of mergers. As our formulas for the final remnant spin
have not been fully extended to the unequal mass case, we consider the
consequence of the mergers of equal-mass binaries. Here, we neglect
the radiation of energy, which would affect the mass-ratio
distribution of subsequent generations. We find that the spin
distribution relaxes to a universal distribution independent of the
initial distribution. We verified this by starting with three
different spin distributions: non-spinning BHs, a spin distribution
uniform in [0,1], and all BHs with unit spin. We first assume that the spin
directions are uniformly distributed on the sphere and then, to model how
accretion disks
torques may affect our results, we also consider the case where the
spins are uniform in the upper hemisphere and uniform in the section
of the sphere $\theta < 30^\circ$.
Figures~\ref{fig:th180_dist}-\ref{fig:th30_dist} show how the
spin distribution changes from merger generation to generation.
The effect of spin alignment is more clearly shown in
Fig.~\ref{fig:EandAdist}, which shows all three final spin distributions.
Using these {\it universal} spin distributions, we find the
distributions of radiated energy, also shown in
Fig.~\ref{fig:EandAdist}.

Note that our distributions of the magnitude of the spin are consistent
with the observed spins in 19 active galactic nuclei (See Fig.~6 in
Ref.~\cite{Reynolds:2013rva}), as most of the observed values lie in
the high spin region.

\begin{figure}[!h]
  \includegraphics[width=0.95\columnwidth]{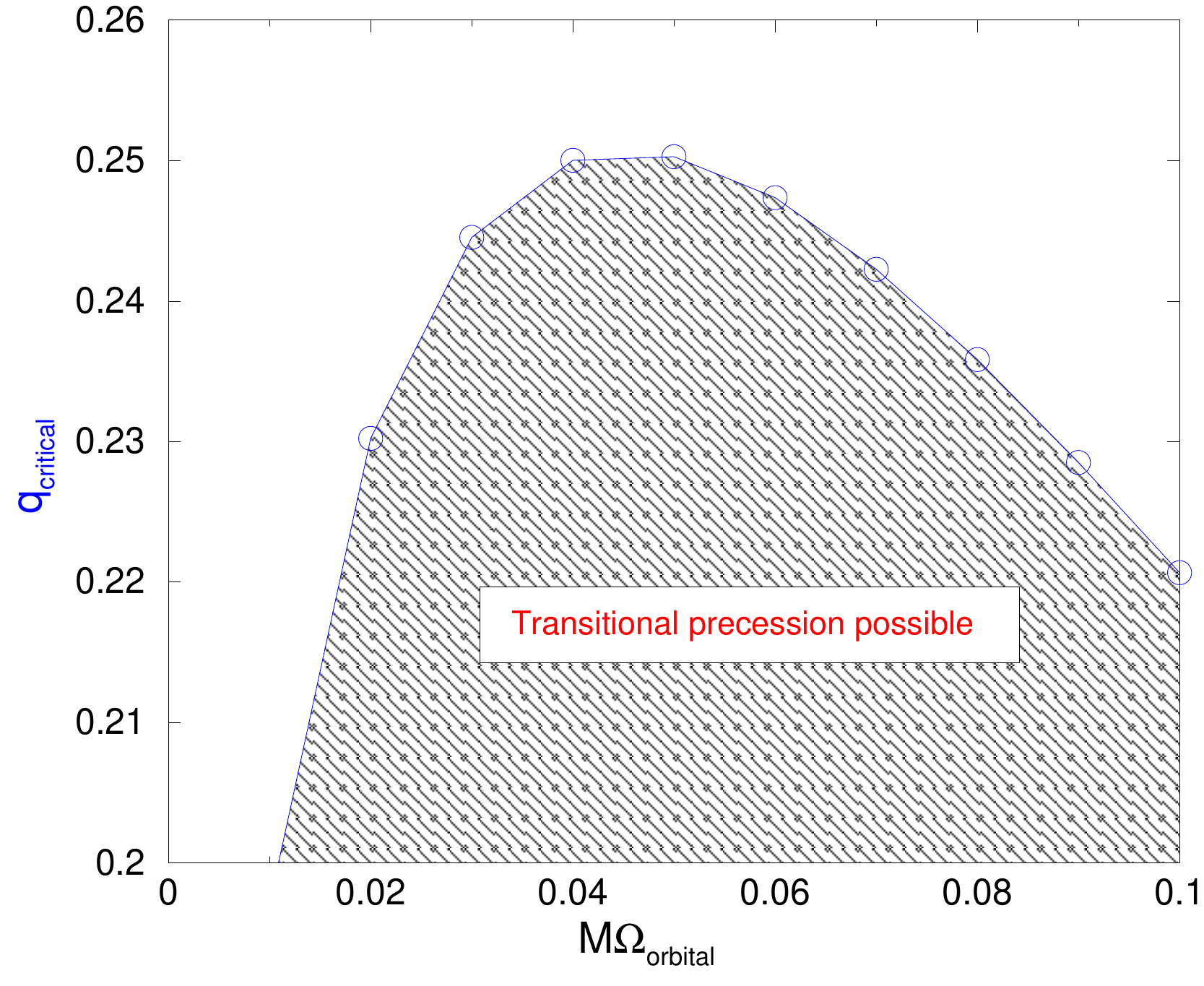}
  \caption{Critical mass ratio for counteraligned, maximally spinning 
black holes according to 3.5PN. Transition only can occur in the parameter
region below this curve.}
  \label{fig:qvsOmega}
\end{figure}

While all our simulations have assumed equal-mass binaries
(we will study explicitly the unequal-mass case in a following paper),
the key aspect of the dynamics of the binary that allows us to model
the direction of the remnant spin, i.e.\ simple precession,
occurs for comparable-mass binaries, as well.
In order
to estimate the critical mass ratio $q=m_1/m_2=q_{critical}$
below which it is possible to
have transitional precession (which can  reverse the direction of the total
angular momentum of the binary system),
we will consider the 
maximally spinning black-hole binaries where the spins are
counteraligned with the orbital
angular momentum $\vec{L}$.
Here, at the transition point, $J_\|$ changes from positive to
negative and net angular momentum rotates by $\pi$ radians.
A good estimate for when this transition occurs can be obtained 
using the 3.5PN
expression for $\vec{L}$ for quasicircular orbits (see Eq.~(4.7) of 
Ref.~\cite{Bohe:2012mr}). 
\begin{eqnarray}
&\frac{L}{m^2}&=\nu\, \left( 1+ \left( \frac{3}{2}+\frac{1}{6}\,\nu \right) x+ \left( -{\frac {35}{6}}
\,s- \frac{5}{2}\,{\it dm}\,\Delta \right) {x}^{3/2}+\right.\nonumber\\
&&\left. \left( {\frac {27}{8}}-{
\frac {19}{8}}\,\nu+\frac{1}{24}\,{\nu}^{2} \right) {x}^{2}+\right.\nonumber\\
&&\left. \left(  \left( -{
\frac {77}{8}}+{\frac {427}{72}}\,\nu \right) s+ \left( -{\frac {21}{8
}}+{\frac {35}{12}}\,\nu \right) {\it dm}\,\Delta \right) {x}^{5/2}+\right.\nonumber\\
&&\left.
 \left( {\frac {135}{16}}+ \left( -{\frac {6889}{144}}+{\frac {41}{24}
}\,\pi  \right) \nu+{\frac {31}{24}}\,{\nu}^{2}+{\frac {7\,{
\nu}^{3}}{1296}} \right) {x}^{3}+\right.\nonumber\\
&&\left. \left(  \left( -{\frac {405}{16}}+{\frac {
1101}{16}}\,\nu-{\frac {29}{16}}\,{\nu}^{2} \right) s+\right.\right.\nonumber\\
&&\left.\left. \left( -{\frac 
{81}{16}}+{\frac {117}{4}}\,\nu-{\frac {15}{16}}\,{\nu}^{2} \right) {
\it dm}\,\Delta \right) {x}^{7/2} \right) {\frac {1}{\sqrt {x}}}
\end{eqnarray}
where
$x=(m \Omega_{\rm orbital})^{2/3}$, 
 $\nu=q/(1+q)^2$, and for our maximally spinning counteraligned configurations
$s=-(1+q^2)/(1+q)^2$ and  ${\it dm}\,\Delta=-(1-q)^2/(1+q)^2$.
Assuming $S$ is conserved, we find the
critical value $q_{\rm critical}$ such that, for a given frequency
$\Omega_{\rm orbit}$, $J$ vanishes.
We plot $q_{\rm critical}$ versus $\Omega_{\rm orbit}$ in
Fig.~\ref{fig:qvsOmega}. The plot covers the inspiral regime down to
the plunge.
Note that transitional precession does not occur if $q\gtrsim1/4$.
This is in good agreement
with our previous estimate \cite{Lousto:2013vpa}, which was
based on considering the complementary regime of a point particle
moving in a Kerr background.

Ultimately, in order to determine the mass ratio dependence of the
remnant mass and spin, one has to perform a systematic study
of binaries in the regime  $1/10\leq q\leq1$ and fit the remaining 
coefficients of the terms in Tables \ref{table:Jp}-\ref{table:Jzm}
(which explicitly depend on $\delta m$). 
This is currently
under investigation by the authors and will provide a unified
phenomenological description for the remnant mass, spin and recoil 
velocities \cite{LZ}.

\section{conclusion}\label{sec:conclusion}

We developed a set of generic empirical formulas to predict the final
remnant mass and spin from the merger of two black holes.  In order to
obtain these empirical formulas, we extended our recent framework for
modeling the remnant recoil, which uses a combination of PN-inspired
variables, fitting parameters, and symmetry
arguments~\cite{Lousto:2012gt}. 
This formalism for modeling remnant
properties with arbitrary mass ratios and spins is summarized in
Tables \ref{table:Jp}-\ref{table:Jzm}.  Note that we used the same
symbolic expansion (\ref{eq:regroupJz})-(\ref{eq:Jzcosphi}) for both
the radiated energy and the magnitude of the remnant spin.  We justify
our assumption that a decomposition of the spins in terms of
components parallel to, and perpendicular to, the orbital angular momentum is
appropriate in light of the results in Fig.~\ref{fig:lsprec},
which shows how well of $\vec S \cdot \vec L$ is
conserved, and those of
Ref.~\cite{Lousto:2013vpa} were the final mass and spins can
be approximated by the nonprecessing hangup formula.
  We also assume that, given the results
in Fig.~\ref{fig:Jprecmax}, which shows how well the direction of the
total angular momentum is conserved, 
the final spin direction agrees with that
of the total angular momentum at large binary separations.

We performed 43 new simulations and combined these results with
over 130 other simulations from Refs.~\cite{Lousto:2012su,Lousto:2012gt}
to fit all relevant equal-mass
coefficients in our phenomenological formulas
(\ref{eq:regroupJz})-(\ref{eq:Jzcosphi}). This large set of
similarly constructed (initial orbital frequencies, number
of orbits, total run times) runs for equal-mass binaries
allowed us to obtain the coefficients for the most
relevant nonlinear terms in our expansion formulas for the radiated
mass and remnant spin to a consistent accuracy. We estimate the relative
errors lie within $2.5\%$.
The natural next step is to extend these formulas to
unequal-mass, generic-spin, BHB mergers. A comparatively large
number of such generic simulations is needed for this task.

Finally, we note that our results lend additional support for the use
of non-precessing waveforms in the modeling of precessing  waveforms
\cite{Boyle:2011gg,Pekowsky:2013ska,Hannam:2013oca}.
That procedure (see Ref.~\cite{Hannam:2013pra} for a recent review)
uses \hangup waveforms (i.e.\ non-precessing, spinning waveforms) in a
dynamically rotating frame to model the waveform from precessing
binaries. Our results show that the component of $\vec S$ along $\vec
L$ is conserved, which is the only component of the spin in the
\hangup case and thus provides a unique \hangup waveform as a
starting point for the modeling.  If the Bondi News $N(t)$  for a
precessing waveform can be described by a time dependent rotation
${\bf R}(t)$ acting on a non-precessing waveform $N_{\rm np}$, then
$\overline{N} N = \overline{{\bf R}(t)N_{\rm np}}\,{\bf R}(t) N_{\rm np}= 
\overline{N}_{\rm np}N_{\rm np}$ (where the {\it bar} denoted complex
conjugation) and the total radiated energy for the precessing
waveforms is equal to that of the hangup waveform.  Again, we see that
the energy radiated by the binary matches, to within an accuracy of
$\lesssim 10\%$, the energy radiated from the equivalent \hangup
waveform.  The fact that we see this agreement for a broad class of binaries
is suggestive.  Our results will also be important for modeling the
late time waveform because the plunge and ringdown part of the waveforms
should have the characteristic frequency of the actual spin of the 
merger remnant, $\alpha_{\rm rem}^2\approx \alpha_{\rm np}^2 +
\alpha_\perp^2$, and thus deviate from the original \hangup waveform.

\acknowledgments 

The authors wish to thank the referee for a very careful reading of
the original manuscript and for many useful suggestions for improving it.
The authors gratefully acknowledge the NSF for financial support from Grants
PHY-1305730, PHY-1212426, PHY-1229173,
AST-1028087, PHY-0969855, OCI-0832606, and
DRL-1136221. Computational resources were provided by XSEDE allocation
TG-PHY060027N, and by NewHorizons and BlueSky Clusters 
at Rochester Institute of Technology, which were supported
by NSF grant No. PHY-0722703, DMS-0820923, AST-1028087, and PHY-1229173.

\bibliographystyle{apsrev}
\bibliography{../../Bibtex/references}

\bigskip\bigskip

\appendix

\begin{widetext}

\section{Results from the SP and UD families of configurations}
\label{app:SPResults}

In this appendix, we give the initial data parameters for all SP and
UD
configurations in Table~\ref{tab:SPID}, the energy and angular momentum
radiated in Table~\ref{tab:SPRadIH}
 (as measured by directly from $\psi_4$ and
inferred from the remnant mass and spin obtained using the 
isolated horizon
formalism), the radiated and remnant angular momentum and the angle
between the initial and final angular momentum in Table~\ref{tab:SPJ},
 the final remnant spin and mass loss (starting from
infinite separation) in Table~\ref{tab:SPRadIH}, and finally, the measured
recoils (calculated using $\ell\leq4$ modes)
for the UD configurations in Table~\ref{tab:udkick}. Note that
we approximate the mass loss from infinity using
$\delta {\cal M} = (M_1 + M_2 - M_{\rm rem}) / (M_1 + M_2)$, where $M_{\rm rem}$ is the
remnant BH's mass and $M_1$ and $M_2$ are the initial masses of the two BHs in
the binary (which is a good approximation for their masses when the
binary was infinitely separated.

Here we also compare the performance of using $\vec L(t) = \vec J(t) -
\vec S(t)$ and $\hat S_{\rm coord}$ to define the component of the
spin
$S_\|$. In Fig.~\ref{fig:Ldefscomp}, we show the angles
 $\{\vec L, \vec S\}$ and $\{\vec L_{\rm coord},\vec S\}$ versus time
for all the N configurations (which are more general than the SP
configurations). A slight secular trend is apparent in
$\{\vec L_{\rm coord},\vec S\}$ that seems to be suppressed in
$\{\vec L, \vec S\}$.

\begin{figure}
  \begin{tabular}{||l||c||r||}
\hline
\hline
  \includegraphics[width=0.32\columnwidth]{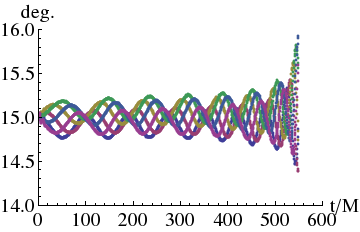}&
  \includegraphics[width=0.32\columnwidth]{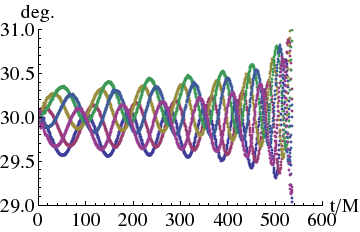}&
  \includegraphics[width=0.32\columnwidth]{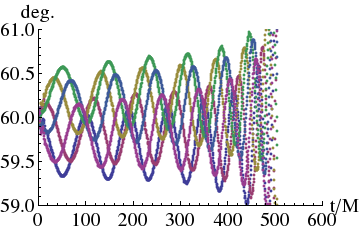}\\
  \includegraphics[width=0.32\columnwidth]{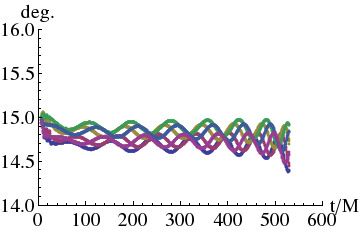}&
  \includegraphics[width=0.32\columnwidth]{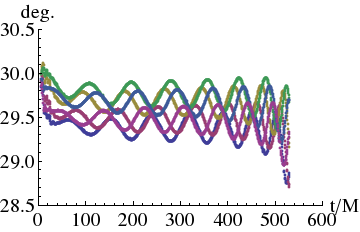}&
  \includegraphics[width=0.32\columnwidth]{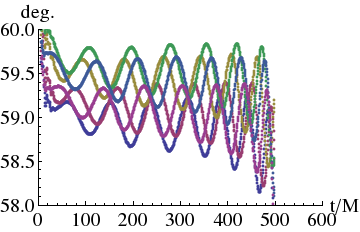}\\
\hline
\hline
  \includegraphics[width=0.32\columnwidth]{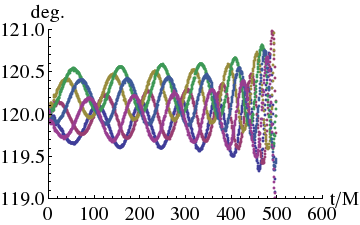}&
  \includegraphics[width=0.32\columnwidth]{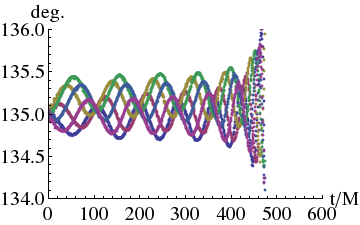}&
  \includegraphics[width=0.32\columnwidth]{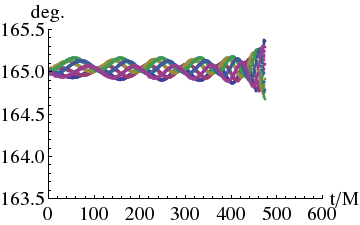}\\
  \includegraphics[width=0.32\columnwidth]{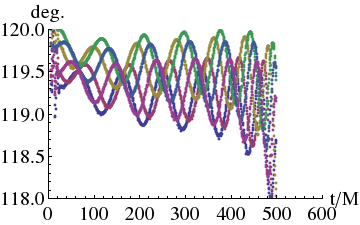}&
  \includegraphics[width=0.32\columnwidth]{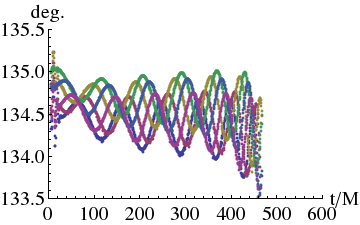}&
  \includegraphics[width=0.32\columnwidth]{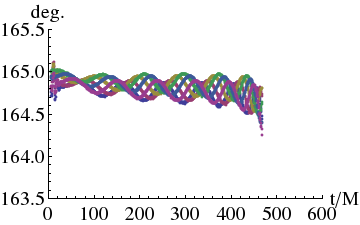}\\
\hline
\hline
  \end{tabular}
\caption{The angles $\{\vec L,\vec S\}$  and $\{\vec L_{coord},\vec
S\}$ versus time for all the N configurations. The panels are ordered
from left to right and top to bottom with the top-left corresponding
the NTH15 and the bottom right to NTH165. For each panel, the top
plot shows  $\{\vec L,\vec S\}$ and the bottom shows $\{\vec
L_{coord},\vec
S\}$ . Note the small secular trends in $\{\vec L_{coord},\vec S\}$
(here a small net negative slope) that are not present
in $\{\vec L,\vec S\}$. Note, however, that the amplitude of sinusoidal
oscillations are slightly larger for $\{\vec L,\vec S\}$. } \label{fig:Ldefscomp}
\end{figure}

\begin{table}
\caption{Initial data parameters. The BHs have initial position
$\vec r = (\pm x, 0,0)$, momenta $(0,\pm p_y,0)$, spins
$\pm (S_x, S_y, S_z)$, and puncture mass parameters $m_p$,
as well as the ADM mass $M_{\rm ADM}$, ADM angular momentum $(J_x, J_y, J_z)$,
initial eccentricity $e_{ini}$, eccentricity just prior to plunge
$e_{fin}$, and total number of orbits $N$. Masses, positions, and
momenta are given in units of an arbitrary mass $M$,
 while angular momenta are given in
units of $M^2$.
}\label{tab:SPID}
\begin{ruledtabular}
\begin{tabular}{l|lllllllllll}
Conf       &  $m_p$ & $x$ & $p_y$ & $S_x$ & $S_y$ & $S_z$ & $M_{\rm ADM}$ & $J_x$ & $J_y$ & $J_z$ & $N^{e_{fin}}_{e_{ini}}$\\
\hline
SPTH0PH0 &  0.30334 &  4.10893 &  0.10524 &  0.00000 &  0.00000 &  0.20534 &  1.00000 &  0.00000 &  0.00000 &  1.27551 & $6^{0.016}_{0.0025}$\\
\hline
SPTH48PH0 &  0.30285 &  4.16101 &  0.10656 &  0.00000 &  0.15294 &  0.13679 &  1.00000 &  0.00000 &  0.30588 &  1.16042 & $5^{0.02}_{0.004}$ \\
SPTH48PH30 &  0.30293 &  4.16101 &  0.10656 & -0.07647 &  0.13245 &  0.13679 &  1.00000 & -0.15294 &  0.26490 &  1.16042  \\
SPTH48PH60 &  0.30309 &  4.16101 &  0.10656 & -0.13245 &  0.07647 &  0.13679 &  1.00000 & -0.26490 &  0.15294 &  1.16042  \\
SPTH48PH90 &  0.30317 &  4.16101 &  0.10656 & -0.15294 &  0.00000 &  0.13679 &  1.00000 & -0.30588 &  0.00000 &  1.16042  \\
SPTH48PH120 &  0.30309 &  4.16101 &  0.10656 & -0.13245 & -0.07647 &  0.13679 &  1.00000 & -0.26490 & -0.15294 &  1.16042  \\
SPTH48PH150 &  0.30293 &  4.16101 &  0.10656 & -0.07647 & -0.13245 &  0.13679 &  1.00000 & -0.15294 & -0.26490 &  1.16042  \\
\hline
SPTH70PH0 &  0.30252 &  4.21471 &  0.10787 &  0.00000 &  0.19331 &  0.06835 &  1.00000 &  0.00000 &  0.38662 &  1.04601 & $ 4.5^{0.02}_{0.005}$\\
SPTH70PH30 &  0.30266 &  4.21471 &  0.10787 & -0.09666 &  0.16741 &  0.06835 &  1.00000 & -0.19331 &  0.33482 &  1.04601  \\
SPTH70PH60 &  0.30294 &  4.21471 &  0.10787 & -0.16741 &  0.09666 &  0.06835 &  1.00000 & -0.33482 &  0.19331 &  1.04601  \\
SPTH70PH90 &  0.30307 &  4.21471 &  0.10787 & -0.19331 &  0.00000 &  0.06835 &  1.00000 & -0.38662 &  0.00000 &  1.04601  \\
SPTH70PH120 &  0.30294 &  4.21471 &  0.10787 & -0.16741 & -0.09666 &  0.06835 &  1.00000 & -0.33482 & -0.19331 &  1.04601  \\
SPTH70PH150 &  0.30266 &  4.21471 &  0.10787 & -0.09666 & -0.16741 &  0.06835 &  1.00000 & -0.19331 & -0.33482 &  1.04601  \\
\hline
SPTH90PH0 &  0.30238 &  4.26949 &  0.10917 &  0.00000 &  0.20488 &  0.00000 &  1.00000 &  0.00000 &  0.40977 &  0.93218  & $4^{0.02}_{0.007}$ \\
SPTH90PH30 &  0.30254 &  4.26949 &  0.10917 & -0.10244 &  0.17744 &  0.00000 &  1.00000 & -0.20488 &  0.35487 &  0.93218  \\
SPTH90PH60 &  0.30287 &  4.26949 &  0.10917 & -0.17744 &  0.10244 &  0.00000 &  1.00000 & -0.35487 &  0.20488 &  0.93218  \\
SPTH90PH90 &  0.30303 &  4.26949 &  0.10917 & -0.20488 &  0.00000 &  0.00000 &  1.00000 & -0.40977 &  0.00000 &  0.93218  \\
SPTH90PH120 &  0.30287 &  4.26949 &  0.10917 & -0.17744 & -0.10244 &  0.00000 &  1.00000 & -0.35487 & -0.20488 &  0.93218  \\
SPTH90PH150 &  0.30254 &  4.26949 &  0.10917 & -0.10244 & -0.17744 &  0.00000 &  1.00000 & -0.20488 & -0.35487 &  0.93218  \\
\hline
SPTH110PH0 &  0.30396 &  5.06743 &  0.09829 &  0.00000 &  0.19250 & -0.06806 &  1.00000 &  0.00000 &  0.38501 &  0.85999 & $5^{0.02}_{0.007}$\\
SPTH110PH30 &  0.30410 &  5.06743 &  0.09829 & -0.09625 &  0.16671 & -0.06806 &  1.00000 & -0.19250 &  0.33342 &  0.85999  \\
SPTH110PH60 &  0.30437 &  5.06743 &  0.09829 & -0.16671 &  0.09625 & -0.06806 &  1.00000 & -0.33342 &  0.19250 &  0.85999  \\
SPTH110PH90 &  0.30450 &  5.06743 &  0.09829 & -0.19250 &  0.00000 & -0.06806 &  1.00000 & -0.38501 &  0.00000 &  0.85999  \\
SPTH110PH120 &  0.30437 &  5.06743 &  0.09829 & -0.16671 & -0.09625 & -0.06806 &  1.00000 & -0.33342 & -0.19250 &  0.85999  \\
SPTH110PH150 &  0.30410 &  5.06743 &  0.09829 & -0.09625 & -0.16671 & -0.06806 &  1.00000 & -0.19250 & -0.33342 &  0.85999  \\
\hline
SPTH132PH0 &  0.30481 &  5.53592 &  0.09375 &  0.00000 &  0.15193 & -0.13589 &  1.00000 &  0.00000 &  0.30386 &  0.76618  & $5^{0.025}_{0.01}$\\
SPTH132PH30 &  0.30490 &  5.53592 &  0.09375 & -0.07596 &  0.13157 & -0.13589 &  1.00000 & -0.15193 &  0.26315 &  0.76618  \\
SPTH132PH60 &  0.30506 &  5.53592 &  0.09375 & -0.13157 &  0.07596 & -0.13589 &  1.00000 & -0.26315 &  0.15193 &  0.76618  \\
SPTH132PH90 &  0.30514 &  5.53592 &  0.09375 & -0.15193 &  0.00000 & -0.13589 &  1.00000 & -0.30386 &  0.00000 &  0.76618  \\
SPTH132PH120 &  0.30506 &  5.53592 &  0.09375 & -0.13157 & -0.07596 & -0.13589 &  1.00000 & -0.26315 & -0.15193 &  0.76618  \\
SPTH132PH150 &  0.30490 &  5.53592 &  0.09375 & -0.07596 & -0.13157 & -0.13589 &  1.00000 & -0.15193 & -0.26315 &  0.76618  \\
\hline
SPTH146PH0 &  0.30559 &  6.07112 &  0.08839 &  0.00000 &  0.11251 & -0.16961 &  1.00000 &  0.00000 &  0.22501 &  0.73403 & $6^{0.02}_{0.005}$\\
SPTH146PH30 &  0.30564 &  6.07112 &  0.08839 & -0.05625 &  0.09743 & -0.16961 &  1.00000 & -0.11251 &  0.19487 &  0.73403  \\
SPTH146PH60 &  0.30572 &  6.07112 &  0.08839 & -0.09743 &  0.05625 & -0.16961 &  1.00000 & -0.19487 &  0.11251 &  0.73403  \\
SPTH146PH90 &  0.30576 &  6.07112 &  0.08839 & -0.11251 &  0.00000 & -0.16961 &  1.00000 & -0.22501 &  0.00000 &  0.73403  \\
SPTH146PH120 &  0.30572 &  6.07112 &  0.08839 & -0.09743 & -0.05625 & -0.16961 &  1.00000 & -0.19487 & -0.11251 &  0.73403  \\
SPTH146PH150 &  0.30564 &  6.07112 &  0.08839 & -0.05625 & -0.09743 & -0.16961 &  1.00000 & -0.11251 & -0.19487 &  0.73403  \\
\hline
SPTH180PH0 &  0.30618 &  6.50419 &  0.08473 &  0.00000 &  0.00000 & -0.20332 &  1.00000 &  0.00000 &  0.00000 &  0.69562 &        $6.5^{0.025}_{0.005}$\\
\hline
UD0.00 &  0.48965 &  4.50114 &  0.10560 &  0.00000 &  0.00000 &
0.00000 &  1.00000 &  0.00000 &  0.00000 &  0.95064  &
$4.5^{0.025}_{0.004}$ \\
UD0.60 &  0.40554 &  4.45038 &  0.10558 &  0.00000 &  0.00000 &
0.15348 &  1.00000 &  0.00000 &  0.00000 &  0.94899
 & $4.5^{0.025}_{0.004}$\\
UD0.70 &  0.36491 &  4.44055 &  0.10558 &  0.00000 &  0.00000 &
0.17906 &  1.00000 &  0.00000 &  0.00000 &  0.94839  &
$4.5^{0.025}_{0.005}$ \\
UD0.80 &  0.30366 &  4.43032 &  0.10557 &  0.00000 &  0.00000 &
0.20465 &  1.00000 &  0.00000 &  0.00000 &  0.94769  &
$4.5^{0.025}_{0.004}$ \\
UD0.85 &  0.25679 &  4.42506 &  0.10557 &  0.00000 &  0.00000 &
0.21744 &  1.00000 &  0.00000 &  0.00000 &  0.94731  &
$4.5^{0.025}_{0.004}$ \\
\end{tabular}
\end{ruledtabular}
\end{table}

\begin{table}
\caption{The radiated energy (in units of $M$) and angular momentum
(in units of $M^2$) as computed directly from
the waveform and by taking the differences between the ADM energy-momentum
and the final remnant mass and spin. Here ``var.'' is the expected magnitude
of the difference between the two measures assuming the no systematic errors
due to finite difference effects.
Note how the actual variation between the isolated
horizon and radiation measures is about an order of magnitude larger.
}\label{tab:SPRadIH}
\begin{ruledtabular}
\begin{tabular}{l|lll|lll|lll|lll|lll}
Conf & $\delta E^{\rm rad}$ & $\delta E^{\rm ih}$  & var. &
$\delta J^{\rm rad}_x$ & $\delta J^{\rm ih}_ z$ & var &
$\delta J^{\rm rad}_y$ & $\delta J^{\rm ih}_y$ & var & $\delta
J^{\rm rad}_z$ & $\delta J^{\rm ih}_z$ & var\\
\hline
SPTH0HP0 & 0.07505 & 0.07794 & 0.00052 & 0.00000 & 0.00854 & 0.00000 & 0.00000 & -0.00877 & 0.00000 & 0.49995 & 0.50437 & 0.00594\\
SPTH0P0 & 0.07463 & 0.07792 & 0.00042 & 0.00000 & 0.00936 & 0.00001 & 0.00000 & -0.00714 & 0.00001 & 0.49307 & 0.50346 & 0.00554\\
\hline
SPTH48P0 & 0.06357 & 0.06624 & 0.00039 & 0.00206 & 0.00931 & 0.00048 & 0.10450 & 0.10282 & 0.00173 & 0.41528 & 0.42519 & 0.00506\\
SPTH48P30 & 0.06139 & 0.06375 & 0.00036 & -0.04957 & -0.03910 & 0.00028 & 0.08900 & 0.08730 & 0.00151 & 0.40904 & 0.41902 & 0.00505\\
SPTH48P60 & 0.06083 & 0.06306 & 0.00035 & -0.08666 & -0.09718 & 0.00100 & 0.05186 & 0.04990 & 0.00130 & 0.40904 & 0.41370 & 0.00514\\
SPTH48P90 & 0.05963 & 0.06169 & 0.00033 & -0.09907 & -0.08870 & 0.00167 & 0.00149 & 0.00777 & 0.00052 & 0.40571 & 0.41589 & 0.00499\\
SPTH48P120 & 0.06060 & 0.06255 & 0.00031 & -0.08760 & -0.09923 & 0.00140 & -0.04674 & -0.05075 & 0.00030 & 0.41053 & 0.41361 & 0.00497\\
SPTH48P150 & 0.06456 & 0.06704 & 0.00035 & -0.05578 & -0.04720 & 0.00107 & -0.08841 & -0.10015 & 0.00102 & 0.42135 & 0.42764 & 0.00503\\
\hline
SPTH70P0 & 0.04970 & 0.05112 & 0.00029 & -0.00744 & -0.01082 & 0.00076 & 0.12241 & 0.13609 & 0.00158 & 0.32988 & 0.32970 & 0.00408\\
SPTH70P30 & 0.04893 & 0.05025 & 0.00027 & -0.06773 & -0.07496 & 0.00014 & 0.10055 & 0.10889 & 0.00175 & 0.32950 & 0.32967 & 0.00415\\
SPTH70P60 & 0.04916 & 0.05050 & 0.00027 & -0.10884 & -0.12516 & 0.00096 & 0.05378 & 0.04803 & 0.00138 & 0.33242 & 0.33224 & 0.00439\\
SPTH70P90 & 0.04889 & 0.05012 & 0.00028 & -0.12017 & -0.11426 & 0.00167 & -0.00843 & -0.01099 & 0.00077 & 0.33341 & 0.33819 & 0.00448\\
SPTH70P120 & 0.04972 & 0.05087 & 0.00026 & -0.09759 & -0.10004 & 0.00165 & -0.06850 & -0.05769 & 0.00026 & 0.33783 & 0.34130 & 0.00437\\
SPTH70P150 & 0.05469 & 0.05650 & 0.00032 & -0.05816 & -0.05297 & 0.00133 & -0.11164 & -0.12214 & 0.00101 & 0.34928 & 0.35278 & 0.00424\\
\hline
SPTH90P0 & 0.04047 & 0.04142 & 0.00023 & -0.00640 & 0.00062 & 0.00069 & 0.13954 & 0.14912 & 0.00133 & 0.25984 & 0.25999 & 0.00352\\
SPTH90P30 & 0.04064 & 0.04156 & 0.00021 & -0.07687 & -0.08625 & 0.00002 & 0.11710 & 0.10770 & 0.00163 & 0.26272 & 0.26759 & 0.00363\\
SPTH90P60 & 0.03979 & 0.04069 & 0.00022 & -0.12369 & -0.12826 & 0.00076 & 0.06163 & 0.05126 & 0.00135 & 0.26248 & 0.26587 & 0.00380\\
SPTH90P90 & 0.03994 & 0.04075 & 0.00022 & -0.13800 & -0.14609 & 0.00149 & -0.01034 & -0.00480 & 0.00126 & 0.26468 & 0.26371 & 0.00390\\
SPTH90P120 & 0.04374 & 0.04474 & 0.00024 & -0.11720 & -0.12652 & 0.00159 & -0.08274 & -0.09683 & 0.00011 & 0.27817 & 0.27492 & 0.00380\\
SPTH90P150 & 0.04226 & 0.04341 & 0.00026 & -0.06518 & -0.06179 & 0.00124 & -0.12172 & -0.11551 & 0.00094 & 0.26804 & 0.27507 & 0.00362\\
SPTH110P0 & 0.03476 & 0.03561 & 0.00025 & -0.01049 & -0.00728 & 0.00139 & 0.13961 & 0.15764 & 0.00219 & 0.25010 & 0.24655 & 0.00544\\
SPTH110P30 & 0.03424 & 0.03506 & 0.00024 & -0.08005 & -0.08585 & 0.00019 & 0.11551 & 0.10988 & 0.00257 & 0.24915 & 0.25343 & 0.00553\\
SPTH110P60 & 0.03484 & 0.03571 & 0.00024 & -0.12670 & -0.14196 & 0.00116 & 0.05866 & 0.05572 & 0.00235 & 0.25499 & 0.25308 & 0.00574\\
SPTH110P90 & 0.03479 & 0.03561 & 0.00025 & -0.13852 & -0.14434 & 0.00228 & -0.01401 & -0.00646 & 0.00142 & 0.25643 & 0.25638 & 0.00587\\
SPTH110P120 & 0.03508 & 0.03585 & 0.00024 & -0.11540 & -0.12814 & 0.00258 & -0.08621 & -0.07900 & 0.00005 & 0.25545 & 0.25441 & 0.00566\\
SPTH110P150 & 0.03796 & 0.03895 & 0.00026 & -0.05974 & -0.06781 & 0.00222 & -0.13264 & -0.13278 & 0.00132 & 0.26541 & 0.26619 & 0.00555\\
\hline
SPTH132P0 & 0.03015 & 0.03085 & 0.00024 & 0.00728 & 0.00430 & 0.00159 & 0.13745 & 0.13642 & 0.00200 & 0.23665 & 0.23966 & 0.00664\\
SPTH132P30 & 0.03078 & 0.03152 & 0.00024 & -0.06180 & -0.06592 & 0.00031 & 0.12304 & 0.12637 & 0.00247 & 0.23997 & 0.24248 & 0.00668\\
SPTH132P60 & 0.03070 & 0.03147 & 0.00023 & -0.11639 & -0.13263 & 0.00093 & 0.07416 & 0.08385 & 0.00239 & 0.24134 & 0.23892 & 0.00685\\
SPTH132P90 & 0.03061 & 0.03137 & 0.00023 & -0.13864 & -0.14073 & 0.00201 & 0.00412 & 0.00175 & 0.00161 & 0.24250 & 0.24408 & 0.00699\\
SPTH132P120 & 0.03030 & 0.03108 & 0.00024 & -0.11869 & -0.13653 & 0.00252 & -0.06261 & -0.05444 & 0.00033 & 0.24046 & 0.24062 & 0.00684\\
SPTH132P150 & 0.02921 & 0.02994 & 0.00024 & -0.07265 & -0.09238 & 0.00233 & -0.11270 & -0.11058 & 0.00104 & 0.23323 & 0.23450 & 0.00669\\
SPTH146P0 & 0.02741 & 0.02807 & 0.00029 & 0.00281 & 0.01577 & 0.00190 & 0.11407 & 0.12002 & 0.00230 & 0.25534 & 0.25609 & 0.00964\\
SPTH146P30 & 0.02737 & 0.02802 & 0.00029 & -0.05545 & -0.03962 & 0.00047 & 0.10104 & 0.10532 & 0.00299 & 0.25519 & 0.25771 & 0.00972\\
SPTH146P60 & 0.02740 & 0.02807 & 0.00029 & -0.09980 & -0.10200 & 0.00109 & 0.05969 & 0.07102 & 0.00289 & 0.25659 & 0.25560 & 0.00983\\
SPTH146P90 & 0.02749 & 0.02819 & 0.00029 & -0.11665 & -0.11353 & 0.00239 & 0.00203 & 0.01466 & 0.00200 & 0.25762 & 0.25960 & 0.00986\\
SPTH146P120 & 0.02807 & 0.02877 & 0.00029 & -0.10245 & -0.10467 & 0.00297 & -0.05609 & -0.05902 & 0.00054 & 0.26024 & 0.26009 & 0.00980\\
SPTH146P150 & 0.02803 & 0.02871 & 0.00029 & -0.05966 & -0.07684 & 0.00278 & -0.09799 & -0.09327 & 0.00106 & 0.25944 & 0.25958 & 0.00969\\
\hline
SPTH180P0 & 0.02549 & 0.02612 & 0.00031 & 0.00000 & -0.00443 & 0.00000 & 0.00000 & -0.00116 & 0.00000 & 0.28783 & 0.29247 & 0.01234\\
\hline
UD0.00C & 0.03697 & 0.03754 & 0.00003 & 0.00000 & 0.00398 & 0.00000 &
0.00000 & 0.00903 & 0.00000 & 0.31222 & 0.31499 & 0.00301\\
UD0.00M & 0.03718 & 0.03754 & 0.00008 & 0.00000 & -0.00049 & 0.00000
& 0.00000 & -0.00702 & 0.00000 & 0.31307 & 0.31495 & 0.00336\\
UD0.60C & 0.03723 & 0.03800 & 0.00008 & 0.00000 & 0.00551 & 0.00000 &
0.00000 & 0.00899 & 0.00000 & 0.31226 & 0.31517 & 0.00339\\
UD0.60M & 0.03744 & 0.03800 & 0.00012 & 0.00000 & -0.00117 &
0.00000 & 0.00000 & -0.00602 & 0.00000 & 0.31314 & 0.31511 & 0.00368\\
UD0.70C & 0.03732 & 0.03828 & 0.00010 & 0.00000 & 0.00303 & 0.00000 &
0.00000 & 0.00716 & 0.00000 & 0.31223 & 0.31534 & 0.00352\\
UD0.70M & 0.03750 & 0.03828 & 0.00013 & 0.00000 & -0.00089 &
0.00000 & 0.00000 & -0.01244 & 0.00000 & 0.31345 & 0.31541 & 0.00458\\
UD0.80C & 0.03733 & 0.03867 & 0.00013 & 0.00000 & 0.00682 & 0.00000 &
0.00000 & 0.00933 & 0.00000 & 0.31048 & 0.31606 & 0.00363\\
UD0.80M & 0.03843 & 0.03873 & 0.00022 & 0.00000 & 0.00659 & 0.00000
& 0.00000 & 0.00607 & 0.00000 & 0.31948 & 0.31571 & 0.00482\\
UD0.80F & 0.03852 & 0.03872 & 0.00015 & 0.00000 & -0.00616 &
0.00000 & 0.00000 & -0.00407 & 0.00000 & 0.31956 & 0.31574 & 0.00454\\
UD0.85C & 0.03642 & 0.03770 & 0.00015 & 0.00000 & 0.00576 & 0.00000 &
0.00000 & 0.00427 & 0.00000 & 0.30303 & 0.31609 & 0.00393\\
UD0.85M & 0.03851 & 0.03901 & 0.00024 & 0.00000 & 0.00612 & 0.00000
& 0.00000 & 0.00369 & 0.00000 & 0.31710 & 0.31650 & 0.00481\\
UD0.85F & 0.03875 & 0.03906 & 0.00017 & 0.00000 &
-0.00563 & 0.00000 & 0.00000 & -0.00473 & 0.00000 & 0.32005 & 0.31601
& 0.00467\\
\end{tabular}
\end{ruledtabular}
\end{table}

\begin{table}
\caption{The magnitude of the ADM angular momentum ($J_{\rm ADM}$) in
units of $M^2$, 
final remnant angular momentum ($J_{\rm rem}$) in units of $M^2$, 
and the difference between
the remnant and ADM angular momentum 
($\delta J = |\vec J_{\rm ADM} - \vec J_{\rm rem}|$) in units of $M^2$,
 as well as the angle
between $\vec J_{\rm ADM}$ and $\vec J_{\rm rem}$, as calculated directly
from the radiated angular momentum and the isolated horizon
formalism.}\label{tab:SPJ}
\begin{ruledtabular}
\begin{tabular}{l|l|ll|ll|ll|ll|}
Conf & $J_{\rm ADM}$ & $J_{\rm rem}^{\rm rad} $ & $J_{\rm
rem}^{\rm ih} $ &
$\delta J^{\rm rad}$ & $\delta J^{\rm ih}$ & $\delta J^{\rm rad}/ J_{\rm ADM}$ & $\delta J^{\rm ih}/ J_{\rm ADM}$ & $\Theta^{\rm rad}$  & $\Theta^{\rm ih}$\\
\hline
SPTH0HPH0 & 1.27551 & 0.77556 & 0.77124 & 0.49995 & 0.50452 & 0.39196 & 0.39554 & 0.00000 & 0.90944\\
SPTH0PH0 & 1.27551 & 0.78245 & 0.77214 & 0.49307 & 0.50360 & 0.38656 & 0.39482 & 0.00000 & 0.87359\\
\hline
SPTH48PH0 & 1.20006 & 0.77188 & 0.76281 & 0.42823 & 0.43755 & 0.35684 & 0.36460 & 0.38753 & 0.97019 \\
SPTH48PH30 & 1.20006 & 0.77859 & 0.77083 & 0.42154 & 0.42980 & 0.35126 & 0.35815 & 0.43966 & 1.31860 \\
SPTH48PH60 & 1.20006 & 0.77882 & 0.77223 & 0.42133 & 0.42788 & 0.35109 & 0.35655 & 0.50045 & 0.39885 \\
SPTH48PH90 & 1.20006 & 0.78254 & 0.77560 & 0.41763 & 0.42532 & 0.34801 & 0.35441 & 0.56778 & 1.60150 \\
SPTH48PH120 & 1.20006 & 0.77785 & 0.77176 & 0.42237 & 0.42836 & 0.35196 & 0.35695 & 0.68470 & 0.45136 \\
SPTH48PH150 & 1.20006 & 0.76604 & 0.75848 & 0.43412 & 0.44174 & 0.36175 & 0.36810 & 0.56784 & 0.71651 \\
\hline
SPTH70PH0 & 1.11517 & 0.76334 & 0.75894 & 0.35194 & 0.35684 & 0.31559 & 0.31999 & 0.55967 & 1.29723 \\
SPTH70PH30 & 1.11517 & 0.76422 & 0.76039 & 0.35109 & 0.35519 & 0.31483 & 0.31850 & 0.63125 & 1.05622 \\
SPTH70PH60 & 1.11517 & 0.76141 & 0.75798 & 0.35389 & 0.35826 & 0.31734 & 0.32126 & 0.60228 & 1.72674 \\
SPTH70PH90 & 1.11517 & 0.76083 & 0.75849 & 0.35450 & 0.35714 & 0.31789 & 0.32025 & 0.67064 & 1.12619 \\
SPTH70PH120 & 1.11517 & 0.75721 & 0.75507 & 0.35825 & 0.36031 & 0.32125 & 0.32309 & 0.90678 & 0.75903 \\
SPTH70PH150 & 1.11517 & 0.74398 & 0.73858 & 0.37127 & 0.37706 & 0.33293 & 0.33812 & 0.48398 & 1.18693 \\
\hline
SPTH90PH0 & 1.01827 & 0.72464 & 0.72096 & 0.29501 & 0.29972 & 0.28971 & 0.29434 & 1.90200 & 2.53544 \\
SPTH90PH30 & 1.01827 & 0.72187 & 0.71892 & 0.29773 & 0.30107 & 0.29239 & 0.29567 & 1.88096 & 2.15371 \\
SPTH90PH60 & 1.01827 & 0.72282 & 0.72036 & 0.29664 & 0.29961 & 0.29131 & 0.29423 & 1.76972 & 2.13355 \\
SPTH90PH90 & 1.01827 & 0.72078 & 0.71861 & 0.29867 & 0.30151 & 0.29331 & 0.29610 & 1.77722 & 2.23547 \\
SPTH90PH120 & 1.01827 & 0.70649 & 0.70414 & 0.31299 & 0.31775 & 0.30737 & 0.31205 & 1.85983 & 3.23635 \\
SPTH90PH150 & 1.01827 & 0.71761 & 0.71384 & 0.30151 & 0.30467 & 0.29610 & 0.29921 & 1.51638 & 0.81087 \\
\hline
SPTH110PH0 & 0.94224 & 0.65749 & 0.65426 & 0.28662 & 0.29273 & 0.30419 & 0.31067 & 2.38164 & 3.83364 \\
SPTH110PH30 & 0.94224 & 0.65822 & 0.65518 & 0.28605 & 0.28926 & 0.30359 & 0.30699 & 2.47981 & 2.59825 \\
SPTH110PH60 & 0.94224 & 0.65320 & 0.65093 & 0.29071 & 0.29548 & 0.30853 & 0.31359 & 2.27745 & 3.61911 \\
SPTH110PH90 & 0.94224 & 0.65210 & 0.64985 & 0.29179 & 0.29429 & 0.30968 & 0.31233 & 2.26623 & 2.44657 \\
SPTH110PH120 & 0.94224 & 0.65138 & 0.64942 & 0.29327 & 0.29561 & 0.31124 & 0.31373 & 2.74427 & 2.97197 \\
SPTH110PH150 & 0.94224 & 0.64146 & 0.63906 & 0.30266 & 0.30510 & 0.32121 & 0.32381 & 2.48309 & 2.52897 \\
\hline
SPTH132PH0 & 0.82423 & 0.55510 & 0.55252 & 0.27377 & 0.27579 & 0.33215 & 0.33461 & 4.25411 & 4.01621 \\
SPTH132PH30 & 0.82423 & 0.55195 & 0.54806 & 0.27666 & 0.28126 & 0.33566 & 0.34124 & 4.16806 & 4.54273 \\
SPTH132PH60 & 0.82423 & 0.55049 & 0.54742 & 0.27801 & 0.28583 & 0.33730 & 0.34679 & 4.13090 & 6.08258 \\
SPTH132PH90 & 0.82423 & 0.54913 & 0.54699 & 0.27937 & 0.28174 & 0.33894 & 0.34183 & 4.14527 & 4.28525 \\
SPTH132PH120 & 0.82423 & 0.55247 & 0.54932 & 0.27537 & 0.28196 & 0.33409 & 0.34208 & 3.77426 & 5.33520 \\
SPTH132PH150 & 0.82423 & 0.55942 & 0.55633 & 0.26902 & 0.27523 & 0.32639 & 0.33392 & 4.00407 & 5.34082 \\
\hline
SPTH146PH0 & 0.76775 & 0.49138 & 0.48959 & 0.27968 & 0.28326 & 0.36428 & 0.36895 & 4.00721 & 5.00522 \\
SPTH146PH30 & 0.76775 & 0.49128 & 0.49012 & 0.28001 & 0.28121 & 0.36471 & 0.36627 & 4.13972 & 4.17585 \\
SPTH146PH60 & 0.76775 & 0.48967 & 0.48912 & 0.28171 & 0.28422 & 0.36694 & 0.37020 & 4.21695 & 5.24828 \\
SPTH146PH90 & 0.76775 & 0.48859 & 0.48757 & 0.28280 & 0.28372 & 0.36835 & 0.36955 & 4.23470 & 4.18955 \\
SPTH146PH120 & 0.76775 & 0.48601 & 0.48540 & 0.28525 & 0.28651 & 0.37154 & 0.37318 & 4.18523 & 4.56914 \\
SPTH146PH150 & 0.76775 & 0.48726 & 0.48652 & 0.28367 & 0.28633 & 0.36948 & 0.37295 & 3.96902 & 5.04595 \\
\hline
SPTH180PH0 & 0.69562 & 0.40780 & 0.40318 & 0.28783 & 0.29251 & 0.41377 & 0.42050 & 0.00000 & 0.65079 \\
\end{tabular}
\end{ruledtabular}
\end{table}

\begin{table}
\caption{The radiated mass and final remnant spin as measured using
the IH formalism and from the radiation. The IH measurement proved
to be more accurate as can be seen by comparing results
from different resolutions.}\label{tab:RemComp}
\begin{ruledtabular}
\begin{tabular}{lllll}
CONF. & $\delta {\cal M}^{\rm IH}$ & $\delta {\cal M}^{\rm rad}$ & $\alpha_{\rm
rem}^{\rm IH} $ & $\alpha_{\rm rem}^{\rm rad} $\\
\hline
SPTH0PH0 & $ 0.08909\pm0.00007 $ &  $ 0.08584\pm0.00041 $ &  $
0.90815\pm0.00070 $ &  $ 0.91374\pm0.00648 $ \\
SPTH0HPH0 & $ 0.08911\pm0.00003 $ &  $ 0.08625\pm0.00052 $ &  $
0.90713\pm0.00047 $ &  $ 0.90652\pm0.00700 $ \\
\hline
SPTH48PH0 & $ 0.07713\pm0.00010 $ &  $ 0.07450\pm0.00037 $ &  $
0.87488\pm0.00155 $ &  $ 0.88024\pm0.00543 $ \\
SPTH48PH30 & $ 0.07467\pm0.00010 $ &  $ 0.07233\pm0.00034 $ &  $
0.87937\pm0.00135 $ &  $ 0.88376\pm0.00542 $ \\
SPTH48PH60 & $ 0.07399\pm0.00010 $ &  $ 0.07178\pm0.00034 $ &  $
0.87968\pm0.00121 $ &  $ 0.88297\pm0.00554 $ \\
SPTH48PH90 & $ 0.07263\pm0.00010 $ &  $ 0.07060\pm0.00031 $ &  $
0.88094\pm0.00120 $ &  $ 0.88493\pm0.00537 $ \\
SPTH48PH120 & $ 0.07346\pm0.00010 $ &  $ 0.07154\pm0.00029 $ &  $
0.87819\pm0.00128 $ &  $ 0.88144\pm0.00532 $ \\
SPTH48PH150 & $ 0.07790\pm0.00009 $ &  $ 0.07545\pm0.00033 $ &  $
0.87140\pm0.00137 $ &  $ 0.87542\pm0.00542 $ \\
\hline
SPTH70PH0 & $ 0.06178\pm0.00001 $ &  $ 0.06037\pm0.00028 $ &  $
0.84291\pm0.00006 $ &  $ 0.84528\pm0.00431 $ \\
SPTH70PH30 & $ 0.06096\pm0.00001 $ &  $ 0.05965\pm0.00026 $ &  $
0.84298\pm0.00010 $ &  $ 0.84488\pm0.00437 $ \\
SPTH70PH60 & $ 0.06120\pm0.00001 $ &  $ 0.05987\pm0.00027 $ &  $
0.84075\pm0.00006 $ &  $ 0.84217\pm0.00459 $ \\
SPTH70PH90 & $ 0.06081\pm0.00001 $ &  $ 0.05959\pm0.00027 $ &  $
0.84065\pm0.00004 $ &  $ 0.84105\pm0.00471 $ \\
SPTH70PH120 & $ 0.06153\pm0.00001 $ &  $ 0.06039\pm0.00026 $ &  $
0.83818\pm0.00004 $ &  $ 0.83852\pm0.00459 $ \\
SPTH70PH150 & $ 0.06708\pm0.00001 $ &  $ 0.06529\pm0.00031 $ &  $
0.82968\pm0.00005 $ &  $ 0.83256\pm0.00450 $ \\
\hline
SPTH90PH0 & $ 0.05181\pm0.00000 $ &  $ 0.05087\pm0.00022 $ &  $
0.78461\pm0.00002 $ &  $ 0.78705\pm0.00361 $ \\
SPTH90PH30 & $ 0.05199\pm0.00000 $ &  $ 0.05108\pm0.00021 $ &  $
0.78262\pm0.00002 $ &  $ 0.78433\pm0.00372 $ \\
SPTH90PH60 & $ 0.05115\pm0.00000 $ &  $ 0.05026\pm0.00021 $ &  $
0.78277\pm0.00002 $ &  $ 0.78397\pm0.00386 $ \\
SPTH90PH90 & $ 0.05121\pm0.00000 $ &  $ 0.05040\pm0.00022 $ &  $
0.78096\pm0.00019 $ &  $ 0.78200\pm0.00398 $ \\
SPTH90PH120 & $ 0.05511\pm0.00000 $ &  $ 0.05413\pm0.00024 $ &  $
0.77164\pm0.00001 $ &  $ 0.77261\pm0.00391 $ \\
SPTH90PH150 & $ 0.05378\pm0.00000 $ &  $ 0.05264\pm0.00026 $ &  $
0.78009\pm0.00001 $ &  $ 0.78233\pm0.00370 $ \\
\hline
SPTH110PH0 & $ 0.04457\pm0.00000 $ &  $ 0.04373\pm0.00025 $ &  $
0.70347\pm0.00000 $ &  $ 0.70570\pm0.00550 $ \\
SPTH110PH30 & $ 0.04406\pm0.00000 $ &  $ 0.04325\pm0.00024 $ &  $
0.70365\pm0.00000 $ &  $ 0.70572\pm0.00559 $ \\
SPTH110PH60 & $ 0.04475\pm0.00000 $ &  $ 0.04389\pm0.00024 $ &  $
0.70003\pm0.00000 $ &  $ 0.70122\pm0.00576 $ \\
SPTH110PH90 & $ 0.04465\pm0.00000 $ &  $ 0.04384\pm0.00024 $ &  $
0.69873\pm0.00001 $ &  $ 0.69996\pm0.00591 $ \\
SPTH110PH120 & $ 0.04485\pm0.00000 $ &  $ 0.04408\pm0.00024 $ &  $
0.69862\pm0.00000 $ &  $ 0.69960\pm0.00572 $ \\
SPTH110PH150 & $ 0.04790\pm0.00000 $ &  $ 0.04692\pm0.00026 $ &  $
0.69191\pm0.00000 $ &  $ 0.69308\pm0.00561 $ \\
\hline
SPTH132PH0 & $ 0.03915\pm0.00000 $ &  $ 0.03846\pm0.00024 $ &  $
0.58826\pm0.00000 $ &  $ 0.59016\pm0.00678 $ \\
SPTH132PH30 & $ 0.03982\pm0.00000 $ &  $ 0.03908\pm0.00023 $ &  $
0.58431\pm0.00000 $ &  $ 0.58756\pm0.00682 $ \\
SPTH132PH60 & $ 0.03980\pm0.00000 $ &  $ 0.03904\pm0.00023 $ &  $
0.58358\pm0.00000 $ &  $ 0.58591\pm0.00698 $ \\
SPTH132PH90 & $ 0.03971\pm0.00000 $ &  $ 0.03896\pm0.00023 $ &  $
0.58300\pm0.00000 $ &  $ 0.58436\pm0.00712 $ \\
SPTH132PH120 & $ 0.03938\pm0.00000 $ &  $ 0.03861\pm0.00023 $ &  $
0.58512\pm0.00000 $ &  $ 0.58754\pm0.00697 $ \\
SPTH132PH150 & $ 0.03825\pm0.00000 $ &  $ 0.03753\pm0.00024 $ &  $
0.59121\pm0.00000 $ &  $ 0.59360\pm0.00678 $ \\
\hline
SPTH146PH0 & $ 0.03578\pm0.00000 $ &  $ 0.03513\pm0.00029 $ &  $
0.51828\pm0.00000 $ &  $ 0.51948\pm0.00995 $ \\
SPTH146PH30 & $ 0.03573\pm0.00000 $ &  $ 0.03509\pm0.00028 $ &  $
0.51878\pm0.00000 $ &  $ 0.51931\pm0.01004 $ \\
SPTH146PH60 & $ 0.03580\pm0.00000 $ &  $ 0.03514\pm0.00029 $ &  $
0.51778\pm0.00001 $ &  $ 0.51765\pm0.01015 $ \\
SPTH146PH90 & $ 0.03592\pm0.00000 $ &  $ 0.03523\pm0.00029 $ &  $
0.51627\pm0.00000 $ &  $ 0.51661\pm0.01019 $ \\
SPTH146PH120 & $ 0.03648\pm0.00000 $ &  $ 0.03578\pm0.00029 $ &  $
0.51459\pm0.00000 $ &  $ 0.51448\pm0.01014 $ \\
SPTH146PH150 & $ 0.03642\pm0.00000 $ &  $ 0.03575\pm0.00029 $ &  $
0.51570\pm0.00001 $ &  $ 0.51577\pm0.01000 $ \\
\hline
SPTH180PH0 & $ 0.03339\pm0.00000 $ &  $ 0.03277\pm0.00031 $ &  $
0.42509\pm0.00000 $ &  $ 0.42941\pm0.01299 $ \\
\hline
UD0.00C & $ 0.04833\pm0.00000 $ &  $ 0.04811\pm0.00006 $ &  $
0.68628\pm0.00000 $ &  $ 0.68247\pm0.00426 $ \\
UD0.00M & $ 0.04833\pm0.00000 $ &  $ 0.04832\pm0.00011 $ &  $
0.68629\pm0.00000 $ &  $ 0.68185\pm0.00450 $ \\
\hline
UD0.60C & $ 0.04862\pm0.00000 $ &  $ 0.04832\pm0.00012 $ &  $
0.68498\pm0.00000 $ &  $ 0.68113\pm0.00462 $ \\
UD0.60M & $ 0.04862\pm0.00000 $ &  $ 0.04855\pm0.00015 $ &  $
0.68498\pm0.00000 $ &  $ 0.68048\pm0.00481 $ \\
\hline
\hline
UD0.70C & $ 0.04874\pm0.00000 $ &  $ 0.04835\pm0.00015 $ &  $
0.68450\pm0.00000 $ &  $ 0.68072\pm0.00480 $ \\
UD0.70M & $ 0.04872\pm0.00000 $ &  $ 0.04857\pm0.00018 $ &  $
0.68451\pm0.00000 $ &  $ 0.68006\pm0.00501 $ \\
\hline
UD0.80C & $ 0.04881\pm0.00000 $ &  $ 0.04819\pm0.00020 $ &  $
0.68358\pm0.00002 $ &  $ 0.68199\pm0.00495 $ \\
UD0.80M & $ 0.04887\pm0.00000 $ &  $ 0.04856\pm0.00022 $ &  $
0.68400\pm0.00000 $ &  $ 0.67943\pm0.00522 $ \\
UD0.80F & $ 0.04886\pm0.00000 $ &  $ 0.04865\pm0.00015 $ &  $
0.68393\pm0.00000 $ &  $ 0.67947\pm0.00492 $ \\
\hline
UD0.85C & $ 0.04804\pm0.00008 $ &  $ 0.04759\pm0.00023 $ &  $
0.68169\pm0.00010 $ &  $ 0.68838\pm0.00528 $ \\
UD0.85M & $ 0.04885\pm0.00003 $ &  $ 0.04834\pm0.00023 $ &  $
0.68311\pm0.00004 $ &  $ 0.68170\pm0.00521 $ \\
UD0.85F & $ 0.04895\pm0.00000 $ &  $ 0.04862\pm0.00017 $ &  $
0.68371\pm0.00000 $ &  $ 0.67885\pm0.00506 $ \\
\end{tabular}
\end{ruledtabular}
\end{table}

\begin{table}
\caption{The recoil velocity for the UD configurations in units of
$\KMS$. Note that
magnitude of the recoil is in good agreement between
resolutions.}\label{tab:udkick}
\begin{ruledtabular}
\begin{tabular}{llll}
Conf & $V_x$ & $V_y$ & $V$\\
\hline
UD0.60C & $ -51.69\pm2.72 $ &  $ -271.61\pm1.71 $ & $ 276.49\pm1.75 $ \\
UD0.60M  & $ -50.04\pm2.26 $ &  $ -274.30\pm2.07 $ & $
278.82\pm2.07 $ \\
\hline
UD0.70C & $ -96.65\pm2.90 $ &  $ -307.53\pm1.23 $ & $ 322.36\pm1.46 $
\\
UD0.70M  & $ -96.86\pm2.74 $ &  $ -310.38\pm1.26 $ & $
325.15\pm1.45 $ \\
\hline
UD0.80C  & $ -244.46\pm3.70 $ &  $ -273.38\pm0.18 $ & $ 366.74\pm2.47 $
\\
UD0.80M & $ -144.65\pm3.74 $ &  $ -341.98\pm0.94 $ & $
371.31\pm1.69 $ \\
UD0.80F  & $ -160.89\pm3.18 $ &  $ -336.12\pm1.25 $ & $
372.65\pm1.78 $ \\
\hline
UD0.85C  & $ -250.52\pm4.25 $ &  $ -314.66\pm1.01 $ & $ 402.21\pm2.76 $
\\
UD0.85M & $ -307.15\pm3.53 $ &  $ -242.73\pm0.12 $ & $
391.48\pm2.77 $ \\
UD0.85F & $ -171.48\pm3.65 $ &  $ -357.11\pm1.76 $ & $
396.14\pm2.24 $ \\

\end{tabular}
\end{ruledtabular}
\end{table}

\end{widetext}

\end{document}